\documentclass[a4paper, 12pt]{article}

\pdfoutput = 1

\usepackage{styleBuding}
\usepackage{bm}
\usepackage{color}
\usepackage[usenames,dvipsnames,svgnames,table]{xcolor}
\usepackage{amsmath}
\usepackage{amssymb}
\usepackage{graphicx}
\usepackage{slashed}
\usepackage{soul}
\usepackage{multirow}
\usepackage{subfigure}
\usepackage{siunitx} 
\usepackage{url} 
\usepackage[utf8]{inputenc}

\newcommand{\rom}[1]{\uppercase\expandafter{\romannumeral #1\relax}}

\let\jnfont=\rm
\def\NPB#1,{{\jnfont Nucl.\ Phys.\ B }{\bf #1},}
\def\PLB#1,{{\jnfont Phys.\ Lett.\ B }{\bf #1},}
\def\EPJC#1,{{\jnfont Eur.\ Phys.\ Jour.\ C }{\bf #1},}
\def\PRD#1,{{\jnfont Phys.\ Rev.\ D }{\bf #1},}
\def\PRL#1,{{\jnfont Phys.\ Rev.\ Lett.\ }{\bf #1},}
\def\MPLA#1,{{\jnfont Mod.\ Phys.\ Lett.\ A }{\bf #1},}
\def\JPG#1,{{\jnfont J.\ Phys.\ G}{\bf #1},}
\def\CTP#1,{{\jnfont Commun.\ Theor.\ Phys.\ }{\bf #1},}
\def\ZPC#1,{{\jnfont Z.\ Phys.\ C }{\bf #1},}
\def\JHEP#1,{{\jnfont JHEP \ }{\bf #1},}

\title{Impact of leptonic unitarity and dark matter direct detection experiments on the NMSSM with inverse seesaw mechanism}
\author{Junjie Cao$^{a,b}$, Yangle He$^a$,  Yusi Pan$^a$, Yuanfang Yue$^a$,  Haijing Zhou$^a$, Pengxuan Zhu$^{a}$}
\affiliation{ $^a$ School of Physics, Henan Normal University, Xinxiang 453007, China}
\affiliation{ $^b$  Center for High Energy Physics, Peking University, Beijing 100871, China}
\emailAdd{junjiec@alumni.itp.ac.cn}
\emailAdd{heyangle@htu.edu.cn}
\emailAdd{panyusi0406@foxmail.com}
\emailAdd{yueyuanfang@htu.edu.cn}
\emailAdd{zhouhaijing0622@163.com}
\emailAdd{zhupx99@icloud.com}
\abstract{In the Next-to-Minimal Supersymmetric Standard Model with the inverse seesaw mechanism to generate neutrino masses, the lightest sneutrino may act as a feasible dark matter candidate in vast parameter space. In this case, the smallness of the leptonic unitarity violation and the recent XENON-1T experiment can limit the dark matter physics. In particular, they set upper bounds of the neutrino Yukawa couplings $\lambda_\nu$ and $Y_\nu$. We study such effects by encoding the constraints in a likelihood function and carrying out elaborated scans over the parameter space of the theory with the Nested Sampling algorithm. We show that these constraints are complementary to each other in limiting the theory, and in some cases, they are very strict. We also study the impact of the future LZ experiment on the theory.
}

\begin{document}
    \maketitle
    \flushbottom

\section{Introduction}
As the most popular ultraviolet-complete Beyond Standard Model, the Minimal Supersymmetric Standard Model (MSSM) with R-parity conservation predicts two kinds of electric neutral, possibly stable and weakly interactive massive particles, namely, sneutrino and neutralino, which may act as dark matter (DM) candidates~\cite{Hagelin:1984wv,Jungman:1995df}. In the 1990s, it was proven that the left-handed sneutrino as the lightest supersymmetric particle (LSP) predicted a much smaller relic abundance than the measured value as well as an unacceptably tremendous DM-nucleon scattering rate due to its interaction with the Z boson~\cite{Falk:1994es,Arina:2007tm}. This fact made the lightest neutralino (usually with the bino field as its dominant component) the only reasonable DM candidate, and consequently, it was studied intensively since then. However, with the rapid progress in DM direct detection (DD) experiments in recent years, the candidate became more and more tightly limited by the experiments~\cite{Baer:2016ucr,Huang:2017kdh,Badziak:2017the,Cao:2019qng} assuming that it was fully responsible for the measured relic density and the higgsino mass $\mu$ was less than 300 GeV, which was favored to predict the Z boson mass naturally~\cite{Baer:2012uy}. These conclusions apply to the Next-to-Minimal Supersymmetric Standard Model (NMSSM)~\cite{Cao:2016cnv,Cao:2018rix,Abdallah:2019znp}, where the sneutrinos are purely left-handed, and the neutralino DM candidate may be either bino- or singlino-dominated~\cite{Ellwanger:2009dp}. In this context, we revived the idea of the sneutrino DM in a series of works~\cite{Cao:2017cjf,Cao:2019qng,Cao:2018iyk,Cao:2019ofo}. In particular, motivated by the phenomenology of the neutrino oscillations, we extended the NMSSM with the inverse seesaw mechanism by introducing two types of gauge singlet chiral superfields $\hat{\nu}_R$ and $\hat{X}$ for each generation matter, which have lepton numbers -1 and 1, respectively, and their fermion components corresponded to the massive neutrinos in literatures~\cite{Cao:2017cjf}. Subsequently, we studied in detail whether  the $\tilde{\nu}_R$ (the scalar component of $\hat{\nu}_R$) or $\tilde{x}$ (the scalar component of $\hat{X}$) dominated sneutrino could act as a feasible DM candidate~\cite{Cao:2017cjf}. We were interested in the inverse seesaw mechanism because it was a TeV scale physics to account for the neutrino oscillations and maybe experimentally testable soon. We showed by both analytic formulas and numerical calculations that the resulting theory (abbreviated as ISS-NMSSM hereafter) was one of the most economic framework to generate the neutrino mass and, meanwhile, to reconcile the DM DD experiments naturally~\cite{Cao:2017cjf,Cao:2019qng}. We add that, besides us, a lot of authors have showed interest in the sneutrino DM in recent years~\cite{An:2011uq,BhupalDev:2012ru,Frank:2017tsm,Araz:2017qcs,Ghosh:2017yeh,Chang:2017qgi,Chatterjee:2017nyx,Ferreira:2017osm,DelleRose:2017uas,Lara:2018rwv,Zhu:2018tzm,Chang:2018agk,
Banerjee:2018uut,Ghosh:2018hly,Ghosh:2018sjz,Choi:2018vdi,Kpatcha:2019gmq,Alonso-Alvarez:2019fym,Moretti:2019yln,Faber:2019mti}, but none of them considered the same theoretical framework as ours.

In the NMSSM, the introduction of the singlet field $\hat{S}$ can solve the $\mu$ problem of the MSSM~\cite{Ellwanger:2009dp}, enhance
the theoretical prediction of the SM-like Higgs boson mass~\cite{Hall:2011aa,Ellwanger:2011aa,Cao:2012fz}, as well as enrich the phenomenology of
the NMSSM significantly (see, for example, Ref.~\cite{Cao:2013gba,Ellwanger:2014hia,Cao:2014efa,Cao:2015loa,Cao:2016nix,Ellwanger:2018zxt}).
In the ISS-NMSSM, the $\hat{S}$ field also plays extraordinary roles in generating the massive neutrino mass by the Yukawa interaction $\lambda_\nu \hat{S} \hat{\nu}_R \hat{X}$ and making the sneutrino DM compatible with various measurements, especially the DM DD experiments~\cite{Cao:2017cjf}. There are at least two aspects in manifesting the latter role. One is that the newly introduced heavy neutrino superfields are singlet under the gauge group of the SM model. Thus, they can interact directly with $\hat{S}$ by the Yukawa couplings~\cite{Cao:2017cjf}.  In this case, the sneutrino DM candidate $\tilde{\nu}_1$, the singlet dominated scalars $h_s$ and $A_s$, and the massive neutrinos $\nu_h$ compose a roughly secluded DM sector where the annihilations $\tilde{\nu}_1 \tilde{\nu}_1^\ast \to A_s A_s, h_s h_s, \nu_h \bar{\nu}_h$ can produce the measured relic density (In the ISS-NMSSM, these annihilations proceed by quartic scalar interactions, $s$-channel exchange of $h_s$ and $t/u$-channel exchange of the sneutrinos or the singlino-dominated neutralino). Since this sector communicates with the SM sector by the small singlet-doublet Higgs mixing (dubbed by Higgs-portal in literatures~\cite{Patt:2006fw}) and/or by the massive neutrinos (neutrino-portal~\cite{Lindner:2010rr,Falkowski:2009yz,Macias:2015cna,Blennow:2019fhy}), the scattering of the DM with nucleons is naturally suppressed, which coincides with current DM DD results. The other aspect is that the singlet-dominated Higgs scalars can mediate the transition between $\tilde{\nu}_1$ pair and the higgsino pair, and consequently, these particles were in thermal equilibrium in early Universe before their freeze-out from the thermal bath. If their mass splitting is less than about $10\%$, the number density of the higgsinos can track that of  $\tilde{\nu}_1$ during the freeze-out~\cite{Coannihilation} (in literatures such a phenomenon was called co-annihilation \cite{Griest:1990kh}). Since, in this case, the couplings of $\tilde{\nu}_1$ with SM particles is usually very weak, the scattering is again naturally suppressed. We emphasize that, in either case, the suppression of the scattering prefers a small higgsino mass that appears in the coupling of  $\tilde{\nu}_1^\ast \tilde{\nu}_1$ state with Higgs bosons, and hence, there is no tension between the DM DD experiments and the naturalness for the mass of the Z boson~\cite{Cao:2017cjf}.

In the ISS-NMSSM, the rates of the DM annihilation and the DM-nucleon scattering depend on the coupling strength of $\tilde{\nu}_1$ interacting with Higgs fields, i.e., the Yukawa couplings $\lambda_\nu$ and $Y_\nu$  (the coefficient for $\hat{\nu}_L \cdot \hat{H}_u \hat{\nu}_R$ interaction) and their corresponding soft-breaking trilinear parameters $A_{\lambda_\nu}$ and $A_{Y_\nu}$. They also depend on the Higgs mass spectrum and the mixing between the Higgs fields that are ultimately determined by the parameters in the Higgs sector~\cite{Cao:2017cjf}. As such, the DM physics is quite complicated and is difficult to understand intuitively. This fact inspired us to study the theory
from different aspects, e.g., from the features of the DM-nucleon scattering~\cite{Cao:2017cjf,Cao:2019qng}, and its capability to explain the muon anomalous magnetic momentum~\cite{Cao:2019evo} or other anomalies at the LHC~\cite{Cao:2019ofo}. In this work, we noted that a large $\lambda_\nu$ or $Y_\nu$ can enhance the DM-nucleon scattering rate significantly, so the recent XENON-1T experiment should limit them~\cite{Aprile:2018dbl}. We also noted that the upper bound on the unitarity violation in neutrino sector sets a specific correlation between the couplings $\lambda_\nu$ and $Y_\nu$~\cite{Baglio:2016bop}, which can limit the parameter space of the ISS-NMSSM. Since these issues were not studied before, we decided to survey the impact of the leptonic unitarity and current and future DM DD experiments on the sneutrino DM sector. We will show that they are complementary to each other in limiting the theory, and in some cases, the constraints are rather tight. It is evident that such a study helps improve the understanding of the theory, and may be treated as a preliminary work before more comprehensive studies in the future.

We organize this work as follows. In section~\ref{Section-Model}, we briefly introduce the theory of the ISS-NMSSM. In section~\ref{Section-Scan}, we describe the
strategy to study the constraints, present numerical results and reveal the underlying physics. Finally, we draw our conclusions in section~\ref{Section-Conclusion}.

\vspace{-0.2cm}

\section{\label{Section-Model}NMSSM with inverse seesaw mechanism}

Since the ISS-NMSSM has been introduced in detail in~\cite{Cao:2017cjf,Cao:2019qng}, we only recapitulate its key features in this section.

\subsection{Basics of the ISS-NMSSM}

The renormalizable superpotential and the soft breaking terms of the ISS-NMSSM take following form~\cite{Cao:2017cjf}
\begin{eqnarray}
 W &= & \left [ W_{\rm MSSM}+\lambda\,\hat{s}\,\hat{H}_u \cdot \, \hat{H}_d\,
 +\frac{1}{3} \kappa \,\hat{s}^3 \right ] +  \left [\frac{1}{2} \mu_X \,\widehat X\,\widehat X\,+  \lambda_\nu \,\hat{s}\,\hat{\nu}_R \,\widehat X\,
 +Y_\nu \,\hat{l} \cdot \hat{H}_u \,\hat{\nu}_R \right ], \nonumber  \\
L^{soft} &=& \left [ L_{\rm MSSM}^{soft} - m_S^2 |S|^2 -  \lambda A_{\lambda} S H_u\cdot H_d - \frac{\kappa}{3} A_{\kappa} S^3 \right ] \nonumber  \\
&& - \left [ m_{\tilde{\nu}}^{2} \tilde{\nu}_{R}\tilde{\nu}^*_{R} +  m_{\tilde{x}}^{2} \tilde{x}\tilde{x}^* + \frac{1}{2} B_{\mu_X} \tilde{x} \tilde{x} +  (\lambda_{\nu} A_{\lambda_\nu} S \tilde{\nu}^*_{R} \tilde{x} + Y_\nu A_{Y_\nu} \tilde{\nu}^*_{R} \tilde{l} H_u + \mbox{h.c.}) \right ], \nonumber
\end{eqnarray}
where $W_{\rm MSSM}$ and $ L_{\rm MSSM}^{soft}$ represent the corresponding terms of the MSSM without the $\mu$-term. The terms in the first brackets on the right side of each equation make up the Lagrangian of the NMSSM that involves the Higgs coupling coefficients $\lambda$ and $\kappa$ and their soft-breaking parameters $A_\lambda$ and $A_\kappa$.  The terms in the second brackets are needed to implement the supersymmetric inverse seesaw mechanism. Coefficients such as the neutrino mass term $\mu_X$, the Yukawa couplings $\lambda_\nu$ and $Y_\nu$, and the soft-breaking parameters $A_{\lambda_\nu}$, $A_{Y_\nu}$, $B_{\mu_X}$, $m_{\tilde{\nu}}$, and $m_{\tilde{x}}$ are all
$3 \times 3 $ matrices in flavor space. Besides, among the parameters in the superpotential, only the matrix  $\mu_X$ is dimensional, and it parameterizes
the effect of lepton number violation (LNV). Since this matric
arises from the integration of massive particles in the  high-energy ultraviolet theory with LNV interactions
(see, for example, \cite{ISS-NMSSM-1,ISS-NMSSM-3,ISS-NMSSM-2}), its magnitude should be small.
Based on a similar perspective, the matrix $B_{\mu_X}$ is also theoretically favored to be suppressed.

It is the same as the NMSSM that the ISS-NMSSM predicts three CP-even Higgs bosons, two CP-odd Higgs bosons, a pair of charged Higgs bosons, and five neutralinos.
Throughout this work, we take $\lambda$, $\kappa$, $\tan \beta \equiv v_u/v_d$, $A_\lambda$, $A_\kappa$, and $\mu \equiv \lambda v_s/\sqrt{2} $ as inputs of the Higgs sector, where $ v_u \equiv \sqrt{2} \langle H_u \rangle $, $ v_d \equiv \sqrt{2} \langle H_d \rangle $,
and $v_s \equiv \sqrt{2} \langle S \rangle $ denote the vacuum expectation values (vev) of the fields $H_{u}$, $H_d$, and $S$, respectively.
The elements of the CP-even Higgs fields' squared mass matrix in the bases ($S_1 \equiv \cos \beta {\rm Re}[H_u^0] - \sin \beta {\rm Re}[H_d^0]$,
$S_2 \equiv \sin \beta {\rm Re}[H_u^0] + \cos \beta {\rm Re}[H_d^0]$, $S_3 \equiv {\rm Re}[S]$)
are given by~\cite{Ellwanger:2009dp,Cao:2012fz}
\begin{eqnarray}
{\cal M}^2_{11}&=& \frac{2 \mu (\lambda A_\lambda + \kappa \mu)}{\lambda \sin 2 \beta} + \frac{1}{2} (2 m_Z^2- \lambda^2v^2)\sin^22\beta, \nonumber \\
{\cal M}^2_{12}&=&-\frac{1}{4}(2 m_Z^2-\lambda^2v^2)\sin4\beta, \nonumber \\
{\cal M}^2_{13}&=&-\sqrt{2} ( \lambda A_\lambda + 2 \kappa \mu) v \cos 2 \beta, \nonumber \\
{\cal M}^2_{22}&=&m_Z^2\cos^22\beta+ \frac{1}{2} \lambda^2v^2\sin^22\beta,\nonumber  \\
{\cal M}^2_{23}&=& \frac{v}{\sqrt{2}} \left[2 \lambda \mu - (\lambda A_\lambda + 2 \kappa \mu) \sin2\beta \right], \nonumber \\
{\cal M}^2_{33}&=& \frac{\lambda A_\lambda \sin 2 \beta}{4 \mu} \lambda v^2   + \frac{\mu}{\lambda} (\kappa A_\kappa +  \frac{4 \kappa^2 \mu}{\lambda} ), \label{Mass-CP-even-Higgs}
\end{eqnarray}
where $S_1$ denotes the heavy doublet Higgs field with a vanishing vev, $S_2$ represents the SM Higgs field with its vev $v \equiv 246~{\rm GeV}$, ${\cal M}_{22}$ is the mass of $S_2$ at tree level without considering its mixing with the other bases, and ${\cal M}_{23}$ characterizes the mixing of $S_2$ with the singlet field $S_3$.

The squared mass matrix in Eq.~(\ref{Mass-CP-even-Higgs}) can be diagonalized by a unitary matrix $U$, and its eigenstates $h_i$ with $i=1,2,3$ are obtained by
\begin{eqnarray}
h_i = \sum_{j=1}^3 U_{ij} S_j,
\end{eqnarray}
where $h_i$ are labelled in an ascending mass order, i.e. $m_{h_1} < m_{h_2} < m_{h_3}$.  Then the couplings of $h_i$ to vector
bosons $W$ and $Z$ and fermions $u$ and $d$ quarks, which are normalized to their SM predictions, take the following form~\cite{Cao:2014kya}
\begin{eqnarray}
\bar{C}_{h_i V^\ast V} = U_{i2}, \quad \quad \bar{C}_{h_i \bar{u} u} = U_{i1} \cot \beta + U_{i2}, \quad \quad \bar{C}_{h_i \bar{d} d} =  - U_{i1} \tan \beta + U_{i2}.
\end{eqnarray}
Obviously, $U_{i2}\simeq 1$ if the components of the particle $h_i$ are far dominated by $S_2$, and consequently, $\bar{C}_{h_i V^\ast V} \simeq \bar{C}_{h_i \bar{u} u} \simeq \bar{C}_{h_i \bar{d} d} \simeq 1$. We call $h_i$ as the SM-like Higgs boson.

Similarly, the elements of the CP-odd Higgs fields' squared mass matrix are~\cite{Ellwanger:2009dp}
\begin{eqnarray}
{\cal M}^2_{P,11}&=& \frac{2 \mu (\lambda A_\lambda + \kappa \mu)}{\lambda \sin 2 \beta}, \nonumber  \\
{\cal M}^2_{P,22}&=& \frac{(\lambda A_\lambda + 3 \kappa \mu) \sin 2 \beta }{4 \mu} \lambda v^2  - \frac{3 \mu}{\lambda} \kappa A_\kappa, \nonumber  \\
{\cal M}^2_{P,12}&=& \frac{v}{\sqrt{2}} ( \lambda A_\lambda - 2 \kappa \mu), \label{Mass-CP-odd-Higgs}
\end{eqnarray}
in the bases ($A \equiv \cos \beta {\rm Im}[H_u^0] + \sin \beta {\rm Im}[H_d^0]$, ${\rm Im}[S]$). As a result, the two CP-odd mass eigenstates $A_1$ and $A_2$ are the mixtures of $A$ and ${\rm Im}[S]$. The charged Higgs are given by $H^\pm = \cos \beta H_u^\pm + \sin \beta H_d^\pm$, and their masses are $m_{H^\pm} = 2 \mu (\lambda A_\lambda + \kappa \mu)/(\lambda \sin 2 \beta) + v^2 (g^2/2 - \lambda^2) $.

Concerning the neutralinos, they are the mixtures of the bino field $\tilde{B}^0$, the wino field $\tilde{W}^0$, the Higgsinos fields $\tilde{H}_{d}^0$ and $\tilde{H}_u^0$, and
the singlino field $\tilde{S}^0$.  In the bases $\psi^0 = (-i \tilde{B}^0, - i \tilde{W}^0, \tilde{H}_{d}^0, \tilde{H}_{u}^0, \tilde{S}^0)$, their mass
matrix is given by~\cite{Ellwanger:2009dp}
\begin{equation}
{\cal M} = \left(
\begin{array}{ccccc}
M_1 & 0 & -\frac{g^\prime v_d}{\sqrt{2}} & \frac{g^\prime v_u}{\sqrt{2}} & 0 \\
  & M_2 & \frac{g v_d}{\sqrt{2}} & - \frac{g v_u}{\sqrt{2}} &0 \\
& & 0 & -\mu & -\lambda v_u \\
& & & 0 & -\lambda v_d\\
& & & & \frac{2 \kappa}{\lambda} \mu
\end{array}
\right), \label{eq:MN}
\end{equation}
where $M_1$ and $M_2$ are soft breaking masses of the gauginos. It can be diagonalized by a rotation matrix $N$ so that the mass eigenstates are
\begin{eqnarray}
\tilde{\chi}_i^0 = N_{i1} \psi^0_1 +   N_{i2} \psi^0_2 +   N_{i3} \psi^0_3 +   N_{i4} \psi^0_4 +   N_{i5} \psi^0_5.
\end{eqnarray}
It is evident that $N_{i3}$ and $N_{i4}$ characterize the $\tilde{H}_d^0$ and $\tilde{H}_u^0$ components in $\tilde{\chi}_i^0$, respectively, and $N_{i5}$ denotes the singlino component.

\begin{table}[t]
\begin{center}
   \resizebox{1 \textwidth}{!}{
    \begin{tabular}{c|c|c|c|c|c|c|c|c|c|c|c}
    \hline \hline
    \multicolumn{6}{c|}{Light $h_s$ scenario with $A_\lambda = 2000~{\rm GeV}$} & \multicolumn{6}{c}{Massive $h_s$ scenario with $A_\lambda = 2000~{\rm GeV}$} \\ \hline
    $\tan \beta $ & 12.38 & $\lambda$ & 0.24 & $\kappa$ & 0.23 & $\tan \beta $ & 28.46 & $\lambda$ & 0.19 & $\kappa$ & 0.60 \\
    $A_t$ & 2433 & $A_\kappa$ & -680.4 & $\mu$ & 195.2 & $A_t$ & 2363 & $A_\kappa$ & -120.4 & $\mu$ & 328.0 \\
    $m_{h_s}$ & 120.4 & $m_{h}$ & 125.1 & $m_{H}$ & 2332 & $m_{h}$ & 125.1 & $m_{h_s}$ & 2042 & $m_{H}$ & 5381 \\
    $m_{A_s}$ & 608.5 & $m_{A_H}$ & 2331 & $m_{H^\pm}$ & 2332 & $m_{A_s}$ & 592.6 & $m_{A_H}$ & 5381 & $m_{H^\pm}$ & 5379 \\
    $m_{{\tilde{\chi}}_1^0}$ & 186.0 & $m_{{\tilde{\chi}}_2^0}$ & -206.1 & $m_{{\tilde{\chi}}_1^{\pm}}$ & 197.9 & $m_{{\tilde{\chi}}_1^0}$ & 318.8 & $m_{{\tilde{\chi}}_2^0}$ & -341.3 & $m_{{\tilde{\chi}}_1^{\pm}}$ & 333.8 \\
    $U_{11}$ & -0.018 & $U_{12}$ & -0.261 & $U_{13}$ & 0.965 & $U_{11}$ & 0.00001 & $U_{12}$ & -0.999 & $U_{13}$ & 0.003 \\
    $U_{21}$ & -0.005 & $U_{22}$ & 0.965 & $U_{23}$ & 0.260 & $U_{21}$ & 0.006 & $U_{22}$ & -0.003 & $U_{23}$ & -0.999 \\
    $U_{31}$ & 0.999 & $U_{32}$ & 0.000 & $U_{33}$ & 0.019 & $U_{31}$ & -0.999 & $U_{32}$ & -0.000 & $U_{33}$ & -0.006  \\
    $\bar{C}_{h_s V^\ast V}$ & -0.261 & $\bar{C}_{h_s \bar{u} u}$ & -0.263 & $\bar{C}_{h_s \bar{d} d}$ & -0.038 & $\bar{C}_{h_s V^\ast V}$ & -0.003 & $\bar{C}_{h_s \bar{u} u}$ & -0.003 & $\bar{C}_{h_s \bar{d} d}$ & -0.162 \\
    $\bar{C}_{h V^\ast V}$ & 0.965 & $\bar{C}_{h \bar{u} u}$ & 0.965 & $\bar{C}_{h \bar{d} d}$ & 1.03 & $\bar{C}_{h V^\ast V}$ & -0.999 & $\bar{C}_{h \bar{u} u}$ & -0.999 & $\bar{C}_{h \bar{d} d}$ & -1.00 \\
    $\bar{C}_{H V^\ast V}$ & 0.000 & $\bar{C}_{H \bar{u} u}$ & 0.081 & $\bar{C}_{H \bar{d} d}$ & -12.4 & $\bar{C}_{H V^\ast V}$ & -0.000 & $\bar{C}_{H \bar{u} u}$ & -0.035 & $\bar{C}_{H \bar{d} d}$ & 28.5 \\
   $\bar{C}_{h_s h_s h_s}$ & 4.014 & $\bar{C}_{h_s h_s h}$ & 1.193 & $\bar{C}_{h_s h h}$ & -0.460 & $\bar{C}_{h_s h_s h_s}$ & -163.4 & $\bar{C}_{h_s h_s h}$ & 0.448 & $\bar{C}_{h_s h h}$ & -1.591 \\
    $N_{13}$ & -0.696 & $N_{14}$ & 0.681 & $N_{15}$ & -0.147 & $N_{13}$ & 0.675 & $N_{14}$ & -0.641 & $N_{15}$ & 0.012 \\
    $N_{23}$ & -0.699 & $N_{24}$ & -0.709 & $N_{25}$ & -0.055 & $N_{23}$ & -0.702 & $N_{24}$ & -0.709 & $N_{25}$ & -0.010 \\
    \hline \hline
    \end{tabular}
    }
\end{center}
\caption{Specific configuration of the Higgs and neutralino sectors for the scenarios discussed in the text, and their prediction on the properties of the Higgs bosons and neutralinos such as the mass spectrum and
the couplings of the Higgs bosons with different particles, $\bar{C}_{ijk}$, which are normalized to their corresponding SM predictions. Parameters with mass dimensions are in the unit of GeV. Other fixed parameters that are not listed in the table include $m_{\tilde{q}} = 2000~{\rm GeV}$ for flavor universal soft-breaking masses of squarks,
$M_1 = M_2 =2000~{\rm GeV}$ and $M_3 = 5000~{\rm GeV}$ for gaugino masses, $A_i = 0$ for all soft-breaking trilinear coefficients except for $A_\lambda$ and
$A_t$, and $[Y_\nu]_{11,22} = 0.01$, $[\lambda_{\nu}]_{11,22} = 0.3$, and $[m_{\tilde{\nu}}]_{11,22} = [m_{\tilde{x}}]_{11,22} = 2000~{\rm  GeV}$ for the parameters of the first two generations of the sneutrinos.
All these parameters are defined at the scale $Q = 1000~{\rm GeV}$. Besides, the Higgs masses and
$U_{ij}$ are obtained by setting $[Y_\nu]_{33} = [\lambda_\nu]_{33} =0$, and sneutrino loop effects
may slightly change them when varying the parameters in the sneutrino DM sector.}
\label{table1}
\end{table}

In this work, we use the following features in the Higgs and neutralino sectors:
\begin{itemize}
\item A CP-even state corresponds to the SM-like Higgs boson discovered at the LHC.  This state is favored to be
$ {\rm Re}[H_u^0]$-dominated by the LHC data when $\tan \beta \gg 1$, and its mass may be significantly affected
by the interaction $\lambda\,\hat{s}\,\hat{H}_u \cdot \, \hat{H}_d$, the
doublet-singlet Higgs mixing as well as the radiative correction from top/stop loops~\cite{Hall:2011aa,Ellwanger:2011aa,Cao:2012fz}.
In the following, we denote this state as $h$.
\item In most cases, the heavy doublet-dominated CP-even state is mainly composed of the field ${\rm Re}[H_d^0]$. It roughly degenerates in mass with the doublet-dominated CP-odd state and also with the charged states. The LHC search for extra Higgs bosons and the $B$-physics measurements requires these states to be heavier than about $500~{\rm GeV}$~\cite{Bagnaschi:2017tru}. We represent them by $H$, $A_H$, and $H^\pm$.
\item Concerning the singlet-dominated states, they may be very light without conflicting with any collider constraint. As we introduced before, these states may appear as the final state of the sneutrino pair annihilation or mediate the annihilation,  and thus, they can play a vital role in the sneutrino DM physics. In this work, we label these states by $h_s$ and $A_s$.
\item The lightest neutralino $\tilde{\chi}_1^0$ is Higgsino dominated if $|\mu| < |M_1|, |M_2|$ and $|2 \kappa/\lambda| <1$. In this case, $|N_{13}| \simeq |N_{14}| \simeq \sqrt{2}/2$.
\end{itemize}

We add that, to study the property of the sneutrino DM, we consider two benchmark scenarios where all the input parameters for the Higgs and neutralino sectors are fixed. The details of the scenarios are presented in Table \ref{table1}. For the first scenario, $h_s$ and the SM-like Higgs boson $h$ correspond to the lightest and the next-to-lightest CP-even Higgs bosons $h_1$ and $h_2$. The $S_2$ component of $h_s$ is measured by the rotation element $U_{12}$, which is determined by the elements ${\cal{M}}_{23}^2$ and ${\cal{M}}_{33}^2$ in Eq.(\ref{Mass-CP-even-Higgs}). We dub this scenario light $h_s$ scenario. By contrast,  we call the second scenario as the massive $h_s$ scenario. It predicts $h=h_1$, $h_s = h_2$, and $U_{22}$ to characterize the $S_2$ component in $h_s$. Besides, we note that triple Higgs interactions may play an essential
role in the sneutrino DM annihilation. So in addition to the couplings $\bar{C}_{h_i V^\ast V}$, $\bar{C}_{h_i \bar{u} u}$ and $\bar{C}_{h_i \bar{d} d}$, we also list in Table \ref{table1} the coupling strengths for $h_s h_s h_s$, $h_s h_s h$ and $h_s h h$ interactions, which are normalized to the triple Higgs coupling in the SM and denoted by $\bar{C}_{h_s h_s h_s}$, $\bar{C}_{h_s h_s h}$ and $\bar{C}_{h_s h h}$, respectively. These strengths are obtained by the formulas in~\cite{Ellwanger:2009dp}. They are characterized by $|\bar{C}_{h_s h_s h_s}| \gg |\bar{C}_{h_s h_s h}|, |\bar{C}_{h_s h h}|$, which is evident by the superpotential and the soft breaking terms of the ISS-NMSSM.

\subsection{Leptonic Unitarity}

In the interaction bases $(\nu_L, \nu_R^\ast, x)$,
the neutrino mass matrix is given by~\cite{Cao:2017cjf}
\begin{eqnarray}
\label{ISSmatrix}
 M_{\mathrm{ISS}}=\left(\begin{array}{c c c} 0 & M_D^T & 0 \\ M_D & 0 & M_R \\ 0 & M_R^T & \mu_X \end{array}\right)\,, \label{Neutrino-mass}
\end{eqnarray}
where both the Dirac mass $M_D = Y_\nu v_u/\sqrt{2} $ and the Majorana mass $M_R = \lambda_\nu v_s/\sqrt{2}$ are $3 \times 3$ matrix in the flavor space.
One can diagonalize this mass matrix by a $9 \times 9$ unitary matrix $U_{\nu}$ to obtain three light neutrinos $\nu_i$ $(i=1,2,3)$ and six massive neutrinos
$\nu_{h}$ as mass eigenstates, i.e., $U^\ast_\nu  M_{\mathrm{ISS}} U^\dag_\nu = {\rm diag}(m_{\nu_i},m_{\nu_{h}})$, and decompose $U_\nu$ into the following blocks:
\begin{eqnarray}
\left ( U_{\nu}^\dag \right )_{9\times 9} = \left(\begin{array}{cc}
\hat{U}_{3\times 3} & X_{3\times 6}\\
Y_{6\times 3} & Z_{6\times 6}
\end{array}\right).  \label{eq:diagfull}
\end{eqnarray}
The sub-matrix $\hat{U}_{3 \times 3}$ encodes the neutrino oscillation information and it is determined by
the neutrino experimental results.

Alternatively,  one can get the analytic expression of the light active neutrinos' mass matrix from Eq.~(\ref{Neutrino-mass})
in the limit $\|\mu_X\| \ll \|M_D\| \ll \|M_R\|$, where $\|M\|$ is defined by $\|M\| \equiv \sqrt{{\rm Tr}(M^\dag M)}$ for an arbitrary matrix $M$. The result is
\begin{eqnarray}
	M_\nu = \left[M_D^T M_R^{T^{-1}}\right]\mu_X \left[(M_R^{-1})M_D\right]+{\cal O}(\mu_{X}^2) \equiv F \mu_X F^T + {\cal O}(\mu_{X}^2) \, .   \label{Active Neutrino mass}
\end{eqnarray}
where $F = M_D^T M_R^{T^{-1}}$, and its elements' magnitude is of the order $\|M_D\|/\|M_R\|$.
This $3 \times 3$ matrix can be diagonalized by the unitary Pontecorvo-Maki-Nakagawa-Sakata (PMNS)
matrix, i.e.,
\begin{eqnarray}
 U_{\rm PMNS}^T M_{\mathrm{\nu}} U_{\rm PMNS} = \mathrm{diag}(m_{\nu_1}\,, m_{\nu_2}\,, m_{\nu_3})\,.
\end{eqnarray}
Due to the mixings among the states $(\nu_L, \nu_R^\ast, x)$,  the matrix $\hat{U}$ in
Eq.(\ref{eq:diagfull}) does not coincide with $U_{\rm PMNS}$. Instead, they are related by~\cite{Fernandez-Martinez:2016lgt}
\begin{eqnarray}
\hat{U} \simeq \left({\bf 1} - \frac{1}{2}F F^\dag \right)U_{\rm PMNS} \equiv
({\bf 1}-\eta) U_{\rm PMNS},  \label{non-unitarity}
\end{eqnarray}
where $\eta=\frac{1}{2}FF^\dag$ is a measure of the non-unitarity for the matrix $\hat{U}$. A recent global fit of the theory to
low energy experimental data reveals that~\cite{Fernandez-Martinez:2016lgt}
\begin{eqnarray}
    \sqrt{2 |\eta|_{ee}} < 0.050,   \sqrt{2 |\eta|_{\mu\mu}} < 0.021, \sqrt{2 |\eta|_{\tau\tau}} < 0.075, \nonumber \\
    \sqrt{2 |\eta|_{e\mu}} < 0.026, \sqrt{2 |\eta|_{e\tau}} < 0.052,  \sqrt{2 |\eta|_{\mu\tau}} < 0.035.
    \label{etaconstraint}
\end{eqnarray}
We call these inequalities as the leptonic unitarity constraint.

Eq.~(\ref{Active Neutrino mass}) indicates that the tininess of the active neutrino masses in the inverse seesaw mechanism is due to the
smallness of the lepton-number violating matrix $\mu_X$ and the suppression factor $\|M_D\|^2/\|M_R\|^2$. For $\|\mu_X \| \lesssim {\cal O}({\rm keV})$
and $\|M_R \| \sim {\cal O}$(TeV), the magnitude of the Dirac Yukawa coupling $Y_\nu$ may reach order one in predicting $m_{\nu_i} \sim 0.1~{\rm eV}$.
However, a large $Y_\nu$ may conflict with the unitarity constraint once the Majorana mass $M_R$
is specified. So the constraint must be taken into account in phenomenological study.

In the following, we will discuss the application of the unitarity constraint in DM physics. After noticing that the neutrino oscillation phenomenon can be explained
by choosing an appropriate $\mu_X$~\cite{Arganda:2014dta,Baglio:2016bop}, we assume flavor diagonal $Y_\nu$ and $\lambda_\nu$ to simplify the DM physics (see discussion below). We then determine $\mu_X$ by the formula~\cite{Arganda:2014dta,Baglio:2016bop}
\begin{eqnarray}
\mu_X=M_R^T ~m_D^{T^{-1}}~ \hat{U}_{\rm 3\times 3}^*  \mathrm{Diag}(m_{\nu_1}\,, m_{\nu_2}\,, m_{\nu_3})   \hat{U}_{\rm 3 \times 3}^\dagger~ {m_D}^{-1} M_R,  \nonumber
\end{eqnarray}
where $m_{\nu_i}$ and $\hat{U} \simeq U_{\rm PMNS}$ take the values extracted from relevant neutrino experiments. With the assumption,  the neutrino oscillation is
solely attributed to the non-diagonality of $\mu_X$, and the unitarity constraint in Eq.~(\ref{etaconstraint}) becomes
\begin{eqnarray}
\left | \frac{[\lambda_{\nu}]_{11} \mu}{{[Y_{\nu}]_{11} \lambda v_u}} \right | > 14.1, \quad \left | \frac{[\lambda_{\nu}]_{22} \mu}{{[Y_{\nu}]_{22} \lambda v_u}} \right | > 33.7, \quad \left | \frac{[\lambda_{\nu}]_{33} \mu}{{[Y_{\nu}]_{33} \lambda v_u}} \right |  > 9.4.  \label{unitaryconstriants}
\end{eqnarray}
These inequalities reveal that the ratio $[\lambda_{\nu}]_{33}/[Y_{\nu}]_{33}$ may be significantly smaller than $[\lambda_{\nu}]_{11}/[Y_{\nu}]_{11}$ and $[\lambda_{\nu}]_{22}/[Y_{\nu}]_{22}$ for fixed $\lambda$, $\mu$, and $v_u$, or equally speaking, $[Y_\nu]_{33}$ may be much larger than $[Y_\nu]_{11}$ and $[Y_\nu]_{22}$ when $\lambda_\nu$ is proportional to identity matrix.

Concerning the LNV coefficients $\mu_X$ and $B_{\mu_X}$, one should note two points. One is that $B_{\mu_X}$ can induce an effective $\mu_X$ through sneutrino-singlino loops to significantly affect the active neutrino masses by Eq.~(\ref{Active Neutrino mass}). We estimate the correction by the mass insertion method, which was widely used in B physics study. We find
\begin{eqnarray}
\delta \mu_X \sim \frac{1}{16 \pi^2} \frac{\lambda_\nu M^2 B_{\mu_X} M^2 \lambda_\nu}{M_{\rm SUSY}^5},
\end{eqnarray}
where $M^2$ parameterizes the mixing of the field $\tilde{\nu}_R^\ast$ with the field $\tilde{x}$, and $M_{\rm SUSY}$ represents the sparticles' mass scale. Under the premise that $Y_\nu$, $\lambda_\nu$, and $B_{\mu_X}$ are flavor diagonal, $M^2$ can be roughly flavor diagonal, too  (see the discussion in the next section). So one can study the correction in one generation case. The result is
\begin{eqnarray}
\delta \mu_X~(\rm keV) \sim 63.3 \times \left ( \frac{\lambda_\nu}{0.1} \right )^2 \left (\frac{M^2}{M_{\rm SUSY}^2} \right )^2 \left (\frac{B_{\mu_X}/GeV^2}{M_{\rm SUSY}/GeV} \right ),
\end{eqnarray}
which indicates that $B_{\mu_X}/{\rm GeV^2}$ may be comparable with $M_{\rm SUSY}/{\rm GeV}$ in getting $\delta \mu_X \sim 1~{\rm keV}$ for $\lambda_\nu = 0.1$ and $M^2 \lesssim 0.1 \times M_{\rm SUSY}^2$. Alternatively, if $\lambda_\nu = 0.5$ and $M^2 = 0.1 \times M_{\rm SUSY}^2$, the approximation requires $B_{\mu_X}/{\rm GeV^2} < 0.1 \times M_{\rm SUSY}/{\rm GeV}$ to get $\delta \mu_X \sim 1~{\rm keV}$. These estimations provide an upper bound on $B_{\mu_X}$'s magnitude. In our study, we limit $B_{\mu_X} \leq 100~{\rm GeV^2}$ for simplicity. The other point is that the LNV coefficients may induce sizable neutrinoless double beta decay since the inverse seesaw scale may be around several hundred GeV and the Yukawa couplings $Y_\mu$ and $\lambda_\nu$ may be moderately large. As indicated in~\cite{Haba:2016lxc}, because $\mu_X$ is related to the active neutrino mass, the decay rate is below current experiment sensitivity when the massive neutrinos are heavier than $1~{\rm GeV}$. So there is no need to consider the constraint in our study.

\subsection{Properties of Sneutrino Dark Matter}

If the sneutrino fields are decomposed into CP-even and CP-odd parts
\begin{eqnarray}
\tilde{\nu}_{L} =  \, \frac{1}{\sqrt{2}} \left ( \phi_1  + i \sigma_1 \right ),~~~~
\tilde{\nu}_{R}^\ast = \frac{1}{\sqrt{2}} \left (\phi_{2}  + i \sigma_{2} \right ), ~~~~ \tilde{x} = \frac{1}{\sqrt{2}}
\left ( \phi_{3}  + i \sigma_{3} \right ),
\end{eqnarray}
the squared mass of the CP-even fields is given by
\begin{eqnarray}
m^2_{\tilde{\nu}} = \left(
\begin{array}{ccc}
m_{11} \quad &m_{12} \quad & m_{13} \\
m_{12}^\ast \quad  &m_{22} \quad &m_{23}\\
m_{13}^\ast \quad &m_{23}^\ast \quad &m_{33} \end{array}
\right),
\label{CP-even elements}
 \end{eqnarray}
in the bases $(\phi_1, \phi_2, \phi_3)$, where
\begin{eqnarray}
m_{11} &=& \frac{1}{4} \left [ 2 v_{u}^{2} {{\rm Re} \Big({Y_{\nu}  Y_\nu^*}\Big)}  + 4 {{\rm Re} \Big(m_{\tilde{l}}^2\Big)} \right ]  + \frac{1}{8} \Big(g_{1}^{2} +
g_{2}^{2}\Big) \Big(- v_{u}^{2}  + v_{d}^{2}\Big) {\bf 1},  \nonumber \\
m_{12} &=& -\frac{1}{2} v_d v_s {{\rm Re} \Big(\lambda Y_\nu^\ast \Big)}  + \frac{1}{\sqrt{2}} v_u {{\rm Re}\Big(Y_\nu A_{Y_\nu} \Big)}, \nonumber  \\
m_{13} &=& \frac{1}{2} v_s v_u {{\rm Re}\Big({Y_{\nu}  \lambda_\nu^\ast}\Big)}, \nonumber \\
m_{22} &=& \frac{1}{4} \left [ 2 v_{s}^{2} {{\rm Re}\Big({\lambda_\nu  \lambda_{\nu}^{\ast}}\Big)}  + 2 v_{u}^{2} {{\rm Re}\Big({Y_\nu  Y_{\nu}^{\ast}}\Big)}
+ 4 {{\rm Re}\Big(m_{\tilde{\nu}}^2\Big)} \right ],  \nonumber \\
m_{23} &=& \frac{1}{8} \left \{ -2 v_d v_u \lambda \lambda_{\nu}  + 2 \left [ \Big(- v_d v_u \lambda  + v_{s}^{2} \kappa \Big)\lambda_{\nu}^{\ast}  + v_{s}^{2} \kappa \lambda_{\nu} \right ] \right . \nonumber \\
&& \quad \quad \quad \quad \quad \quad \quad \quad \quad \quad \quad \quad \left .  + \sqrt{2} v_s \left [ -4 {{\rm Re}\Big({\mu_X  \lambda_{\nu}^{\ast}}\Big)}
+ 4 {{\rm Re}\Big(A_{\lambda_\nu}^{\ast} \lambda_\nu \Big)} \right ] \right \}, \nonumber \\
m_{33} &=& \frac{1}{8} \Big [ 4 v_{s}^{2} {{\rm Re}\Big({\lambda_{\nu}  \lambda_\nu^\ast}\Big)} - 8 {{\rm Re}\Big(B_{\mu_X}\Big)}  + 8 {{\rm Re}\Big({\mu_X  \mu_X^\ast}\Big)}  + 8 {{\rm Re}\Big(m_{\tilde{x}}^2\Big)} \Big ].  \label{Matrix-elements}
\end{eqnarray}
These formulas indicate the following facts:
\begin{itemize}
\item The squared mass is a $9 \times 9$ matrix in three-generation $(\phi_1, \phi_2, \phi_3)$ bases.  It involves a series of $3 \times 3$ matrices in the flavor space, such as $Y_\nu$, $\lambda_\nu$, $A_{Y_\nu}$, $A_{\lambda_\nu}$, $\mu_X$, $B_{\mu_X}$, $m_{\tilde{l}}$, $m_{\tilde{\nu}}$, and $m_{\tilde{x}}$. Among these matrices, only $\mu_X$  must be flavor non-diagonal to account for the neutrino oscillations, but since its magnitude is less than 10~keV~\cite{Arganda:2014dta}, it can be neglected. Thus, if there is no flavor mixings for the other matrices, the squared mass is flavor diagonal, and one can adopt one-generation  $(\phi_1, \phi_2, \phi_3)$ bases in studying the mass. In this work, we only consider the third generation sneutrinos as DM candidates. This is motivated by that both the unitarity bound and the LHC constraint in sparticle search
    are weakest for the third generation~\cite{Cao:2017cjf}. When we mention the sneutrino parameters hereafter, we are actually referring to their 33 elements. Under the assumption, the squared mass is diagonalized by a $3 \times 3 $ unitary matrix $V$, which parameterizes the chiral mixings between the fields $\phi_1$, $\phi_2$ and $\phi_3$. Consequently, the sneutrino mass eigenstates are given by $\tilde{\nu}_{R,i} = V_{ij} \phi_j$ with $i,j=1$, 2, and 3. we add that $Y_\nu$ and $\lambda_\nu$ are  real and positive numbers after properly rotating the phase of the fields $\hat{\nu}_R$ and $\hat{X}$.

\item The mixing of $\phi_1$ with the other fields is determined by $Y_{\nu}$ and $A_\nu$. As $Y_{\nu}$ approaches zero, $|m_{12}|$ and $|m_{13}|$ diminish monotonically, and so is $|V_{11}|$ which represents the $\tilde{\nu}_L$ component in the lightest sneutrino state $\tilde{\nu}_{R,1}$. In the extreme case $Y_\nu = 0$,  all these quantities vanish and $\tilde{\nu}_{R,1}$ is merely the mixture of $\phi_2$ and $\phi_3$. Furthermore, if $\lambda_\nu/\lambda$ is moderately large, the first term in $m_{22}$ and $m_{33}$ may be far dominant over the other contributions so that $m_{22} \simeq m_{33}$. This results in maximal mixing between $\phi_2$ and $\phi_3$ and $\tilde{\nu}_{R, 1} \simeq 1/\sqrt{2} [ \phi_2 - {\rm sgn}(m_{23}) \phi_3 ] $~\cite{Cao:2017cjf}. This is a case encountered frequently in our study.
\end{itemize}

Similarly, one may adopt the one-generation $(\sigma_1,\sigma_2,\sigma_3)$ bases to study the CP-odd sneutrino's mass, which is the same as Eq.~(\ref{CP-even elements}) except for the substitution $B_{\mu_X} \to - B_{\mu_X}$. The mass eigenstates are then given by $\tilde{\nu}_{I,i} = V_{ij}^\prime \sigma_j$ ($i,j=1, 2 , 3$) where $V^\prime$ denotes the rotation of the CP-odd fields. Given that $B_{\mu_X}$ represents the degree of the LNV and is theoretically small, we are particularly interested in the following two cases:
\begin{itemize}
\item The extreme case of $B_{\mu_X} = 0$ where any CP-even sneutrino state is accompanied by a mass-degenerate CP-odd state. In this case, any sneutrino mass eigenstate corresponds to a complex field and it has an anti-particle~\cite{MSSM-ISS-1}. Concerning the sneutrino DM $\tilde{\nu}_1$, we have $\tilde{\nu}_{R,1} \equiv {\rm Re}[\tilde{\nu}_1]$, $\tilde{\nu}_{I,1} \equiv {\rm Im}[\tilde{\nu}_1]$, $V_{ij} = V^\prime_{ij}$, and $\tilde{\nu}_1$ and its anti-particle $\tilde{\nu}_1^\ast$ contribute equally to the relic density. This case is actually a two-component DM theory. It is notable that the $\tilde{\nu}_1^\ast \tilde{\nu}_1 Z$ coupling is proportional to $|V_{11}|^2$ and it contributes to the scattering of $\tilde{\nu}_1$ with nucleons. This effect is important when $|V_{11}| \sim 0.01$ (discussed below). It is also notable that the $\tilde{\nu}_1^\ast \tilde{\nu}_1 A_i$ coupling vanishes since it is induced only by the LNV effect.
\item A more general case satisfying $|B_{\mu_X}| \lesssim 100~{\rm GeV^2}$. It has four distinctive features. First, since $m_{\tilde{\nu}_{R,i}} > m_{\tilde{\nu}_{I,i}}$ when $B_{\mu_X} > 0 $, the DM candidate $\tilde{\nu}_1$ is identified as the $\tilde{\nu}_{I,1}$ state with a definite CP number -1. The opposite conclusion applies to $B_{\mu_X} < 0 $ case. Second, any CP-even state is slightly different from its CP-odd partner in mass, e.g., $m_{\tilde{\nu}_{R,1}} - m_{\tilde{\nu}_{I,1}}  \simeq 0.2~{\rm GeV}$ when $B_{\mu_X} = 100~{\rm GeV^2}$ and $m_{\tilde{\nu}_{R,1}} = 100~{\rm GeV}$, and so are the rotations $V$ and $V^\prime$. These sneutrino states compose a pseudo-complex particle ~\cite{MSSM-ISS-2,ISS-NMSSM-3,MSSM-ISS-10}. Third, given the approximate mass degeneracy, $\tilde{\nu}_{R,1}$ and $\tilde{\nu}_{I,1}$ always co-annihilated in early universe to affect the DM density. We will discuss this issue later. Finally, $Z$ boson does not mediate the DM-nucleon scattering any more since it couples only to a pair of sneutrino states with opposite CP numbers. It also contributes little to the DM annihilation because the $\tilde{\nu}_{R,1} \tilde{\nu}_{I,1} Z$ coupling is suppressed by a factor $V_{11}^\ast V^\prime_{11} \simeq |V_{11}|^2$.
\end{itemize}

We fix $B_{\mu_X} = 0$ or $B_{\mu_X} = - 100~{\rm GeV^2}$ in this work. In either case, the $\tilde{\nu}_1^\ast \tilde{\nu}_1 h_i$  ($h_i = h_s, h, H$) coupling coefficient is given by
\begin{eqnarray}
C_{\tilde{\nu}_1^\ast \tilde{\nu}_1 h_i} =  C_{\tilde{\nu}_1^\ast \tilde{\nu}_1 {\rm Re}[H_d^0]} \tilde{U}_{i1}  +  C_{\tilde{\nu}_1^\ast \tilde{\nu}_1 {\rm Re}[H_u^0]} \tilde{U}_{i2} +  C_{\tilde{\nu}_1^\ast \tilde{\nu}_1 {\rm Re}[S]} \tilde{U}_{i3},  \nonumber
\end{eqnarray}
where $\tilde{U}$ diagonalizes the CP-even Higgs fields' squared mass in $({\rm Re}[H_d^0], {\rm Re}[H_u^0], {\rm Re}[S])$ bases~\cite{Ellwanger:2011aa,Cao:2012fz}, and $C_{\tilde{\nu}_1^\ast \tilde{\nu}_1 s}$ on the right side denotes the sneutrino coupling to the scalar field $s$. For the one-generation sneutrino case, $C_{\tilde{\nu}_1^\ast \tilde{\nu}_1 s}$ is given by
\begin{eqnarray}
C_{\tilde{\nu}_1^\ast \tilde{\nu}_1 {\rm Re}[H_d^0]} &=& \lambda Y_\nu v_s V_{11} V_{12} + \lambda \lambda_\nu v_u V_{12} V_{13} - \frac{1}{4} (g_1^2 + g_2^2) v_d V_{11} V_{11}, \nonumber \\
C_{\tilde{\nu}_1^\ast \tilde{\nu}_1 {\rm Re}[H_u^0]} &=&  \lambda \lambda_\nu v_d V_{12} V_{13} - \sqrt{2} Y_\nu A_{Y_\nu} V_{11} V_{12} - Y_\nu^2 v_u  V_{11} V_{11} - \lambda_\nu Y_\nu v_s V_{11} V_{13} \nonumber \\ && \quad \quad - Y_\nu^2 v_u V_{12} V_{12} + \frac{1}{4} (g_1^2 + g_2^2) v_u V_{11} V_{11},  \nonumber \\
C_{\tilde{\nu}_1^\ast \tilde{\nu}_1 {\rm Re}[S]} &=& \lambda Y_\nu v_d V_{11} V_{12} - 2 \kappa \lambda_\nu v_s  V_{12} V_{13} - \sqrt{2} \lambda_\nu A_{\lambda_\nu} V_{12} V_{13}  + \sqrt{2} \lambda_\nu \mu_X V_{12} V_{13} \nonumber \\ && \quad \quad -  \lambda_\nu Y_\nu v_u V_{11} V_{13} - \lambda_\nu^2 v_s (V_{12} V_{12} + V_{13} V_{13} ).   \label{Coupling-expression}
\end{eqnarray}
These formulas indicate that the parameters $Y_\nu$, $\lambda_\nu$, $A_{Y_\nu}$, and $A_{\lambda_\nu}$ affect not only the sneutrino interactions but also their mass spectrum and mixing.  In particular, a large $\lambda_\nu$ or $Y_\nu$ can enhance the coupling significantly. Instead, the soft-breaking masses $m_{\tilde{\nu}}^2$ and $m_{\tilde{x}}^2$ affect only the latter property. For typical values of the parameters in Eq.~(\ref{Coupling-expression}),  e.g.,  $\tan \beta \gg 1$, $|V_{11}| < 0.1$, $Y_\nu, \kappa, \lambda, \lambda_\nu \sim {\cal{O}}(0.1)$ and $\lambda_\nu v_s, \lambda v_s, A_{Y_\nu}, A_{\lambda_\nu} \sim {\cal{O}}({\rm 100~GeV})$, $C_{\tilde{\nu}_1 \tilde{\nu}_1 s}$ is approximated by
\begin{eqnarray}
C_{\tilde{\nu}_1^\ast \tilde{\nu}_1 {\rm Re}[H_d^0]} &\simeq & \lambda Y_\nu v_s V_{11} V_{12} + \lambda \lambda_\nu v_u V_{12} V_{13}, \nonumber \\
C_{\tilde{\nu}_1^\ast \tilde{\nu}_1 {\rm Re}[H_u^0]} &\simeq &  - \sqrt{2} \lambda_\nu A_{Y_\nu} V_{11} V_{12} - \lambda_\nu Y_\nu v_s V_{11} V_{13} - Y_\nu^2 v_u V_{12} V_{12},  \nonumber \\
C_{\tilde{\nu}_1^\ast \tilde{\nu}_1 {\rm Re}[S]} &\simeq & - 2 \kappa \lambda_\nu v_s  V_{12} V_{13} - \sqrt{2} \lambda_\nu A_{\lambda_\nu} V_{12} V_{13} - \lambda_\nu^2 v_s.  \label{approximation-1}
\end{eqnarray}
It is estimated that $|C_{\tilde{\nu}_1^\ast \tilde{\nu}_1 {\rm Re}[H_d^0]}|, |C_{\tilde{\nu}_1^\ast \tilde{\nu}_1 {\rm Re}[H_u^0]}| \lesssim 10~{\rm GeV}$ and $C_{\tilde{\nu}_1^\ast \tilde{\nu}_1 {\rm Re}[S]} \lesssim 100~{\rm GeV}$ in most cases, which reflects that $|C_{\tilde{\nu}_1^\ast \tilde{\nu}_1 {\rm Re}[S]}|$ may be much larger than the other two couplings. The basic reason is that $\tilde{\nu}_1$ is a singlet-dominated scalar, so it can couple directly to the field $S$ and the mass dimension of $C_{\tilde{\nu}_1^\ast \tilde{\nu}_1 {\rm Re}[S]}$ is induced by $v_s$ or $A_{\lambda_\nu}$. By contrast, the other couplings emerge only after the electroweak symmetry breaking when $V_{11} = 0$, and their mass dimension originates from $v_u$.

\subsection{Relic Density of Sneutrino Dark Matter}

In the $B_{\mu_X}=0$ case, both $\tilde{\nu}_1$ and $\tilde{\nu}_1^\ast$ act as the DM candidate. Their annihilation includes those initiated by $\tilde{\nu}_1 \tilde{\nu}_1^\ast$, $\tilde{\nu}_1 \tilde{\nu}_1$, and $\tilde{\nu}_1^\ast \tilde{\nu}_1^\ast$ state, and the co-annihilation of $\tilde{\nu}_1$ and $\tilde{\nu}_1^\ast$ with the other sparticles. Considering the numerousness of the  annihilation channels and the complexity of this issue, we will only discuss the channels frequently met in our study (see footnote 2 of this work for more details), which are~\cite{Cao:2017cjf,Cerdeno:2009dv}:
\begin{itemize}
\item[(1)] $\tilde{\nu}_1 \tilde{H}, \tilde{\nu}_1^\ast \tilde{H} \rightarrow X Y$ and $\tilde{H} \tilde{H}^\prime \rightarrow X^\prime Y^\prime$, where $\tilde{H}$ and $\tilde{H}^\prime$ denote Higgsino-dominated neutralinos or charginos, and $X^{(\prime)}$ and $Y^{(\prime)}$ represent any possible SM particles, the massive neutrinos or the Higgs bosons if the kinematics are accessible. More specifically, the channels  $\tilde{\nu}_1 \tilde{H} \rightarrow W l, Z \nu, h \nu$ ($l$ and $\nu$ denote any possible lepton and neutrino, respectively) proceed by the $s$-channel exchange of neutrinos, and the $t/u$ channel exchange of sleptons or sneutrinos. The processes
 $\tilde{H} \tilde{H}^\prime \rightarrow f \bar{f}^\prime, V V^\prime, h V$ ($f$ and $f^\prime$ denote quarks or leptons, and $V$ and $V^\prime$ represent SM vector bosons) proceed by the $s$-channel exchange of vector bosons or Higgs bosons, and the $t/u$ channel exchange of sfermions, neutralinos or charginos. This annihilation mechanism is called co-annihilation~\cite{Coannihilation,Griest:1990kh}.
\item[(2)] $\tilde{\nu}_1 \tilde{\nu}_1^\ast \rightarrow s s^\ast $ ($s$ denotes a light Higgs boson), which proceeds through any relevant quartic scalar coupling, the $s$-channel exchange of CP-even Higgs bosons, and the $t/u$-channel exchange of sneutrinos.
\item[(3)]  $\tilde{\nu}_1 \tilde{\nu}_1^\ast \rightarrow \nu_h \bar{\nu}_h$ via the $s$-channel exchange of CP-even Higgs bosons or the $t/u$-channel exchange of neutralinos, where $\nu_h$ denotes a massive neutrino.
\item[(4)]  $\tilde{\nu}_1 \tilde{\nu}_1 \rightarrow \nu_h \nu_h$ and $\tilde{\nu}_1^\ast \tilde{\nu}_1^\ast \rightarrow \bar{\nu}_h \bar{\nu}_h$, which mainly proceed through the $t/u$-channel exchange of a singlino-dominated neutralino due to its majorana nature.
\item[(5)]  $\tilde{\nu}_1 \tilde{\nu}_1^\ast \rightarrow V V^\ast$, $Vs $, $f \bar{f} $, which proceeds mainly by the $s$-channel exchange of CP-even Higgs bosons. They are important if one of the bosons is at resonance.
\end{itemize}

Under specific parameter configurations, these channels can be responsible for the DM density precisely measured by the Planck experiment~\cite{Aghanim:2018eyx}. In this aspect, we have the following observations (see footnote 2 for more explanations):
\begin{itemize}
\item In most cases, the DMs annihilated mainly through the co-annihilation to get the measured density. This mechanism works only when the mass splitting between $\tilde{H}$ and $\tilde{\nu}_1$ is less than about $10\%$, and a specific channel's contribution to the density depends not only on its cross-section but also on the mass splitting. To illustrate this point, we assume that the DM annihilations comprise those initiated by  $\tilde{\nu}_1 \tilde{\nu}_1$, $\tilde{\nu}_1 \tilde{\nu}_1^\ast$, $\tilde{\nu}_1^\ast \tilde{\nu}_1^\ast$, $\tilde{\nu}_1 \tilde{\chi}_1^0$, $\tilde{\nu}_1^\ast \tilde{\chi}_1^0$, and $\tilde{\chi}_1^0 \tilde{\chi}_1^0$ states, and denote the cross-sections of these channels by $\sigma_{AB}$ with $A, B= \tilde{\nu}_1, \tilde{\nu}_1^\ast,\tilde{\chi}_1^0$. The effective annihilation rate at temperature $T$ is then given by~\cite{Griest:1990kh}
    \begin{eqnarray}
    \sigma_{eff} &&= \frac{1}{4} \left( \frac{1}{ 1 + ( 1 + \Delta )^{3/2} e^{-x\Delta} }\right)^2 \times \left \{ \sigma_{\tilde\nu_1 \tilde\nu_1} + 2 \sigma_{\tilde\nu_1 \tilde\nu_1^\ast} + \sigma_{\tilde\nu_1^\ast \tilde\nu_1^\ast} \right. \nonumber\\
    &&\left.  +  4 \left( \sigma_{\tilde\nu_1 \tilde\chi^0_1}+\sigma_{\tilde\nu_1^\ast \tilde\chi^0_1} \right) \left( 1+ \Delta \right)^{3/2} e^{-x\Delta} +  4 \sigma_{\tilde\chi^0_1 \tilde\chi^0_1} \left( 1+ \Delta \right)^{3} e^{-2 x \Delta} \right \}, \label{Co-annihilation experession}
\end{eqnarray}
where $\Delta \equiv (m_{\tilde{\chi}_1^0} - m_{\tilde{\nu}_1^0})/m_{\tilde{\nu}_1^0} $ parameterizes the mass splitting and $x \equiv m_{\tilde{\nu}_1^0}/T$. This formula indicates that the $\tilde{\nu}_1 \tilde{\chi}_1^0$ and $\tilde{\chi}_1^0 \tilde{\chi}_1^0$ channel's contributions are suppressed by factors $e^{-x \Delta}$ and $e^{-2 x \Delta}$, respectively. So they become less and less critical as $m_{\tilde{\nu}_1^0}$ deviates from $m_{\tilde{\chi}_1^0}$. Besides,  the formulae of the density in~\cite{Griest:1990kh} indicate that the density depends on the sneutrino parameters only through $m_{\tilde{\nu}_1}$ and $\sigma_{eff}$. In the extreme case of $\sigma_{\tilde\nu_1 \tilde\nu_1} \simeq \sigma_{\tilde\nu_1 \tilde\nu_1^\ast} \simeq \sigma_{\tilde\nu_1 \tilde\chi^0_1} \simeq 0$ realized when $\lambda_\nu$ and $Y_\nu$ are sufficiently small, the $\tilde{\chi}_1^0 \tilde{\chi}_1^0$ annihilation is solely responsible for the measured density through tuning the value of $m_{\tilde{\nu}_1}$.  This situation was intensively studied in~\cite{Cao:2019qng}. We will present such examples in Section III.

\item Barring the co-annihilation, $\tilde{\nu}_1 \tilde{\nu}_1 \rightarrow s s^\ast $ is usually the most crucial channel in affecting the density if the kinematics are accessible. In particular, the process $\tilde{\nu}_1 \tilde{\nu}_1 \rightarrow h_s h_s$ can be solely responsible for the measured density if the Yukawa coupling $\lambda_\nu$ is moderately large. We exemplify this point by considering the light $h_s$ scenario in Table \ref{table1}. From the Higgs boson and sneutrino mass spectrum and the $\tilde{\nu}_1$'s couplings to $h_s$, one can learn that the annihilation proceeds mainly by the $s$-channel exchange of $h_s$, $t/u$-channel exchange of $\tilde{\nu}_1$, and $\tilde{\nu}_1 \tilde{\nu}_1^\ast h_s h_s$ quartic scalar coupling. As a result, the cross-section of the annihilation near the freeze-out temperature is approximated by~\cite{Cao:2017cjf,Cerdeno:2009dv}
    \begin{eqnarray}
    \sigma v \simeq a + b v^2,  \label{relic density}
    \end{eqnarray}
where
\begin{eqnarray}
    a &=&  \frac{\sqrt{1- m_{h_s}^2/m_{\tilde{\nu}_1}^2}}{64 \pi m_{\tilde{\nu}_1}^2} \left | C_{\tilde{\nu}_1 \tilde{\nu}_1 h_s h_s} -
    \frac{C_{\tilde{\nu}_1\tilde{\nu}_1 h_s } C_{h_i h_s h_s}}{ 4 m_{\tilde{\nu}_1}^2 -m_{h_s}^2} +
    \frac{2 C_{\tilde{\nu}_1 \tilde{\nu}_1 h_s}^2}{2 m_{\tilde{\nu}_1}^2 - m_{h_s}^2} \right |^2,  \nonumber \\
    b &=& \left ( -\frac{1}{4} + \frac{m_{h_s}^2}{8 (m_{\tilde{\nu}_1}^2 - m_{h_s}^2 )} \right ) \times a -
    \frac{\sqrt{1- m_{h_s}^2/m_{\tilde{\nu}_1}^2}}{64 \pi} \times \nonumber \\
    && \left \{ \frac{C^2_{\tilde{\nu}_1 \tilde{\nu}_1 h_s} C^2_{h_s h_s h_s}}{(4 m_{\tilde{\nu}_1}^2 - m_{h_s}^2)^3 } - \frac{2 C_{\tilde{\nu}_1 \tilde{\nu}_1 h_s} C_{h_s h_s h_s} C_{\tilde{\nu}_1 \tilde{\nu}_1 h_s}^2 (10 m_{\tilde{\nu}_1}^2 - 3 m_{h_s}^2) } {(4 m_{\tilde{\nu}_1}^2 - m_{h_s}^2)^2 (2 m_{\tilde{\nu}_1}^2 - m_{h_s}^2)^2 } \right . \nonumber \\
    && \left . + \frac{2 C_{\tilde{\nu}_1 \tilde{\nu}_1 h_s}^4 }{ (2 m_{\tilde{\nu}_1}^2 - m_{h_s}^2)^3 } - 2 C_{\tilde{\nu}_1 \tilde{\nu}_1 h_s h_s} \left ( \frac{ C_{\tilde{\nu}_1 \tilde{\nu}_1 h_s} C_{h_s h_s h_s} } {(4 m_{\tilde{\nu}_1}^2 - m_{h_s}^2)^2 } - \frac{ C_{\tilde{\nu}_1 \tilde{\nu}_1 h_s}^2 } {(2 m_{\tilde{\nu}_1}^2 - m_{h_s}^2)^2 } \right ) \right \}. \nonumber
\end{eqnarray}
The measured density then requires $a + 3 b /25 \simeq 4.6 \times 10^{-26} {\rm cm}^2$ because we are considering a two-component DM theory~\cite{Chang:2013oia,Berlin:2014tja}. This requirement limits the $\tilde{\nu}_1$'s couplings to $h_s$ or for the fixed parameters in Table \ref{table1}, ultimately the Yukawa coupling $\lambda_\nu$ since the cross-section is very sensitive to $\lambda_\nu$. We estimate that $\lambda_\nu \sim 0.4$ for $m_{\tilde{\nu}_1} = 130~{\rm GeV}$ can account for the measured density.

\item The process $\tilde{\nu}_1 \tilde{\nu}_1^\ast \to \nu_h \bar{\nu}_h$ could be responsible for the density when $m_{\tilde{\nu}_1} > \nu_h$, $m_{\tilde{\nu}_1} < h_s$, and the co-annihilation mechanism did not work. This process proceeded mainly by the $s$-channel exchange of $h_s$, and consequently, the cross-section at the freeze-out temperature $T_f$ takes the following form:
\begin{eqnarray}
\langle \sigma v \rangle_{T_f} \sim \left ( \frac{C_{\tilde{\nu}_1 \tilde{\nu}_1^\ast h_s} C_{\bar{\nu}_h \nu_h h_s}}{4 m_{\tilde{\nu}_1}^2 - m_{h_s}^2} \right )^2,
\end{eqnarray}
which implies that the density limits non-trivially $\lambda_\nu$, $m_{\tilde{\nu}_1}$ and $m_{h_s}$.

\item About the other channels, they usually played a minor role in determining the density. So we leave the discussion of them in our future works.
\end{itemize}

Concerning the $B_{\mu_X} \neq 0$ case, either $\tilde{\nu}_{R,1}$ or $\tilde{\nu}_{I,1}$ acts as the DM candidate. Since the mass splitting between the DM $\tilde{\nu}_1$ and $\tilde{\nu}_1^\prime$ (the partner of $\tilde{\nu}_1$ with a different CP number) is small, $\tilde{\nu}_1$ always co-annihilated with $\tilde{\nu}_1^\prime$ to get the measured density. The relevant annihilation included $\tilde{\nu}_1 \tilde{\nu}_1^\prime$ and $\tilde{\nu}_1^\prime \tilde{\nu}_1^\prime$ initiated processes, and they proceeded in a way similar to the previous discussion. We confirmed that the density is insensitive to $B_{\mu_X}$ for $|B_{\mu_X}| \leq 100~{\rm GeV^2}$, which can be inferred from Eq.~(\ref{Co-annihilation experession}). We also verified that the cross-section of the DM annihilation today is insensitive to $B_{\mu_X}$.

\subsection{DM-nucleon Scattering}

In the $B_{\mu_X} \neq 0$ case, the scattering of $\tilde{\nu}_1$ with nucleon $N$ ($N=p,n$) proceeds by the $t/u$-channel exchange of the CP-even Higgs bosons. Consequently, the spin independent (SI) cross-section is given by~\cite{Cao:2017cjf}
\begin{eqnarray}
\sigma^{\rm SI}_{\tilde{\nu}_1-N} & = &  \frac{F^{(N)2}_u g^2 \mu^2_{\rm red} m_N^2  }{16 \pi m_W^2} \times \left \{\sum_i \left [ \frac{C_{\tilde{\nu}_1^\ast \tilde{\nu}_1 h_i}}{m_{h_i}^2 m_{\tilde{\nu}_1}} (\frac{U_{i2}}{\sin\beta} + \frac{U_{i1}}{\cos \beta} \frac{F^N_d}{F^N_u} ) \right ] \right \}^2, \nonumber
\end{eqnarray}
where $\mu_{\rm red}= m_N/( 1+ m_N^2/m_{\tilde{\nu}_1}^2) $ ($N=p, n$) represents the reduced mass of the nucleon with $m_{\tilde{\nu}_1}$, $F^{N}_u=f^{N}_u+\frac{4}{27}f^{N}_G$ and $F^{N}_d=f^{N}_d+f^{N}_s+\frac{2}{27}f^{N}_G$ are nucleon form factors with $f^{N}_q=m_N^{-1}\left<N|m_qq\bar{q}|N\right>$ and $f^{N}_G=1-\sum_{q}f^{N}_q$ for $q=u,d,s$~\cite{Jungman:1995df}. With the default setting of the package micrOMEGAs~\cite{micrOMEGAs-1,micrOMEGAs-2,micrOMEGAs-3} for nucleon sigma terms, i.e., $\sigma_{\pi N} = 34~{\rm MeV}$ and $\sigma_0 = 42~{\rm MeV}$~\cite{Ellis:2008hf}\footnote{It is notable that $\sigma_0$ was replaced by the strangeness-nucleon sigma term, $\sigma_s \equiv m_s/(m_u + m_d) \times (\sigma_{\pi N} - \sigma_0) \simeq 12.4 \times (\sigma_{\pi N} - \sigma_0)$, in recent calculation of the nucleon form factor~\cite{micrOMEGAs-2}. Compared with the previous calculation, this treatment changes significantly the strange quark content in nucleon $N$, $f_s^{N}$, but it change little $F_u^{N}$ and $F_d^{N}$. }, one can conclude $F_u^{p} \simeq 0.15$ and  $F_d^{p} \simeq 0.14$ for protons. Instead, if $\sigma_{\pi N} = 59~{\rm MeV}$~\cite{Alarcon:2011zs,Ren:2014vea,Ling:2017jyz} and $\sigma_0 = 57~{\rm MeV}$~\cite{Alarcon:2012nr} are adopted, the form factors become $F_u^{p} \simeq 0.16$ and  $F_d^{p} \simeq 0.13$. These results reflect that different choices of $\sigma_{\pi N}$ and $\sigma_0$ can induce uncertainties of ${\cal{O}} (10\%)$ in $F_u^{p}$ and $F_d^{p}$, and it does not drastically change the cross-section. Besides, the default setting also predicts $F_u^{n} \simeq 0.15$ and  $F_d^{n} \simeq 0.14$ for neutrons, which implies the relation $\sigma^{\rm SI}_{\tilde{\nu}_1-p} \simeq \sigma^{\rm SI}_{\tilde{\nu}_1-n}$ for the Higgs-mediated contribution.

To clarify the features of the cross-section, we assume $m_{H^\pm} \gtrsim 1 {\rm TeV}$ and integrate out the heavy doublet Higgs field. As a result, the CP-even Higgs sector at the electroweak scale contains only the SM Higgs field $S_2 = {\rm \sin \beta} {\rm Re}[H_u^0] + {\rm \cos \beta} {\rm Re}[H_d^0]$ and the singlet field ${\rm Re}[S]$. We then  calculate the scattering amplitude by the mass insertion method to get the following result:
\begin{eqnarray}
\sigma^{\rm SI}_{\tilde{\nu}_1-N} & \simeq & \frac{F^{(N)2}_u g^2 \mu^2_{\rm red} m_N^2  }{16 \pi m_W^2 (125~{\rm GeV})^4} \times \left ( \frac{125~{\rm GeV}}{m_h} \right )^4 \times \left ( \frac{ C_{\tilde{\nu}_1^\ast \tilde{\nu}_1 {\rm Re}[S]}}{m_{\tilde{\nu}_1}} \times \delta \sin \theta \cos \theta \right . \nonumber  \\
& & \left . - \frac{\cos \beta C_{\tilde{\nu}_1^\ast \tilde{\nu}_1 {\rm Re}[H_d^0]} + \sin \beta C_{\tilde{\nu}_1^\ast \tilde{\nu}_1 {\rm Re}[H_u^0]}}{m_{\tilde{\nu}_1}} \times (1 + \delta \sin^2 \theta ) \right )^2 \nonumber \\
& \simeq & 4.2 \times 10^{-44}~{\rm cm^2} \times \left ( \frac{125~{\rm GeV}}{m_h} \right )^4 \times \left ( \frac{ C_{\tilde{\nu}_1^\ast \tilde{\nu}_1 {\rm Re}[S]}}{m_{\tilde{\nu}_1}} \times \delta \sin \theta \cos \theta \right . \nonumber  \\
& & \left . - \frac{\cos \beta C_{\tilde{\nu}_1^\ast \tilde{\nu}_1 {\rm Re}[H_d^0]} + \sin \beta C_{\tilde{\nu}_1^\ast \tilde{\nu}_1 {\rm Re}[H_u^0]}}{m_{\tilde{\nu}_1}} \times (1 + \delta \sin^2 \theta ) \right )^2,  \label{Approximation}
\end{eqnarray}
where $\delta = m_h^2/m_{h_s}^2 - 1 $, and $\theta$ is the mixing angle of the $S_2$ field with ${\rm Re}[S]$ to form mass eigenstates.  This formula reveals that if the terms in the second brackets are on the order of 0.1, which can be achieved if $\lambda_\nu$ and/or $Y_\nu$ in Eq.~(\ref{approximation-1}) are sufficiently large,
the cross-section may reach the sensitivity of the recent XENON-1T experiment~\cite{Aprile:2018dbl}. We will discuss this issue later.

Concerning the $B_{\mu_X} = 0$ case, where the DM corresponds to a complex field, the $Z$-boson also mediates the elastic scattering
of the DM with nucleons. Since  the total SI cross-section in this case is obtained by averaging over $ \tilde{\nu}_1 N$
and  $\tilde{\nu}_1^\ast N$ scatterings and the interferences between the $Z$- and the Higgs-exchange diagrams for the two scatterings have opposite signs~\cite{Dumont:2012ee}, the SI cross-section is given by~\cite{Arina:2007tm}
\begin{eqnarray}
\sigma_N^{\rm SI} \equiv \frac{\sigma_{\tilde{\nu}_1-N}^{\rm SI} + \sigma_{\tilde{\nu}_1^\ast - N}^{\rm SI} }{2}  =  \sigma_N^h + \sigma_N^Z,
\end{eqnarray}
where $\sigma_N^h$ is the same as before and the $Z$-mediated contributions are
\begin{eqnarray}
\sigma_n^Z \equiv \frac{G_F^2 V_{11}^4}{2 \pi} \frac{m_n^2}{(1 + m_n/m_{\tilde{\nu}_1})^2}, \quad \sigma_p^Z \equiv \frac{G_F^2 V_{11}^4 (4 \sin^2 \theta_W - 1)^2}{2 \pi} \frac{m_p^2}{(1 + m_p/m_{\tilde{\nu}_1})^2},  \label{DM-neutron}
\end{eqnarray}
with $G_F$ and $\theta_W$ denoting the Fermi constant and the weak angle, respectively. Since $\sigma_n^Z$ is larger than $\sigma_p^Z$ by a factor around 100, $\sigma_n^{\rm SI}$ may differ significantly from $\sigma_p^{\rm SI}$. Correspondingly, one may define the effective cross-section for the coherent scattering of the DMs with xenon nucleus as $\sigma_{\rm eff}^{\rm SI} = (\sigma^{\rm SI}_{\tilde{\nu}_1-Xe} + \sigma^{\rm SI}_{\tilde{\nu}_1^\ast-Xe})/(2 A^2)$, where $A$ denotes the mass number of the xenon nucleus, and calculate it by
\begin{eqnarray}
\sigma_{\rm eff}^{\rm SI} = 0.169 \sigma^{\rm SI}_p  +  0.347 \sigma^{\rm SI}_n  + 0.484 \sqrt{ \sigma^{\rm SI}_p  \sigma^{\rm SI}_n },   \label{Effective-Cross-Section}
\end{eqnarray}
where the three coefficients on the right side are obtained by averaging the abundance of different xenon isotopes in nature. It is evident that the effective cross-section is identical to $\sigma^{\rm SI}_p$ if $\sigma_p^{\rm SI} = \sigma_n^{\rm SI}$, and it is related directly with the bound of the XENON-1T experiment~\cite{Aprile:2018dbl}.

Before concluding the introduction of the sneutrino DM, we add that its spin dependent cross-section is always zero, and its SI cross-section is usually much smaller than that of the neutralino DM in the MSSM and NMSSM, which was discussed in detail in Ref.\cite{Cao:2017cjf,Cao:2019qng}. As a result, the extension is readily consistent with the XENON-1T experiment except for large $\lambda_\nu$ and/or $Y_\nu$ case studied in this work.

\section{\label{Section-Scan} Constraints on sneutrino DM sector}

In this section, we clarify the impact of the leptonic unitarity and current and future DM DD experiments on the sneutrino DM sector  under the premise that
the theory predicts the right density and the photon spectrum from the DM annihilation in dwarf galaxies is compatible with the Fermi-LAT observation.
Since the singlet-dominated Higgs boson, $h_s$, plays a vital role in the density and the DM-nucleon scattering, we study the DM physics
in both the light and the massive $h_s$ scenarios in Table~\ref{table1}.  We emphasize that fixing the parameters in the Higgs and neutralino
sectors can simplify greatly the analysis of the impact and make the underlying physics clear. We also emphasize that the two scenarios were
obtained by scanning intensively the parameters in the Higgs and DM sectors\footnote{With the parameter scan strategy reported in~\cite{Cao:2018iyk} for the Type-I NMSSM, we explored the parameter space of the ISS-NMSSM which takes $\tan \beta$, $\lambda$, $\kappa$, $A_t$, $A_\kappa$, $\mu$, $\lambda_\nu$, $Y_\nu$, $A_{\lambda_\nu}$, $A_{Y_\nu}$, $m_{\tilde{\nu}}$, and $m_{\tilde{x}}$ as inputs. For either the light or the massive $h_s$ scenario, we have studied more than fifty million samples. The settings in Table \ref{table1} were chosen from the samples that best fit the experimental data. We will present the analysis of these samples elsewhere.}. They agree well with the latest Higgs data of the LHC if the exotic decays $h \to \nu_h \bar{\nu}_h, \tilde{\nu}_1 \tilde{\nu}_1^\ast $ are kinematically forbidden. This was confirmed by the packages \textsf{HiggsSignal-2.4.0}~\cite{HiggsSignal} and \textsf{HiggsBounds-5.7.0}~\cite{HiggsBounds}.

\subsection{Research strategy}

The procedure of our study is as follows. We constructed a likelihood function of the DM physics to guide sophisticated scans over the sneutrino parameters for either scenario. With the samples obtained in the scans, we plotted the profile likelihood map in different two-dimensional planes to illustrate its features and underlying physics. We express the likelihood function as
\begin{eqnarray}
	\mathcal{L}_{\rm DM} = \mathcal{L}_{\Omega_{\tilde{\nu}_1}} \times  \mathcal{L}_{\rm DD} \times \mathcal{L}_{\rm ID} \times  \mathcal{L}_{Unitary},   \label{DM-profile}
\end{eqnarray}
where $\mathcal{L}_{\Omega_{\tilde{\nu}_1}}$, $\mathcal{L}_{\rm DD}$, $\mathcal{L}_{\rm ID}$, and $\mathcal{L}_{Unitary}$ describe the relic density, the current XENON-1T experiment~\cite{Aprile:2018dbl} or the future LZ experiment~\cite{Akerib:2018dfk}, the Fermi-LAT observation of dwarf galaxies, and the unitarity constraint, respectively. They are given by
\begin{itemize}
\item $\mathcal{L}_{\Omega_{\tilde{\nu}_1}}$ is Gaussian distributed, i.e.,
\begin{equation}
\mathcal{L}_{\Omega_{\tilde{\nu}_1}}=e^{-\frac{[\Omega_{\rm th}-\Omega_{\rm obs}]^2}{2\sigma^2}},
\end{equation}
where $\Omega_{th}$ denotes the theoretical prediction of the density $\Omega_{\tilde{\nu}_1} h^2$, $\Omega_{\rm obs}=0.120$ represents
its experimental central value~\cite{Aghanim:2018eyx}, and $\sigma = 0.1 \times \Omega_{\rm obs}$ is the total (including both theoretical and experimental) uncertainty of the density.
\item $\mathcal{L}_{\rm DD}$ takes a Gaussian distributed form with a mean value of zero~\cite{Matsumoto:2016hbs}:
\begin{equation}
\mathcal{L}_{\rm DD}=e^{-\frac{1}{2} \left ( \frac{\sigma^{\rm SI}_{\rm eff}}{\delta_{\sigma}} \right )^2}.
\end{equation}
In this formula, $\sigma^{\rm SI}_{\rm eff}$ is defined in Eq.~(\ref{Effective-Cross-Section}) and its error bar $\delta_{\sigma}$ is evaluated by $\delta_{\sigma} = \sqrt{UL_\sigma^2/1.64^2 + (0.2 \sigma^{\rm SI}_{\rm eff})^2}$, where $UL_\sigma$ denotes experimental upper limits on the scattering cross-section at $90\%$ C.L. and $0.2 \sigma^{\rm SI}_{\rm eff}$ parameterizes the theoretical uncertainty of $\sigma^{\rm SI}_{\rm eff}$.
\item $\mathcal{L}_{\rm ID}$ is calculated by the likelihood function proposed in~\cite{Likelihood-dSph,Zhou} with the data of the Fermi-LAT collaboration taken from~\cite{Ackermann:2015zua,Fermi-Web}.

\item The likelihood function of the unitarity constraint in Eq.~(\ref{unitaryconstriants}) is as follows:
    \begin{equation}
    \mathcal{L}_{Unitary}=\left \{ \begin{aligned}
    exp[-\frac{1}{2} \left ( \frac{r - 9.4 }{0.2 r} \right )^2] \ \  {\rm if} \ r \leq 9.4 \\  1 \ \  {\rm if} \ r > 9.4
    \end{aligned} \right.
    \end{equation}
    where $r \equiv \lambda_{\nu} \mu/(Y_{\nu} \lambda v_u)$ and $0.2 r$ parameterizes total uncertainties.
\end{itemize}
In addition, we abandoned samples that open up the decays $h \to \nu_h \bar{\nu}_h, \tilde{\nu}_1 \tilde{\nu}_1^\ast $\footnote{In fitting the ISS-NMSSM to experimental data, the total likelihood function is calculated by $\mathcal{L}_{\rm tot} = \mathcal{L}_{\rm Higgs} \times \mathcal{L}_{\rm DM}$, where $\mathcal{L}_{\rm Higgs}$ represents the Higgs physics function. Given $\chi^2 \equiv -2 \ln \mathcal{L}$, one can infer that $\chi^2_{\rm tot} =  \chi^2_{\rm Higgs} + \chi^2_{\rm DM}$, and the $2\sigma$ confidence interval defined below Eq.~(\ref{stat:Frequen}) satisfies
\begin{eqnarray}
\chi^2_{\rm tot} -  \chi^2_{\rm tot, min} \equiv \delta \chi^2_{\rm Higgs} + \delta \chi^2_{\rm DM}  \leq 6.18,  \nonumber 
\end{eqnarray}
where $ \delta \chi^2_{\rm Higgs} \equiv \chi^2_{\rm Higgs} -  \chi^2_{\rm Higgs, min}$ and  $\delta \chi^2_{\rm DM} \equiv \chi^2_{\rm DM} -  \chi^2_{\rm DM, min}$ . For the Higgs parameter settings in Table~\ref{table1}, $\delta \chi^2_{\rm Higgs}$ vanishes  because the settings correspond to the scenarios' best points when the decays are forbidden. It increases if the decays are open and their branching ratios gradually increase. Since we hope to determine the interval only by the DM physics in the following study, we kinematically shut down the decays.  }. In practice, this was completed by setting the likelihood value to be $e^{-100}$ if any of the decays were kinematically accessible.

To make the conclusions in this study complete, we adopt the MultiNest algorithm~\cite{Feroz:2008xx,Feroz:2013hea} to implement the scans. We take the prior probability density function (PDF) of the input parameters uniformly distributed and set {\it nlive} parameter of the algorithm to be 10000. This parameter represents the number of active or live points used to determine the iso-likelihood contour in each scan's iteration~\cite{Feroz:2008xx,Feroz:2013hea}. The larger it is, the more elaborated the scan becomes. The output of the scans includes the Bayesian evidence defined by
	\begin{eqnarray}
	Z(D|M) \equiv \int{P(D|O(M,\Theta)) P(\Theta|M) \prod d \Theta_i}, \nonumber
	\end{eqnarray}
where $P(\Theta|M)$ represents the prior PDF of the inputs $\Theta = (\Theta_1,\Theta_2,\cdots)$ in a model $M$, and $P(D|O(M, \Theta))\equiv \mathcal{L}(\Theta)$ denotes the likelihood function involving theoretical predictions of observables $O$ and their experimental measurements $D$. Computationally, the evidence is an averaged likelihood that depends on the priors of the theory's input. In comparing different scenarios of the theory, the larger $Z$ is, the more readily the corresponding scenario is consistent with the data.

The output of the scan also includes the profile likelihood (PL) defined in frequentist statistics as the most significant likelihood value~\cite{Fowlie:2016hew,Cao:2018iyk}. For example, two-dimensional (2D) PL is defined by
\begin{eqnarray}
\mathcal{L}(\Theta_A,\Theta_B)=\mathop{\max}_{\Theta_1,\cdots,\Theta_{A-1},\Theta_{A+1},\cdots, \Theta_{B-1}, \Theta_{B+1},\cdots}\mathcal{L}(\Theta), \label{stat:Frequen}
\end{eqnarray}
where the maximization is obtained by varying the parameters other than $\Theta_A$ and $\Theta_B$.  The PL reflects the preference of the theory on the parameter $(\Theta_A, \Theta_B)$, or in other words, the capability of the parameter to account for experimental data. Sequentially, one can introduce the concept of confidence interval (CI) to classify the parameter region by how well the points in it fit the data. For example, the $1\sigma$ and $2\sigma$ CIs for the 2D PL are defined by satisfying $\chi^2 - \chi^2_{\rm min} \leq 2.3$ and $\chi^2 - \chi^2_{\rm min} \leq 6.18$, respectively, where $\chi^2 \equiv -2 \ln \mathcal{L} (\Theta_A,\Theta_B)$ and $\chi^2_{\rm min}$ is the minimal value of $\chi^2$ for the samples obtained in the scan.

In this work, we utilized the package \textsf{SARAH-4.11.0}~\cite{sarah-1,sarah-2,sarah-3} to build the model file of the ISS-NMSSM, the \textsf{SPheno-4.0.3} \cite{spheno} code
to generate its particle spectrum, and the package \textsf{MicrOMEGAs 4.3.4}~\cite{Belanger:2013oya,micrOMEGAs-1,micrOMEGAs-3}
to calculate the DM observables.

\begin{figure*}[t]
		\centering
		\resizebox{0.92\textwidth}{!}{
        \includegraphics[width=0.90\textwidth]{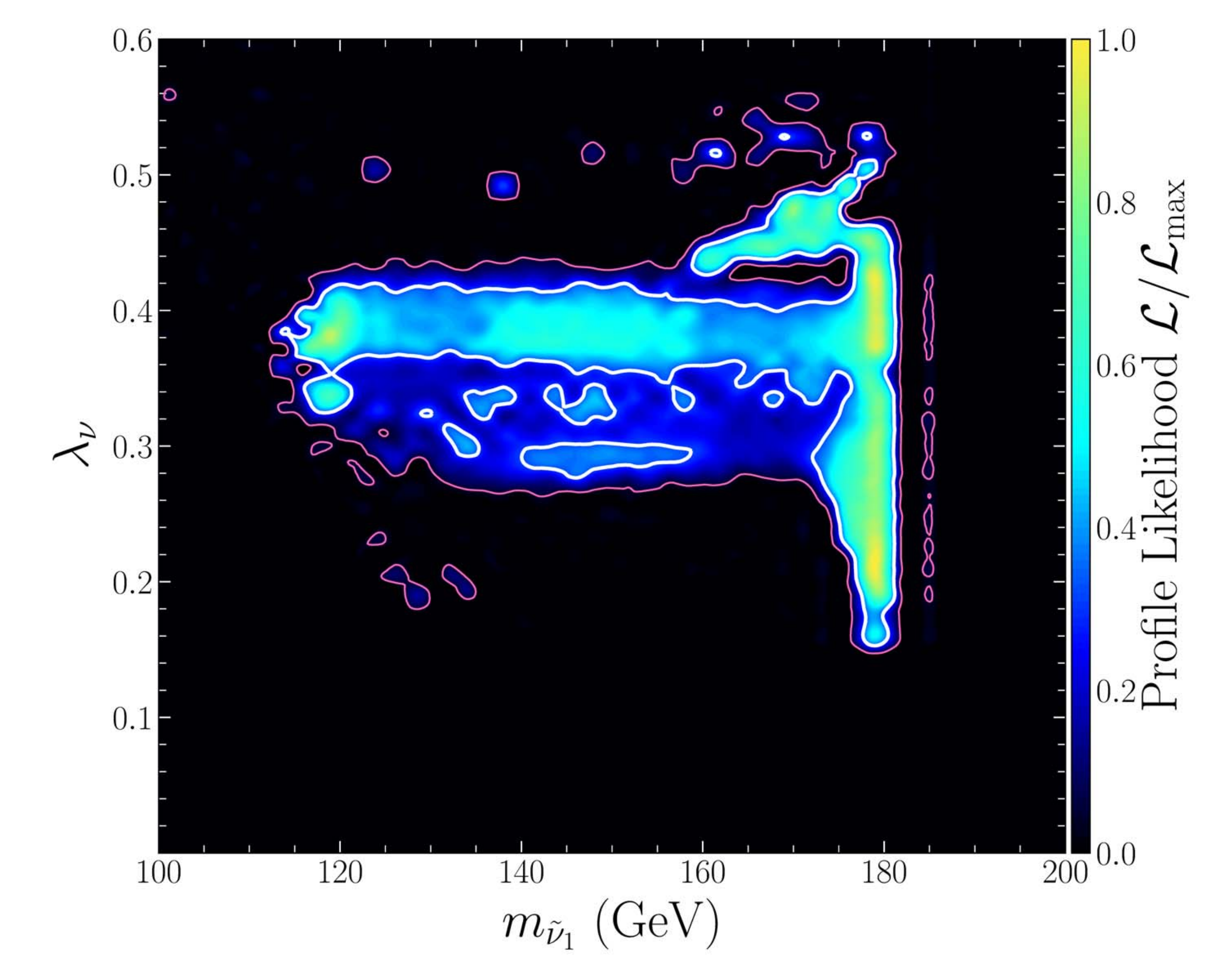}
        \includegraphics[width=0.90\textwidth]{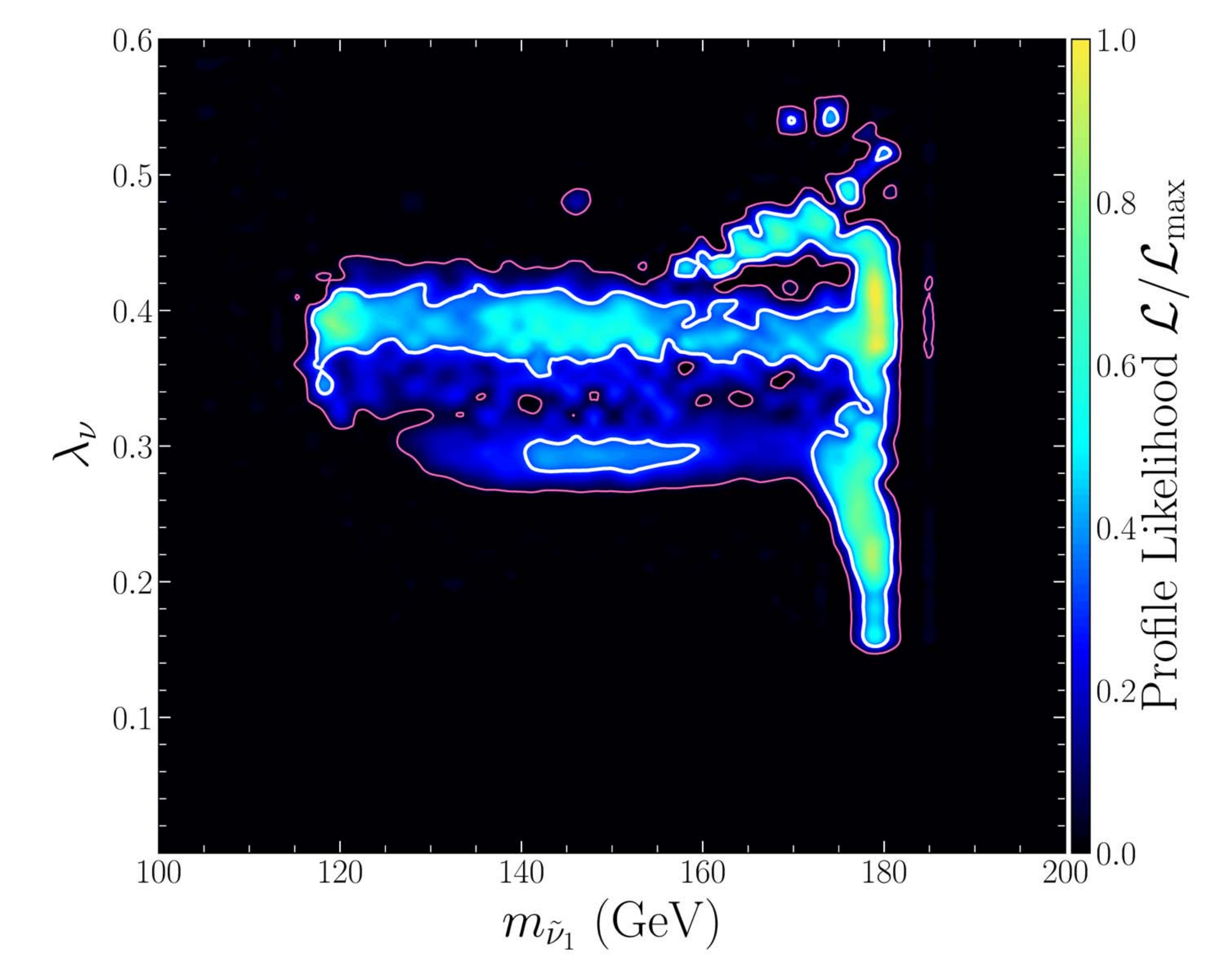}
        }

		\resizebox{0.92\textwidth}{!}{
        \includegraphics[width=0.90\textwidth]{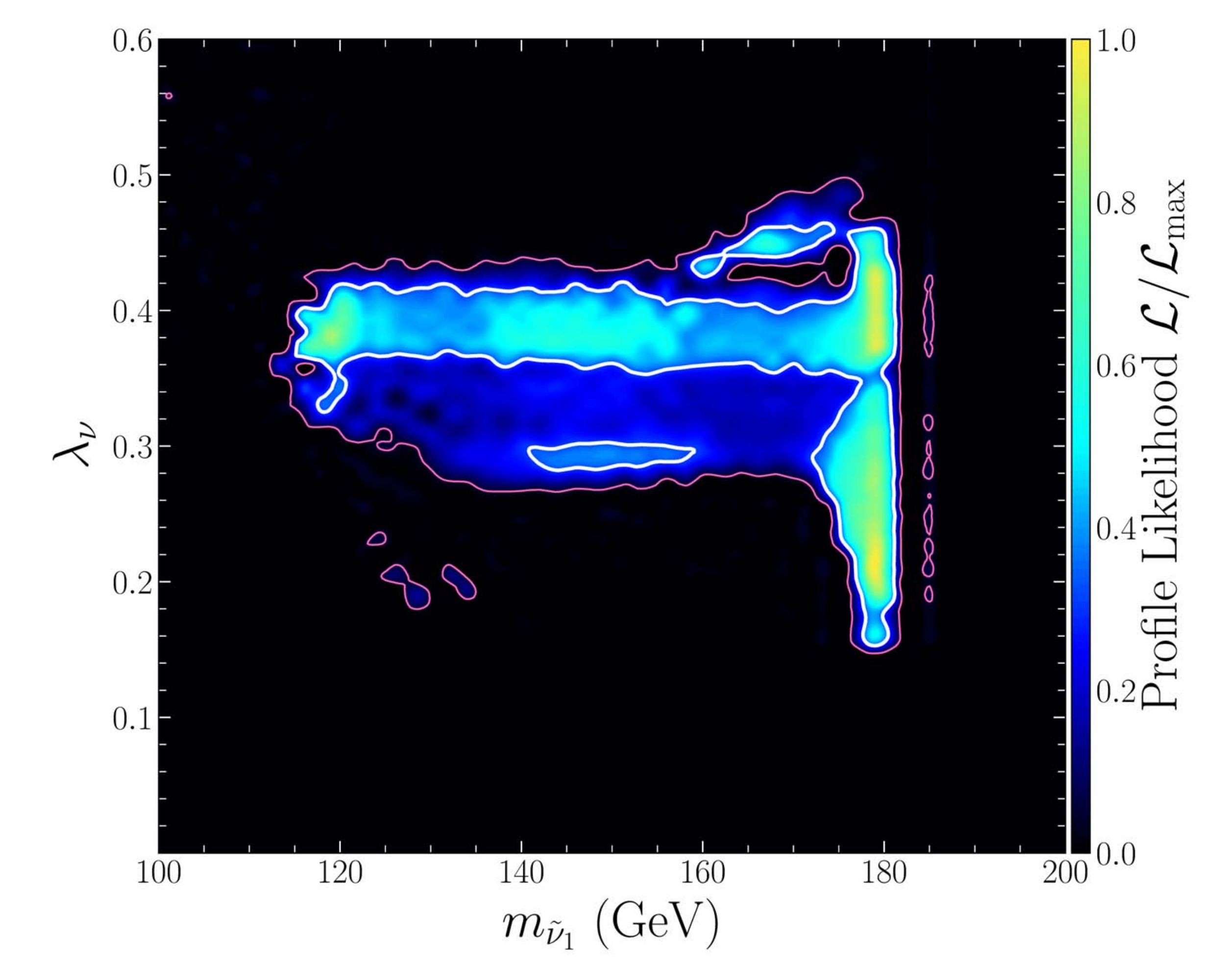}
        \includegraphics[width=0.90\textwidth]{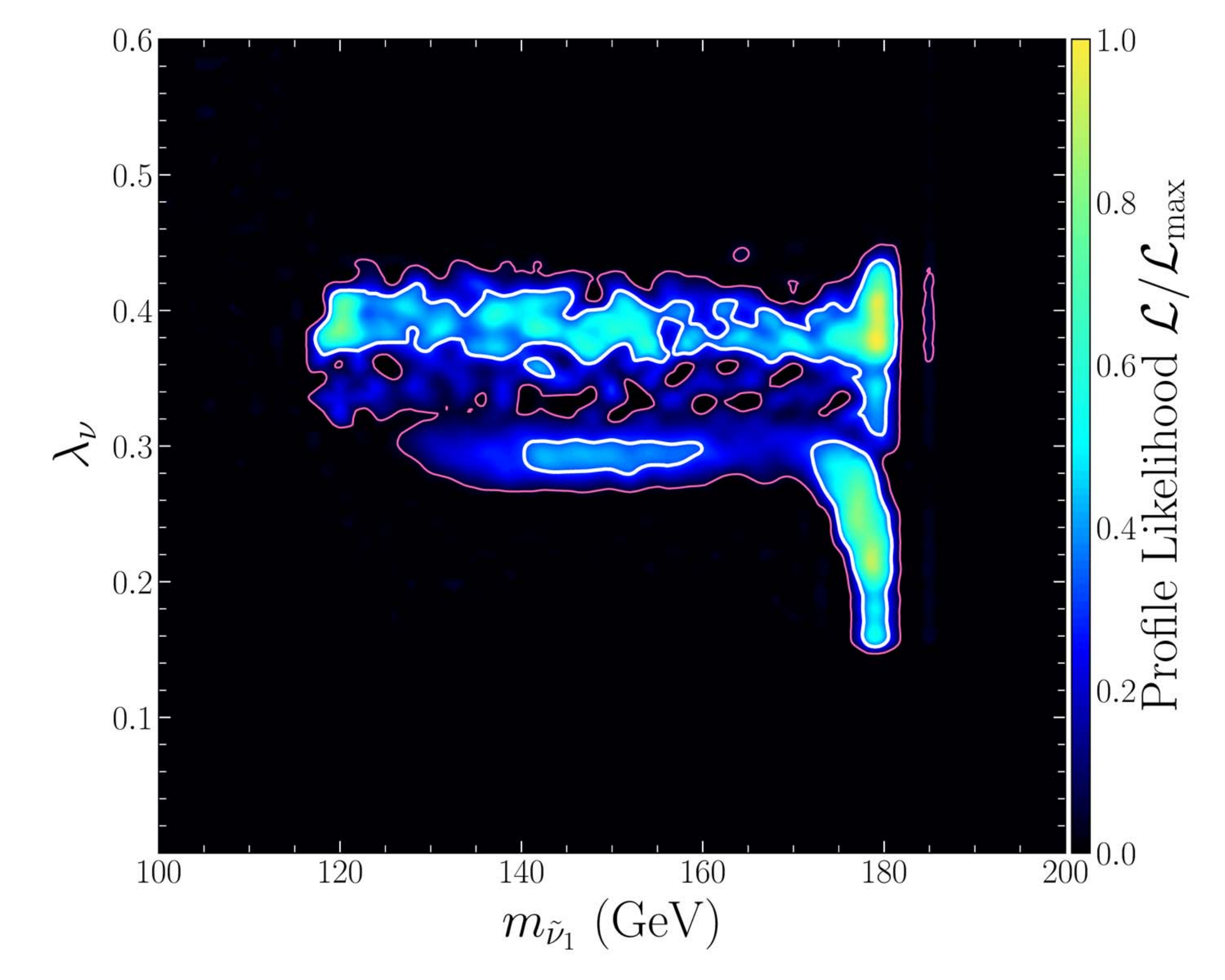}
        }
\caption{The profile likelihoods of the function $\mathcal{L}_{\rm DM}$ in Eq.~(\ref{DM-profile}) for the light $h_s$ scenario, projected onto $\lambda_\nu-m_{\tilde{\nu}_1}$ plane. The upper panels are the results for  the $B_{\mu_X} \neq 0$ case, where the bounds of the DM-nucleon scattering cross-section were taken from the XENON-1T (2018) experiment (left panel) and the future LZ experiment (right panel), respectively. The lower panels are same as the upper panels except that they are for the $B_{\mu_X} = 0$ case. Since $\chi^2_{\rm min} \simeq 0$ for the best point of the scans, the boundaries of the $1 \sigma$ and $2 \sigma$ confidence interval satisfy $\chi^2 \simeq 2.3$ and $\chi^2 \simeq 6.18$ and are marked with white and red solid line, respectively. This figure reflects the preference of the DM measurements on the parameters $\lambda_\nu$ and $m_{\tilde{\nu}_1}$.  \label{fig1}}
\end{figure*}	

\begin{figure*}[t]
		\centering
		\resizebox{0.92\textwidth}{!}{
        \includegraphics[width=0.90\textwidth]{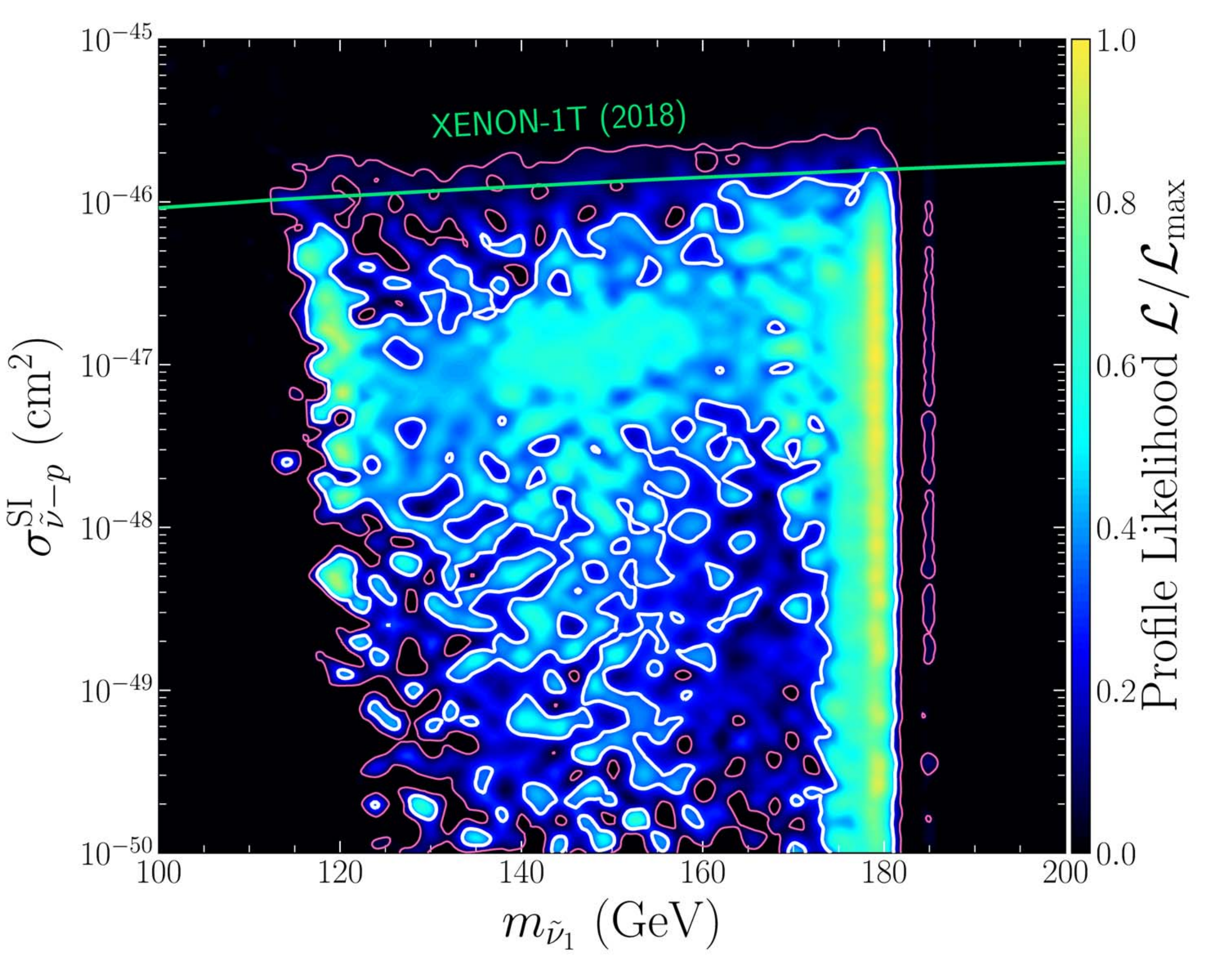}
        \includegraphics[width=0.90\textwidth]{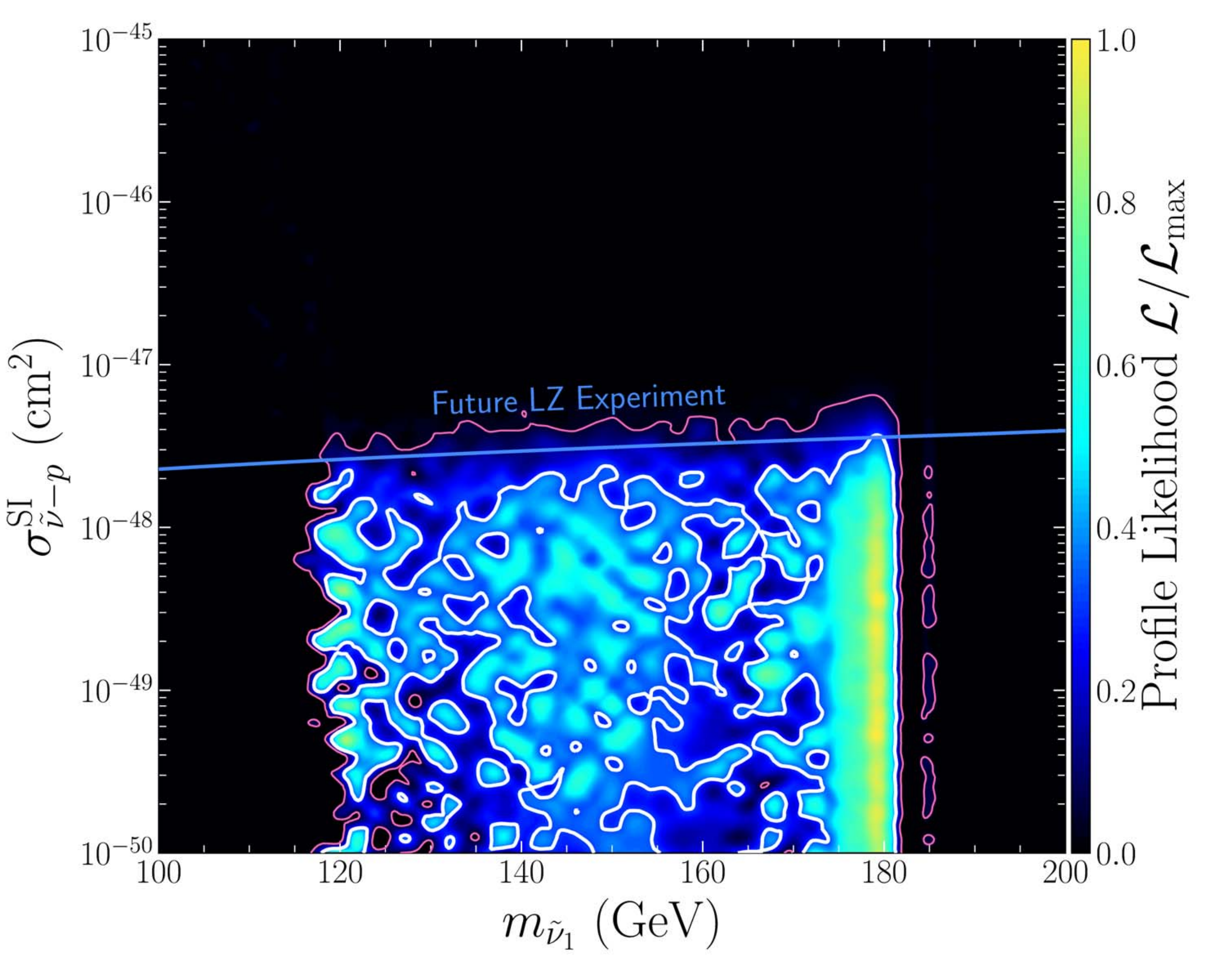}
        }

		\resizebox{0.92\textwidth}{!}{
        \includegraphics[width=0.90\textwidth]{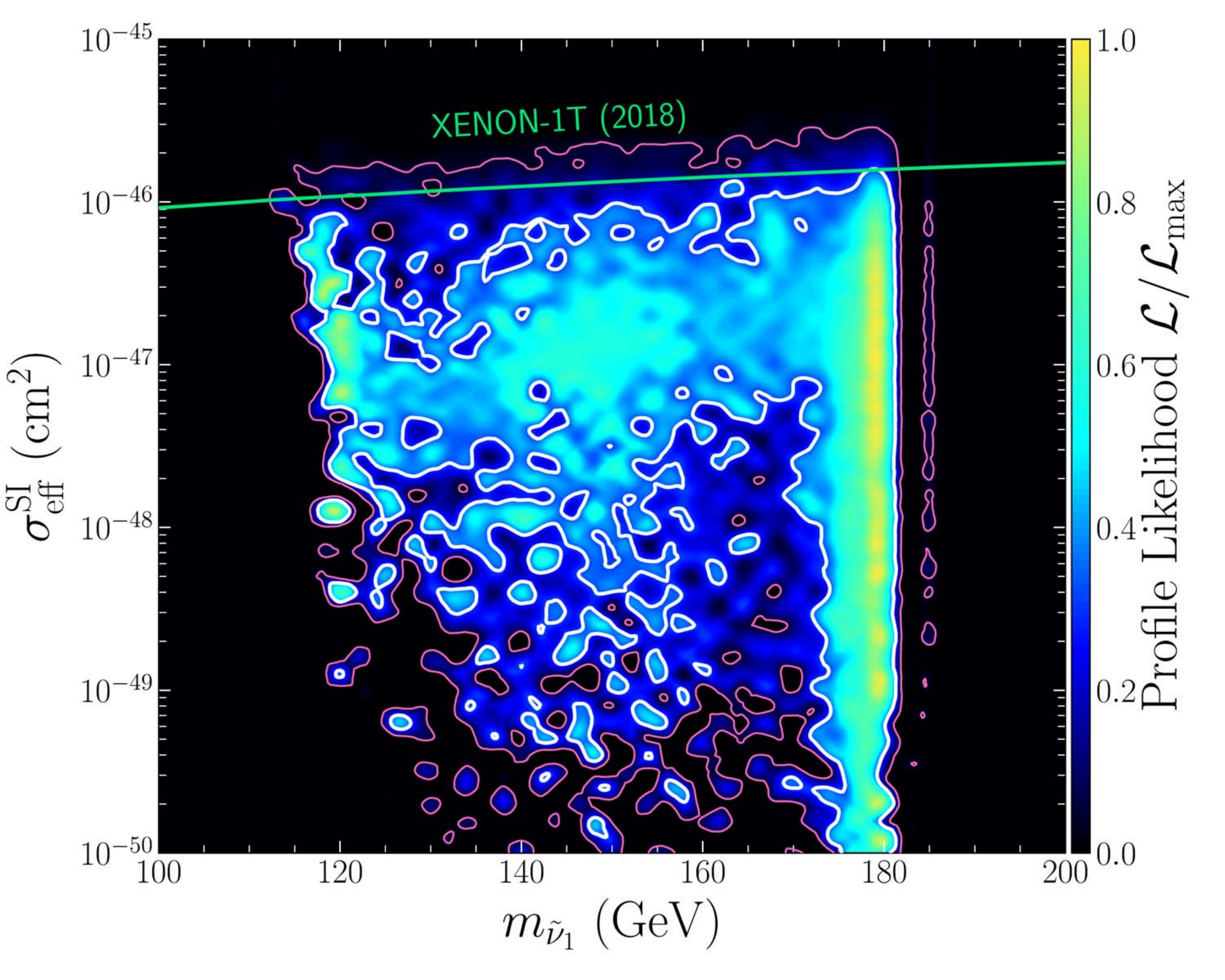}
        \includegraphics[width=0.90\textwidth]{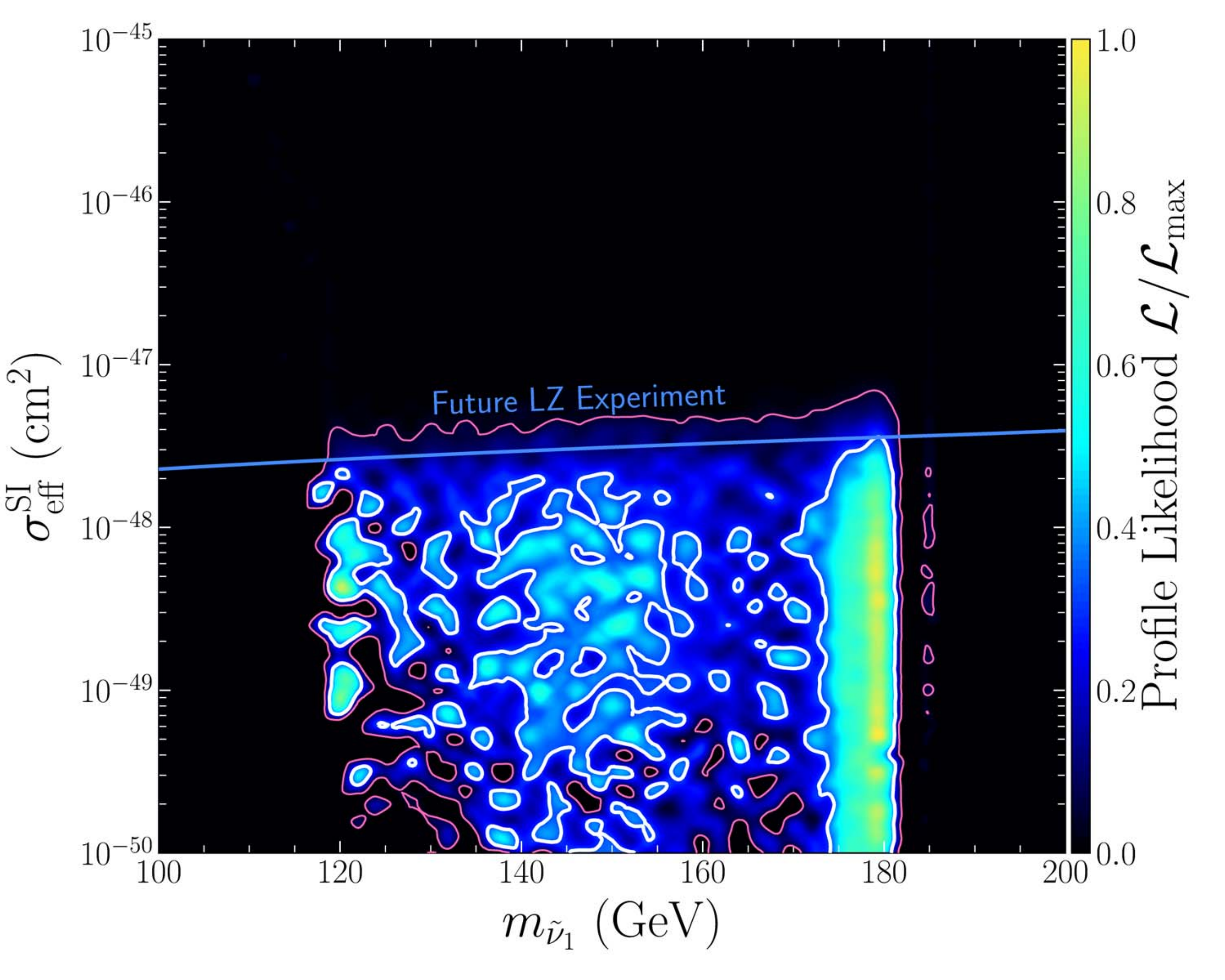}
        }
\caption{Same as Fig.\ref{fig1}, but for the profile likelihood projected onto $\sigma^{\rm SI}_{\tilde{\nu}-p} - m_{\tilde{\nu}_1} $ plane. \label{fig2}}
\end{figure*}

\begin{figure*}[t]
		\centering
		\resizebox{0.92\textwidth}{!}{
        \includegraphics[width=0.90\textwidth]{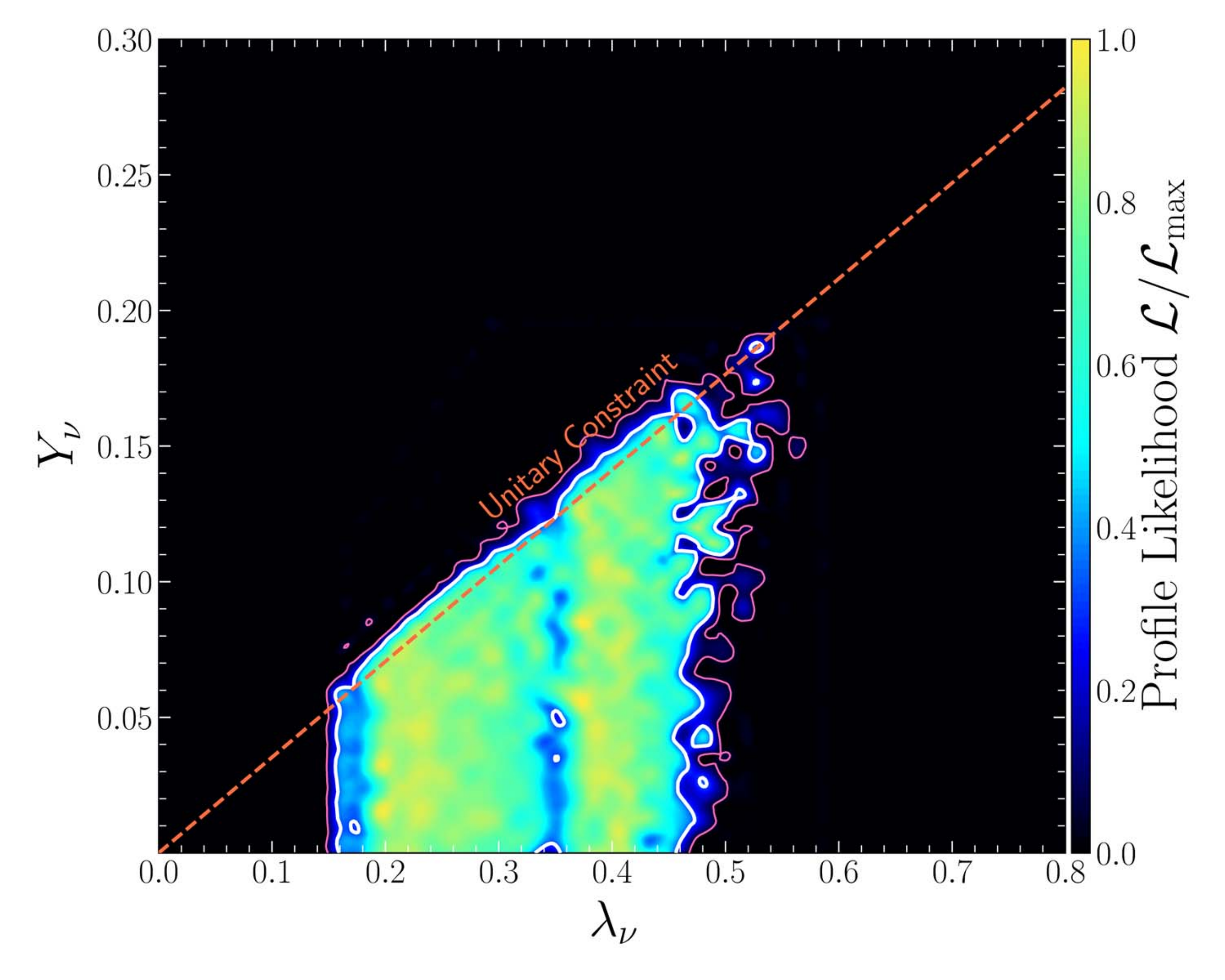}
        \includegraphics[width=0.90\textwidth]{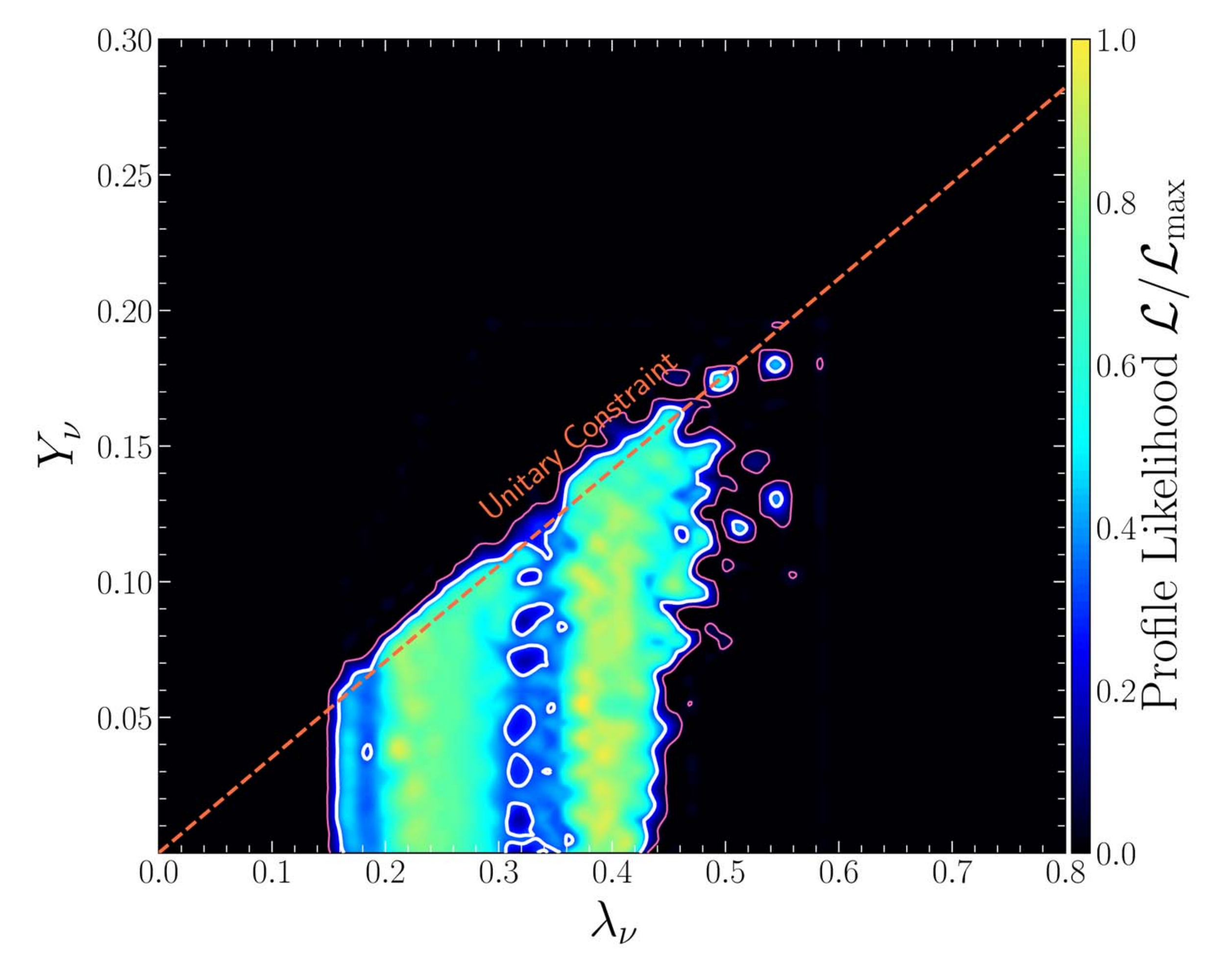}
        }

		\resizebox{0.92\textwidth}{!}{
        \includegraphics[width=0.90\textwidth]{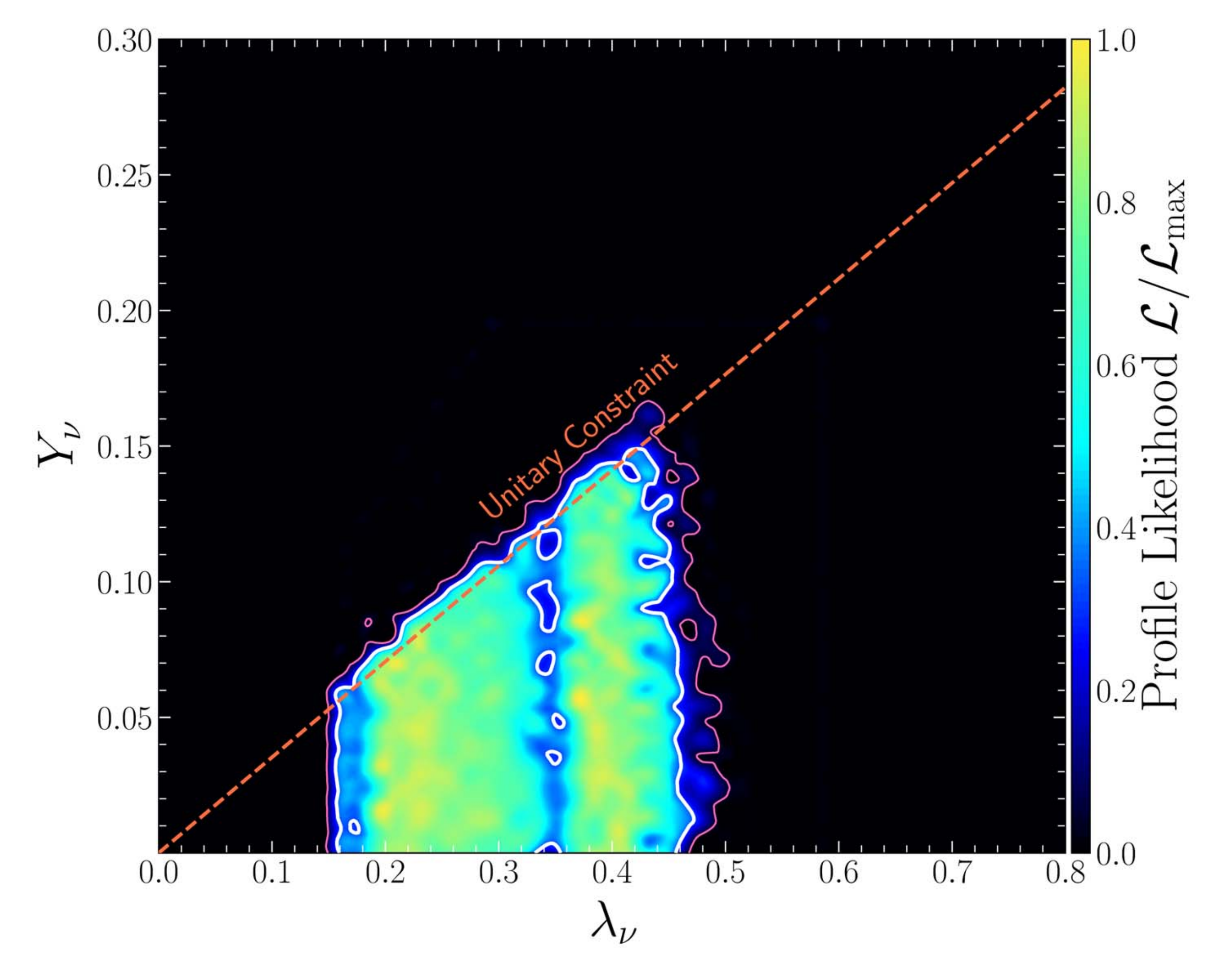}
        \includegraphics[width=0.90\textwidth]{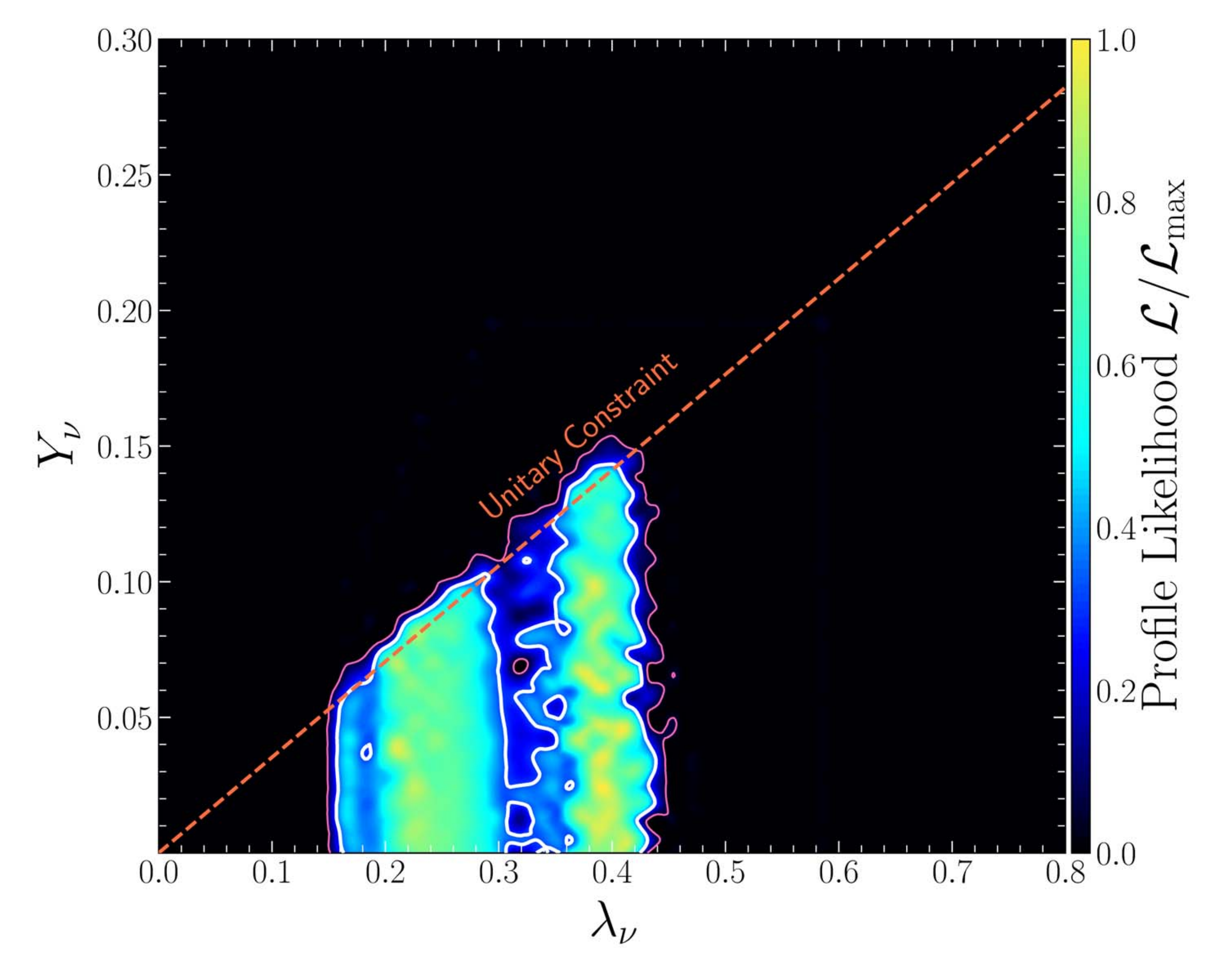}
        }
\caption{Same as Fig.\ref{fig1}, but for the profile likelihood projected onto $Y_\nu - \lambda_\nu $ plane, where
the red line denotes the leptonic unitarity bound. \label{fig3} }
\end{figure*}	
\begin{figure*}[t]
		\centering
		\resizebox{0.92 \textwidth}{!}{
        \includegraphics[width=0.90\textwidth]{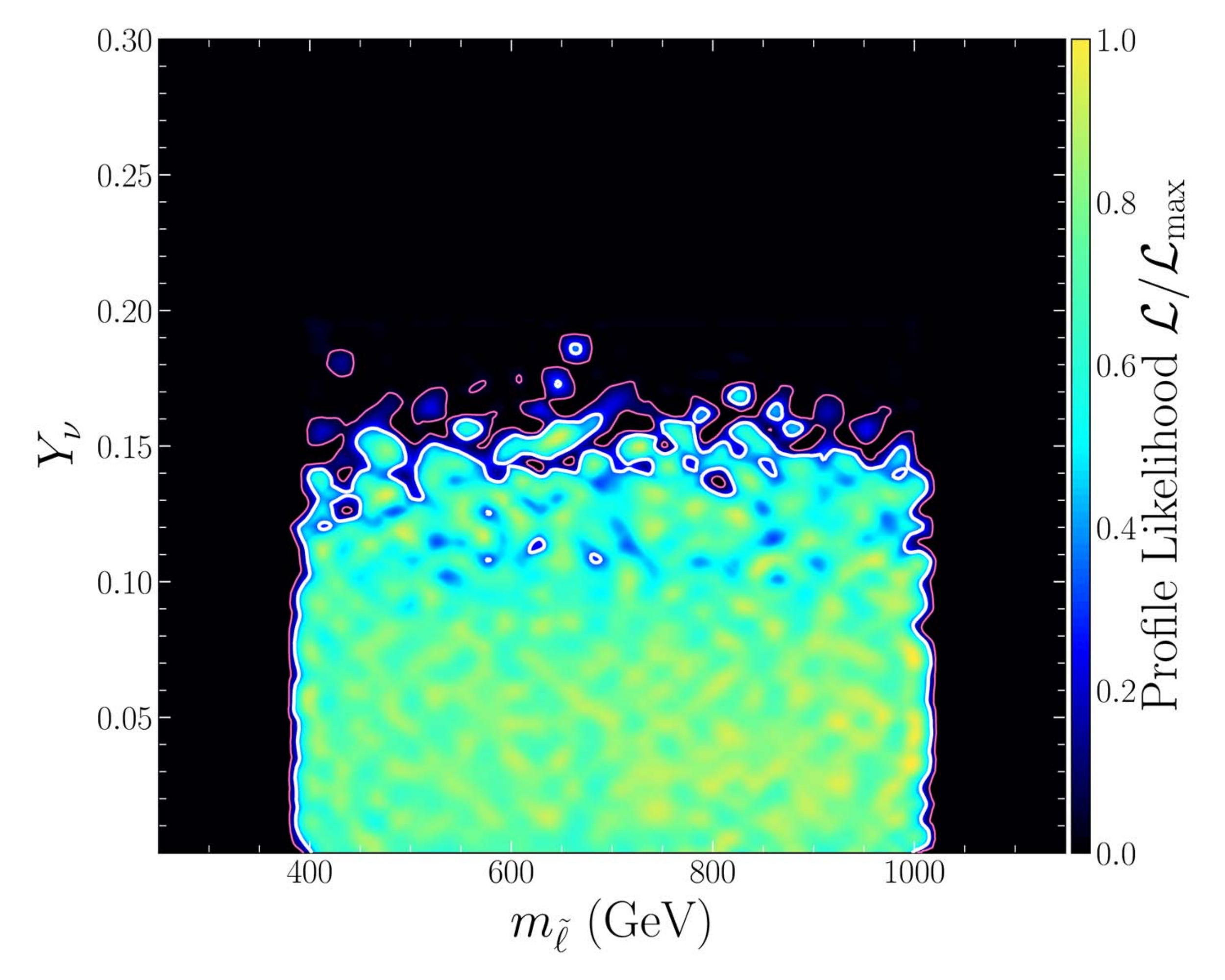}
        \includegraphics[width=0.90\textwidth]{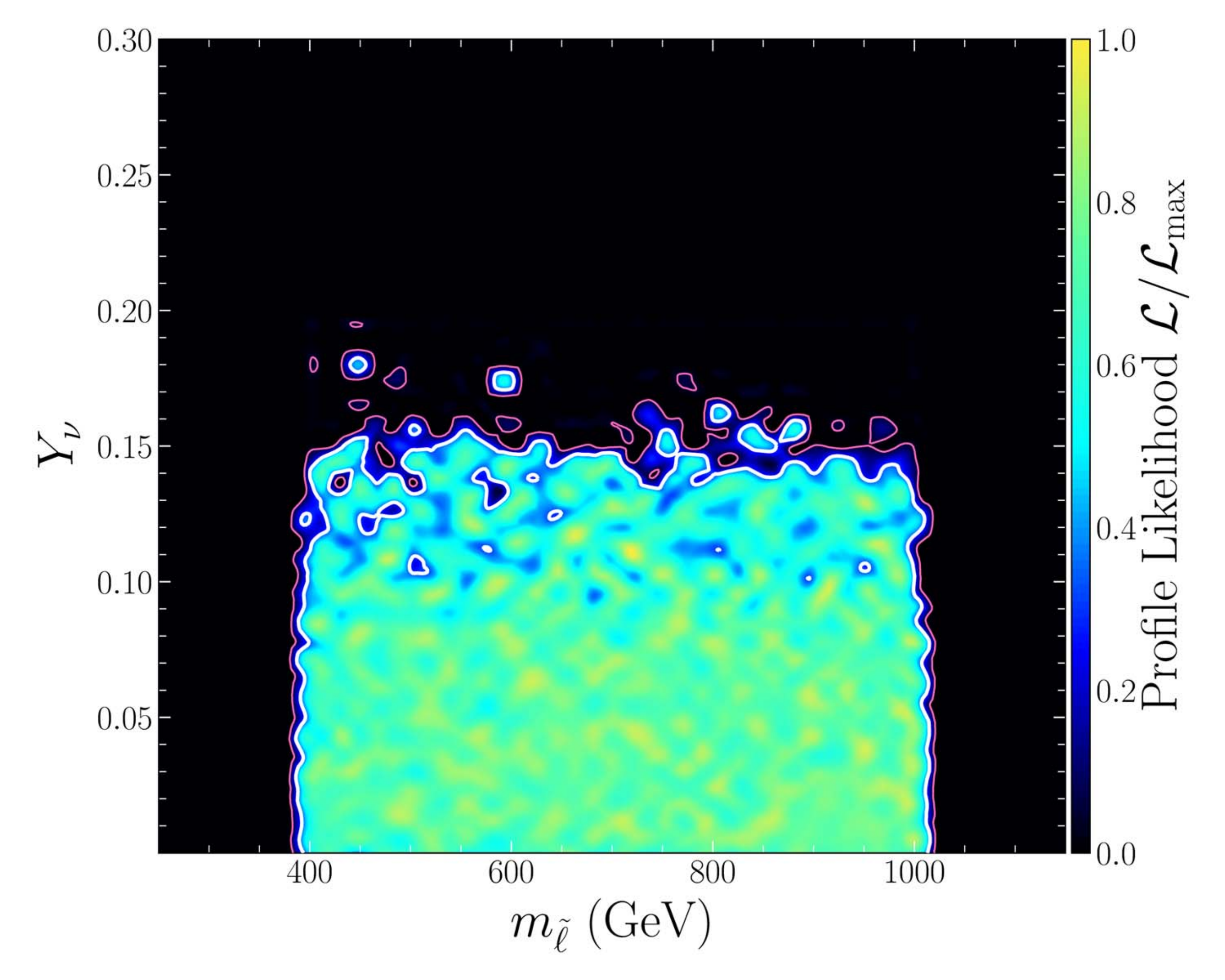}
        }

		\resizebox{0.92 \textwidth}{!}{
        \includegraphics[width=0.90\textwidth]{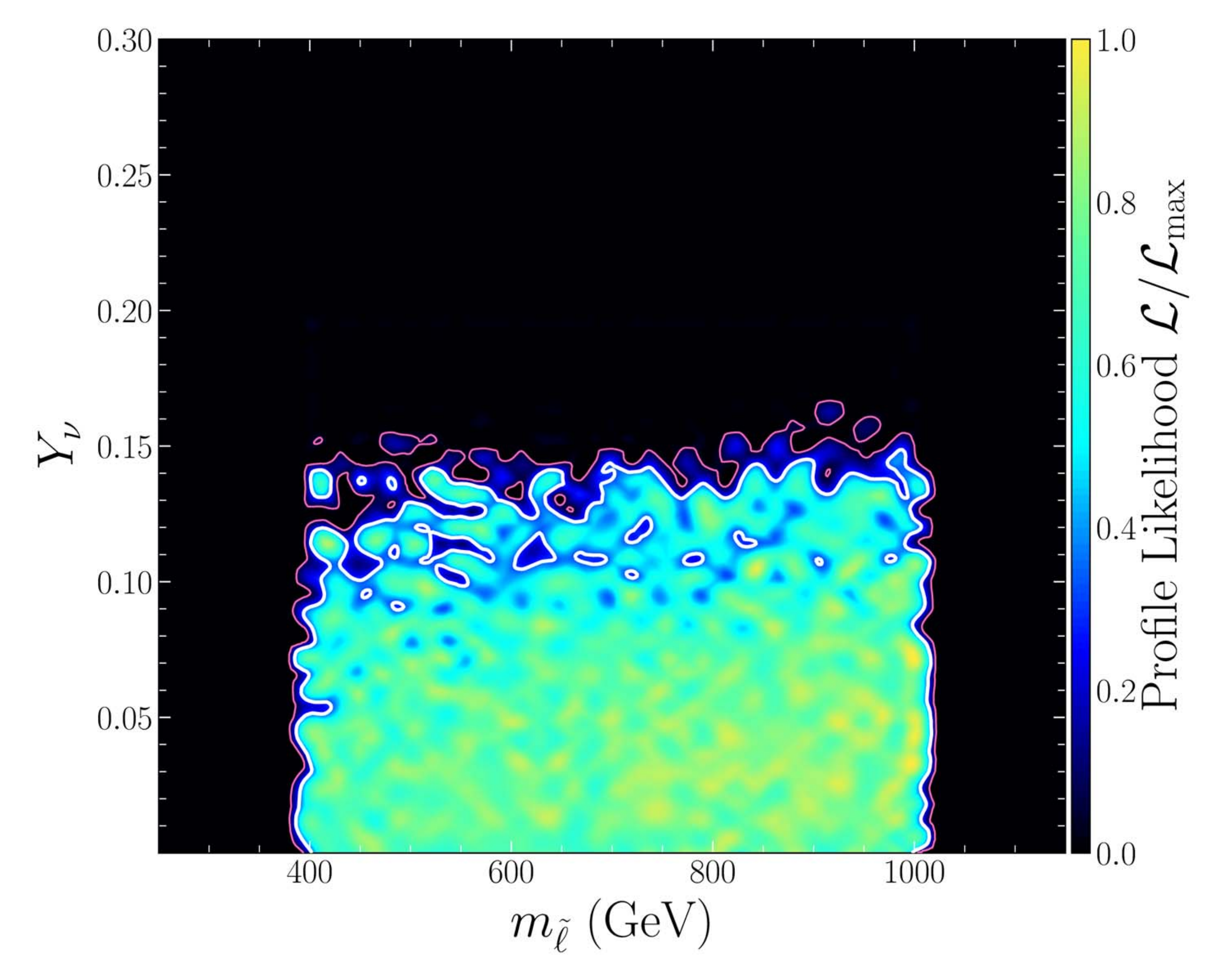}
        \includegraphics[width=0.90\textwidth]{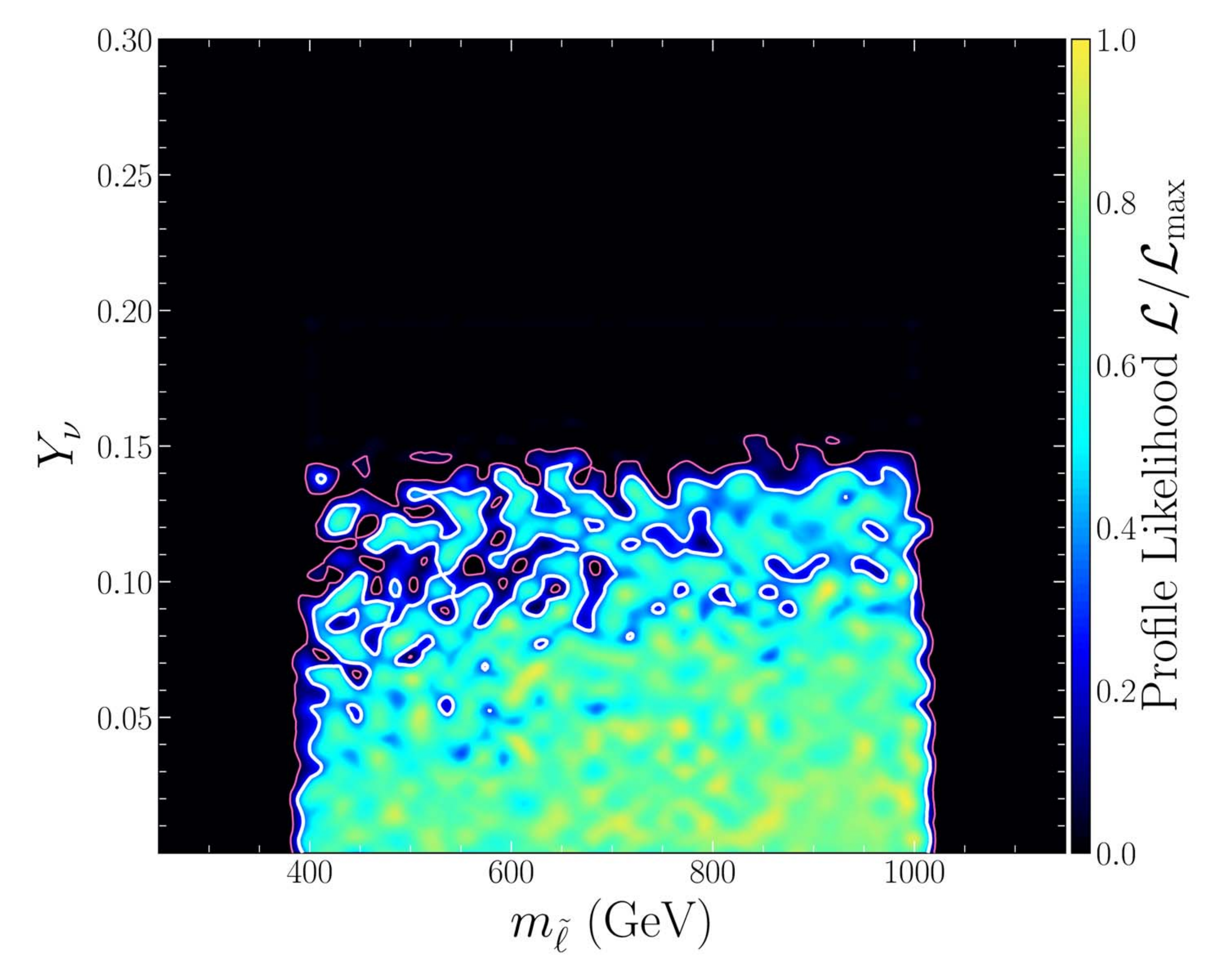}
        }
\caption{Same as Fig.\ref{fig1}, but for the profile likelihood projected onto $Y_\nu - m_{\tilde{l}} $ plane. \label{fig4}}
\end{figure*}

\begin{figure*}[t]
		\centering
		\resizebox{0.92\textwidth}{!}{
        \includegraphics[width=0.90\textwidth]{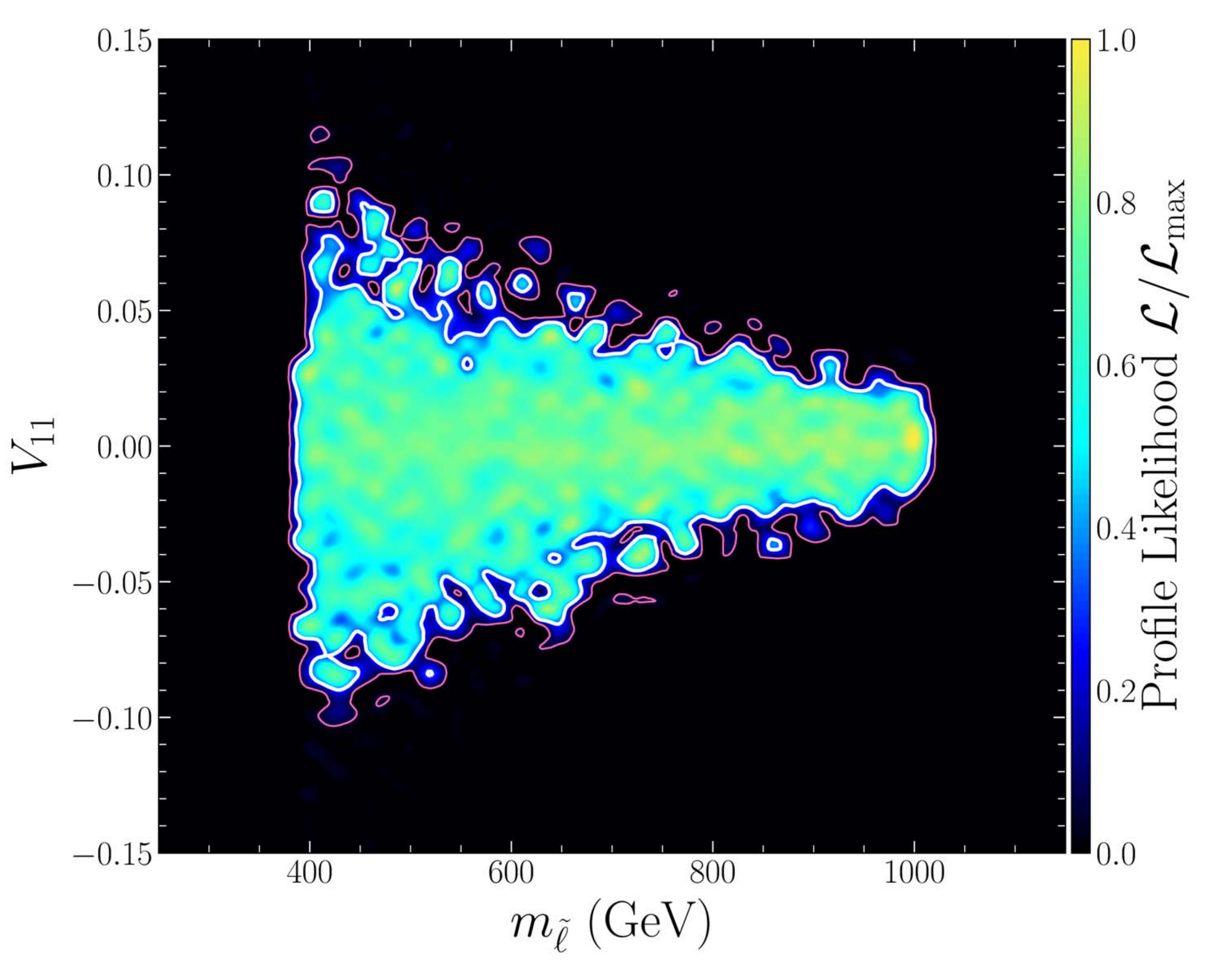}
        \includegraphics[width=0.90\textwidth]{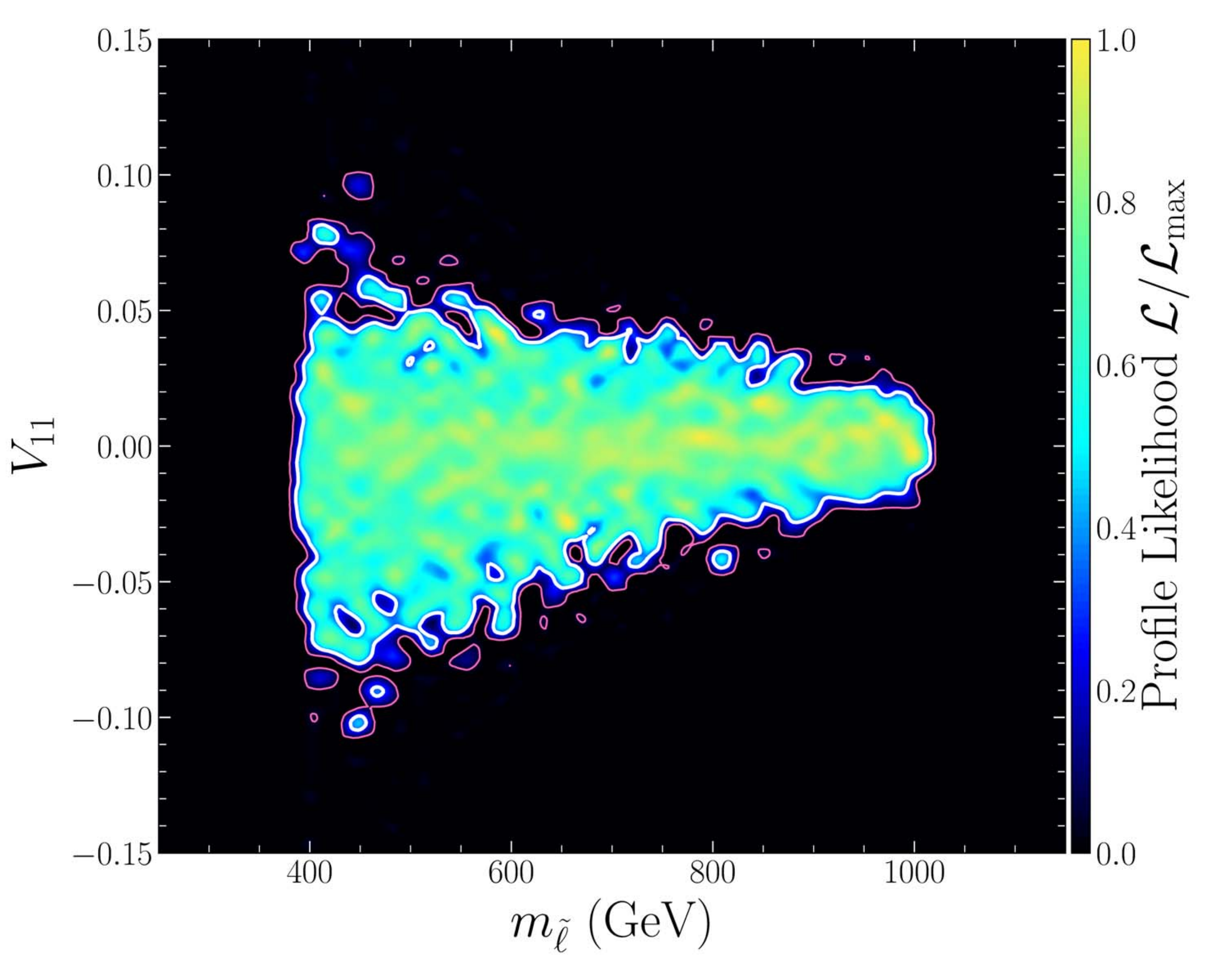}
        }

		\resizebox{0.92\textwidth}{!}{
        \includegraphics[width=0.90\textwidth]{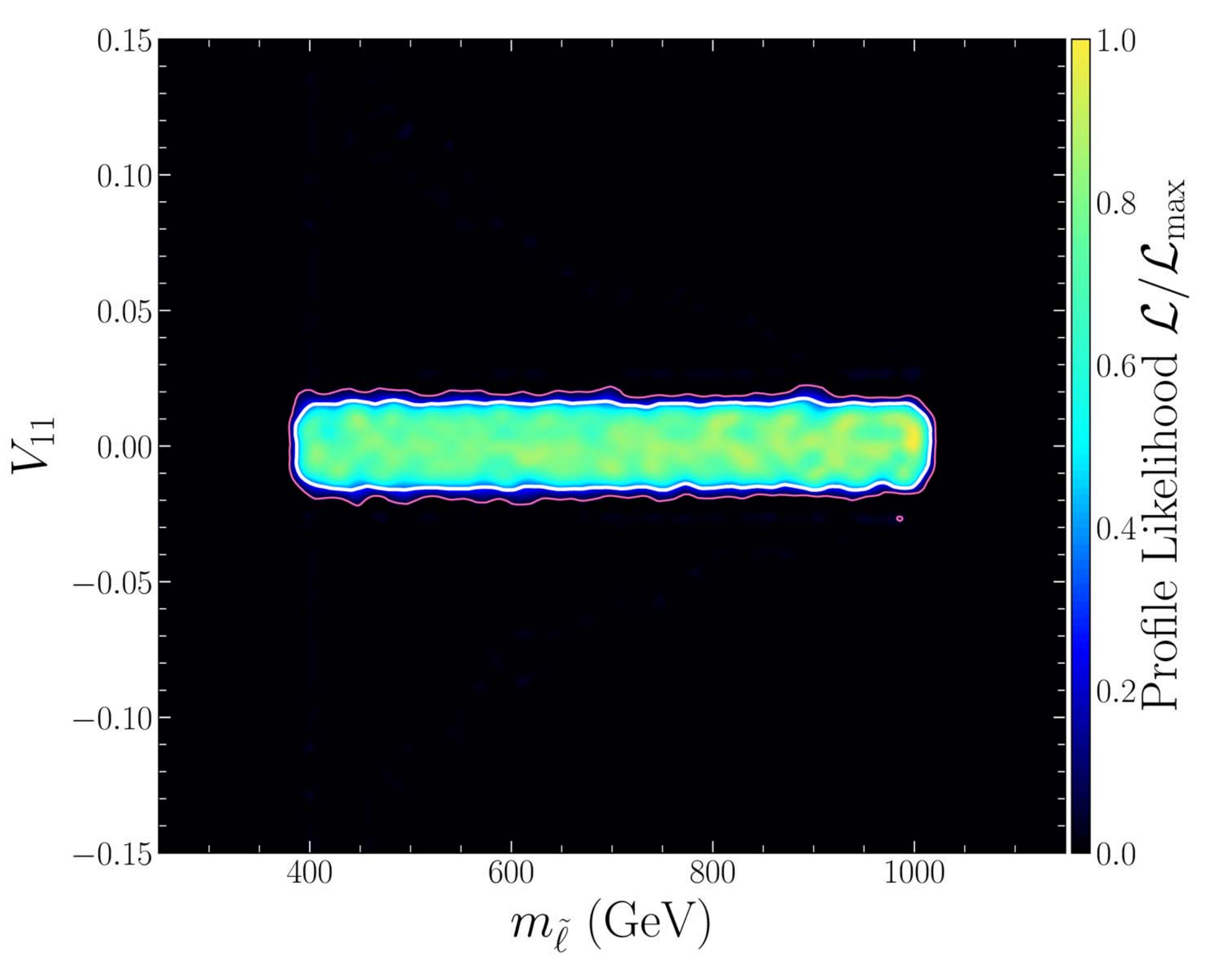}
        \includegraphics[width=0.90\textwidth]{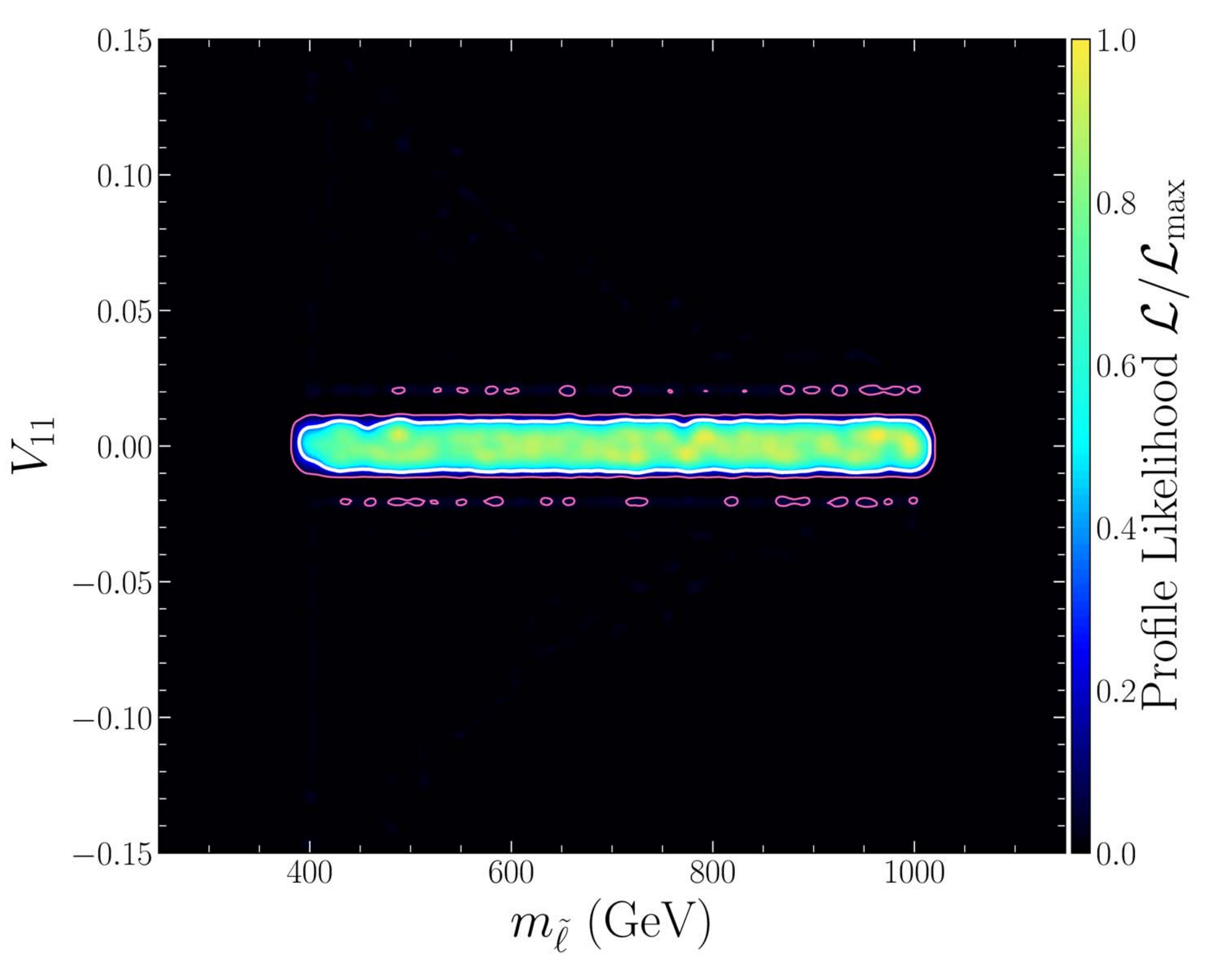}
        }
\caption{Same as Fig.\ref{fig1}, but for the profile likelihood projected onto $V_{11} - m_{\tilde{l}} $ plane. \label{fig5}}
\end{figure*}

\begin{table}[]
\scalebox{0.65}{
\begin{tabular}{|cccccccccc|}
\hline\hline
\multicolumn{5}{|c|}{Light $h_s$ scenario}   & \multicolumn{5}{c|}{Massive $h_s$ scenario}    \\ \hline
    \multicolumn{1}{|c|}{} & \multicolumn{2}{c|}{$P_1$} & \multicolumn{2}{c|}{$P_2$} & \multicolumn{1}{c|}{} & \multicolumn{2}{c|}{$P_3$} & \multicolumn{2}{c|}{$P_4$}    \\ \hline
    \multicolumn{1}{|c|}{$Y_\nu$} & \multicolumn{2}{c|}{0.060} & \multicolumn{2}{c|}{0.035} & \multicolumn{1}{c|}{$Y_\nu$} & \multicolumn{2}{c|}{0.178} & \multicolumn{2}{c|}{0.227}    \\
    \multicolumn{1}{|c|}{$\lambda_\nu$} & \multicolumn{2}{c|}{0.384} & \multicolumn{2}{c|}{0.411} & \multicolumn{1}{c|}{$\lambda_\nu$} & \multicolumn{2}{c|}{0.473} & \multicolumn{2}{c|}{0.364}    \\
    \multicolumn{1}{|c|}{$A_{Y_\nu}$} & \multicolumn{2}{c|}{731.6} & \multicolumn{2}{c|}{-722.1} & \multicolumn{1}{c|}{$A_{Y_\nu}$} & \multicolumn{2}{c|}{297.8} & \multicolumn{2}{c|}{225.2}    \\
    \multicolumn{1}{|c|}{$A_{\lambda_\nu}$} & \multicolumn{2}{c|}{463.3} & \multicolumn{2}{c|}{396.3} & \multicolumn{1}{c|}{$A_{\lambda_\nu}$} & \multicolumn{2}{c|}{-288.8} & \multicolumn{2}{c|}{-295.0}    \\
    \multicolumn{1}{|c|}{$m_{\tilde\nu}$} & \multicolumn{2}{c|}{495.7} & \multicolumn{2}{c|}{348.4} & \multicolumn{1}{c|}{$m_{\tilde\nu}$} & \multicolumn{2}{c|}{191.8} & \multicolumn{2}{c|}{375.9}    \\
    \multicolumn{1}{|c|}{$m_{\tilde x}$} & \multicolumn{2}{c|}{217.6} & \multicolumn{2}{c|}{333.1} & \multicolumn{1}{c|}{$m_{\tilde x}$} & \multicolumn{2}{c|}{269.6} & \multicolumn{2}{c|}{442.7}    \\
    \multicolumn{1}{|c|}{$m_{\tilde l}$} & \multicolumn{2}{c|}{857.2} & \multicolumn{2}{c|}{816.5} & \multicolumn{1}{c|}{$m_{\tilde l}$} & \multicolumn{2}{c|}{875.3} & \multicolumn{2}{c|}{568.9}    \\
    \multicolumn{1}{|c|}{$m_{\tilde\nu_1}$} & \multicolumn{2}{c|}{132.6} & \multicolumn{2}{c|}{179.4} & \multicolumn{1}{c|}{$m_{\tilde\nu_1}$} & \multicolumn{2}{c|}{313.0} & \multicolumn{2}{c|}{299.6}    \\
    \multicolumn{1}{|c|}{$V_{11}$} & \multicolumn{2}{c|}{-0.002} & \multicolumn{2}{c|}{-0.007} & \multicolumn{1}{c|}{$V_{11}$} & \multicolumn{2}{c|}{-0.014} & \multicolumn{2}{c|}{-0.031}    \\
    \multicolumn{1}{|c|}{$V_{12}$} & \multicolumn{2}{c|}{0.526} & \multicolumn{2}{c|}{-0.697} & \multicolumn{1}{c|}{$V_{12}$} & \multicolumn{2}{c|}{-0.717} & \multicolumn{2}{c|}{-0.726}    \\
    \multicolumn{1}{|c|}{$V_{13}$} & \multicolumn{2}{c|}{-0.851} & \multicolumn{2}{c|}{0.717} & \multicolumn{1}{c|}{$V_{13}$} & \multicolumn{2}{c|}{0.697} & \multicolumn{2}{c|}{0.687}    \\
    \multicolumn{1}{|c|}{$\Omega h^2$} & \multicolumn{2}{c|}{0.108} & \multicolumn{2}{c|}{0.120} & \multicolumn{1}{c|}{$\Omega h^2$} & \multicolumn{2}{c|}{0.119} & \multicolumn{2}{c|}{0.126}    \\
    \multicolumn{1}{|c|}{$C_{\tilde\nu_1 \tilde\nu_1^\ast h_s}$} & \multicolumn{2}{c|}{34.80} & \multicolumn{2}{c|}{30.95} & \multicolumn{1}{c|}{$C_{\tilde\nu_1 \tilde\nu_1^\ast h_s}$} & \multicolumn{2}{c|}{-53.54} & \multicolumn{2}{c|}{-134.9}    \\
    \multicolumn{1}{|c|}{$C_{\tilde\nu_1  \tilde\nu_1^\ast h}$} & \multicolumn{2}{c|}{8.097} & \multicolumn{2}{c|}{7.481} & \multicolumn{1}{c|}{$C_{\tilde\nu_1  \tilde\nu_1^\ast h}$} & \multicolumn{2}{c|}{3.694} & \multicolumn{2}{c|}{4.927}    \\
    \multicolumn{1}{|c|}{$\sigma^{SI}_{\tilde\nu_1-p}$} & \multicolumn{2}{c|}{$3.64 \times 10^{-48}$} & \multicolumn{2}{c|}{$1.90 \times 10^{-48}$} & \multicolumn{1}{c|}{$\sigma^{SI}_{\tilde\nu_1-p}$} & \multicolumn{2}{c|}{$2.28 \times 10^{-48}$} & \multicolumn{2}{c|}{$4.91 \times 10^{-47}$}    \\
    \multicolumn{1}{|c|}{$\sigma^{SI}_{\tilde\nu_1-n}$} & \multicolumn{2}{c|}{$4.34 \times 10^{-48}$} & \multicolumn{2}{c|}{$8.00 \times 10^{-48}$} & \multicolumn{1}{c|}{$\sigma^{SI}_{\tilde\nu_1-n}$} & \multicolumn{2}{c|}{$3.94 \times 10^{-46}$} & \multicolumn{2}{c|}{$7.64 \times 10^{-45}$}    \\
    \multicolumn{1}{|c|}{} & \multicolumn{1}{c}{} & \multicolumn{1}{c|}{} & \multicolumn{1}{c}{} & \multicolumn{1}{c|}{} & \multicolumn{1}{c|}{} & \multicolumn{1}{c}{} & \multicolumn{1}{c|}{} & \multicolumn{1}{c}{} & \multicolumn{1}{c|}{} \\
    \multicolumn{1}{|c|}{} & \multicolumn{1}{l}{} & \multicolumn{1}{l|}{} & \multicolumn{1}{l}{$16\%$} & \multicolumn{1}{l|}{$\tilde\chi^0_1 \tilde\chi^0_1 \rightarrow W^+ W^-$} & \multicolumn{1}{l|}{} & \multicolumn{1}{l}{$11\%$} & \multicolumn{1}{l|}{$\tilde\chi^0_1 \tilde\chi^-_1 \rightarrow d \bar{u}$} & \multicolumn{1}{l}{$42\%$} & \multicolumn{1}{l|}{$\tilde\nu_1 \tilde\chi^0_1 \rightarrow W^+ \tau^-$} \\
    \multicolumn{1}{|c|}{}  & \multicolumn{1}{l}{} & \multicolumn{1}{l|}{} & \multicolumn{1}{l}{$13\%$} & \multicolumn{1}{l|}{$\tilde\chi^0_1 \tilde\chi^-_1 \rightarrow d \bar{u}$} & \multicolumn{1}{l|}{} & \multicolumn{1}{l}{$11\%$} & \multicolumn{1}{l|}{$\tilde\chi^0_1 \tilde\chi^-_1 \rightarrow s \bar{c}$} & \multicolumn{1}{l}{$21\%$} & \multicolumn{1}{l|}{$\tilde\nu_1 \tilde\chi^0_1 \rightarrow Z \nu_\tau$} \\
    \multicolumn{1}{|c|}{annihilation} & \multicolumn{1}{l}{$89\%$} & \multicolumn{1}{l|}{$\tilde\nu_1 \tilde\nu_1^\ast \rightarrow h_s h_s$} & \multicolumn{1}{l}{$13\%$} & \multicolumn{1}{l|}{$\tilde\chi^0_1 \tilde\chi^-_1 \rightarrow s \bar{c}$} & \multicolumn{1}{c|}{annihilation} & \multicolumn{1}{l}{$11\%$} & \multicolumn{1}{l|}{$\tilde\chi^0_1 \tilde\chi^0_1 \rightarrow W^+ W^-$} & \multicolumn{1}{l}{$19\%$} & \multicolumn{1}{l|}{$\tilde\nu_1 \tilde\chi^0_1 \rightarrow h \nu_\tau$} \\
    \multicolumn{1}{|c|}{processes} & \multicolumn{1}{l}{$10\%$} & \multicolumn{1}{l|}{$\tilde\nu_1 \tilde\nu_1^\ast \rightarrow h_s h$} & \multicolumn{1}{l}{$11\%$} & \multicolumn{1}{l|}{$\tilde\chi^0_1 \tilde\chi^0_1 \rightarrow Z Z$} & \multicolumn{1}{c|}{processes} & \multicolumn{1}{l}{$8.9\%$} & \multicolumn{1}{l|}{$\tilde\chi^0_1 \tilde\chi^-_1 \rightarrow b \bar{t}$} & \multicolumn{1}{l}{$1.6\%$} & \multicolumn{1}{l|}{$\tilde\chi^0_1 \tilde\chi^-_1 \rightarrow d \bar{u}$} \\
    \multicolumn{1}{|c|}{} & \multicolumn{1}{l}{} & \multicolumn{1}{l|}{} & \multicolumn{1}{l}{$8.9\%$} & \multicolumn{1}{l|}{$\tilde\chi^0_1 \tilde\chi^-_1 \rightarrow b \bar{t}$} & \multicolumn{1}{c|}{} & \multicolumn{1}{l}{$7.4\%$} & \multicolumn{1}{l|}{$\tilde\chi^0_1 \tilde\chi^0_1 \rightarrow Z Z$} & \multicolumn{1}{l}{$1.6\%$} & \multicolumn{1}{l|}{$\tilde\chi^0_1 \tilde\chi^-_1 \rightarrow s \bar{c}$} \\
    \multicolumn{1}{|c|}{} & \multicolumn{1}{l}{} & \multicolumn{1}{l|}{} & \multicolumn{1}{c}{...} & \multicolumn{1}{l|}{\qquad.~.~.} & \multicolumn{1}{c|}{} & \multicolumn{1}{c}{...} & \multicolumn{1}{l|}{\qquad.~.~.} & \multicolumn{1}{c}{...} & \multicolumn{1}{l|}{\qquad.~.~.} \\
    \multicolumn{1}{|l|}{}                  & \multicolumn{1}{l}{} & \multicolumn{1}{l|}{} & \multicolumn{1}{l}{} & \multicolumn{1}{l|}{} & \multicolumn{1}{l|}{}                  & \multicolumn{1}{l}{} & \multicolumn{1}{l|}{} & \multicolumn{1}{l}{} & \multicolumn{1}{l|}{} \\
    \hline\hline
    \end{tabular}}
    \caption{Detailed information of the points in the light $h_s$ scenario (left side of the table) and the massive $h_s$ scenario (right side of the table) with the setting $B_{\mu_X} = 0$.  The number before each annihilation process represents the fraction of its contribution to the total DM annihilation cross section at the freeze-out temperature. Parameters in mass dimension are in unit of GeV, and the DM-nucleon scattering cross section are in unit of ${\rm cm^2}$.     } \label{table2}
    \end{table}

\begin{figure*}[htpb]
		\centering
		\resizebox{0.92\textwidth}{!}{
        \includegraphics[width=0.90\textwidth]{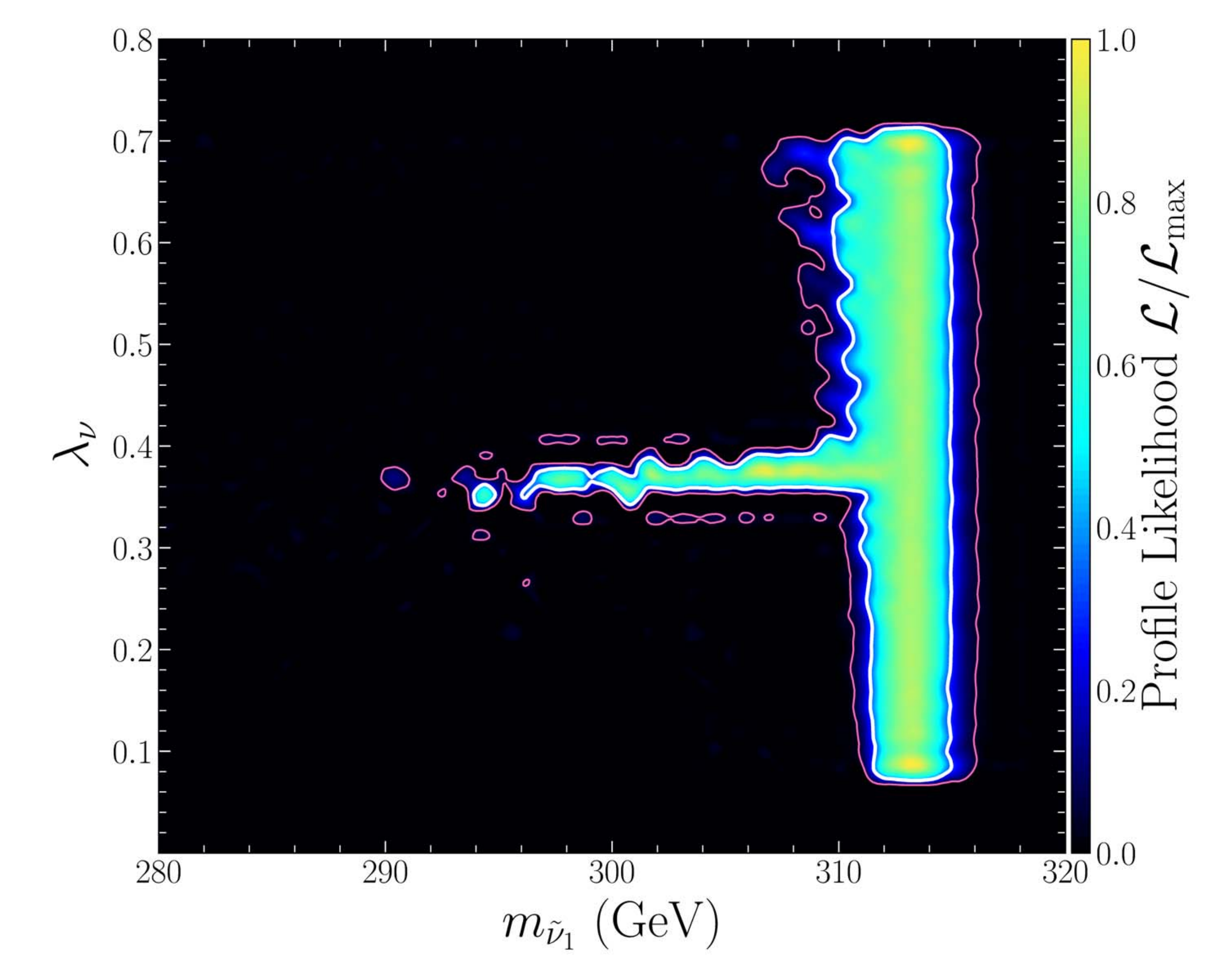}
        \includegraphics[width=0.90\textwidth]{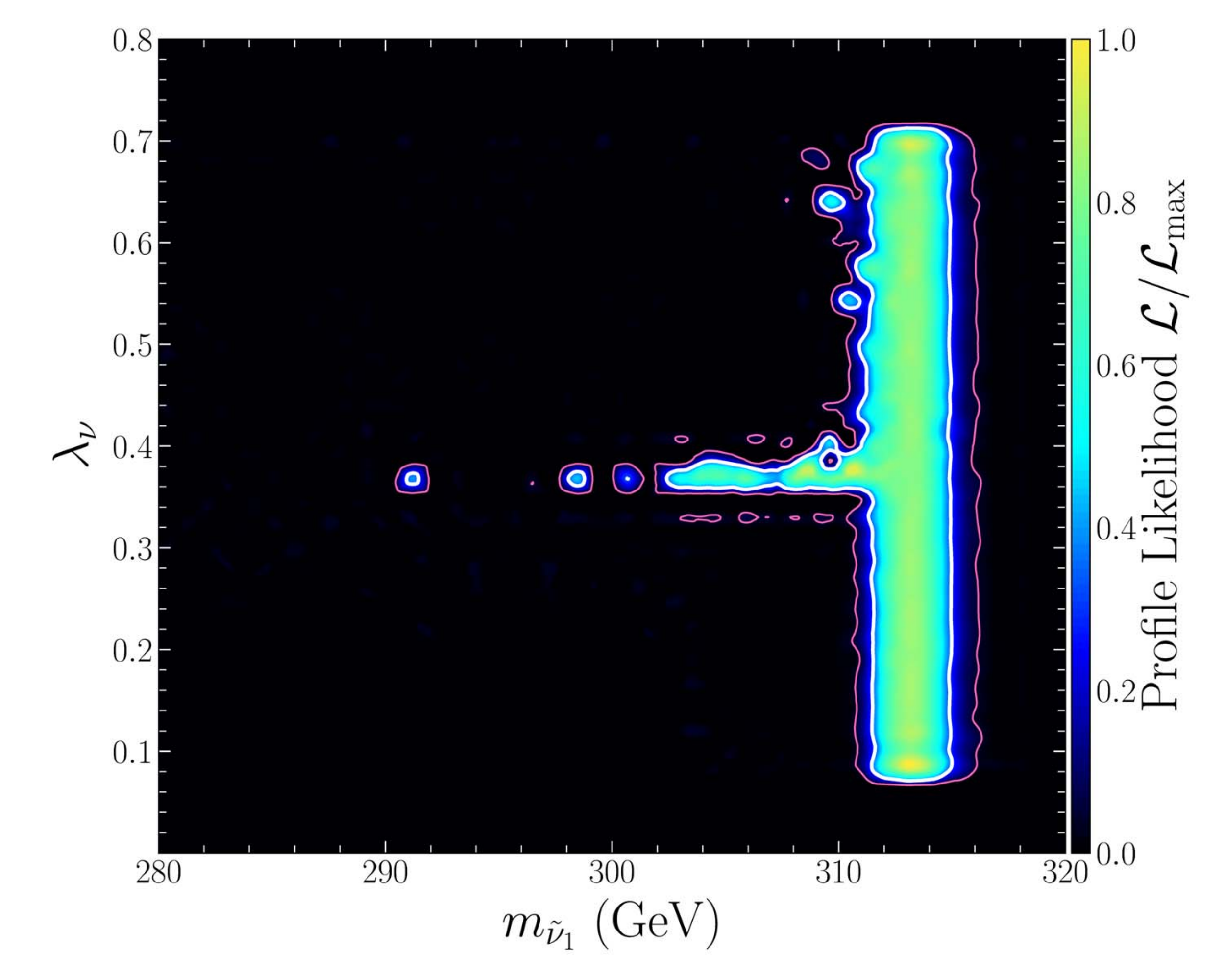}
        }

		\resizebox{0.92\textwidth}{!}{
        \includegraphics[width=0.90\textwidth]{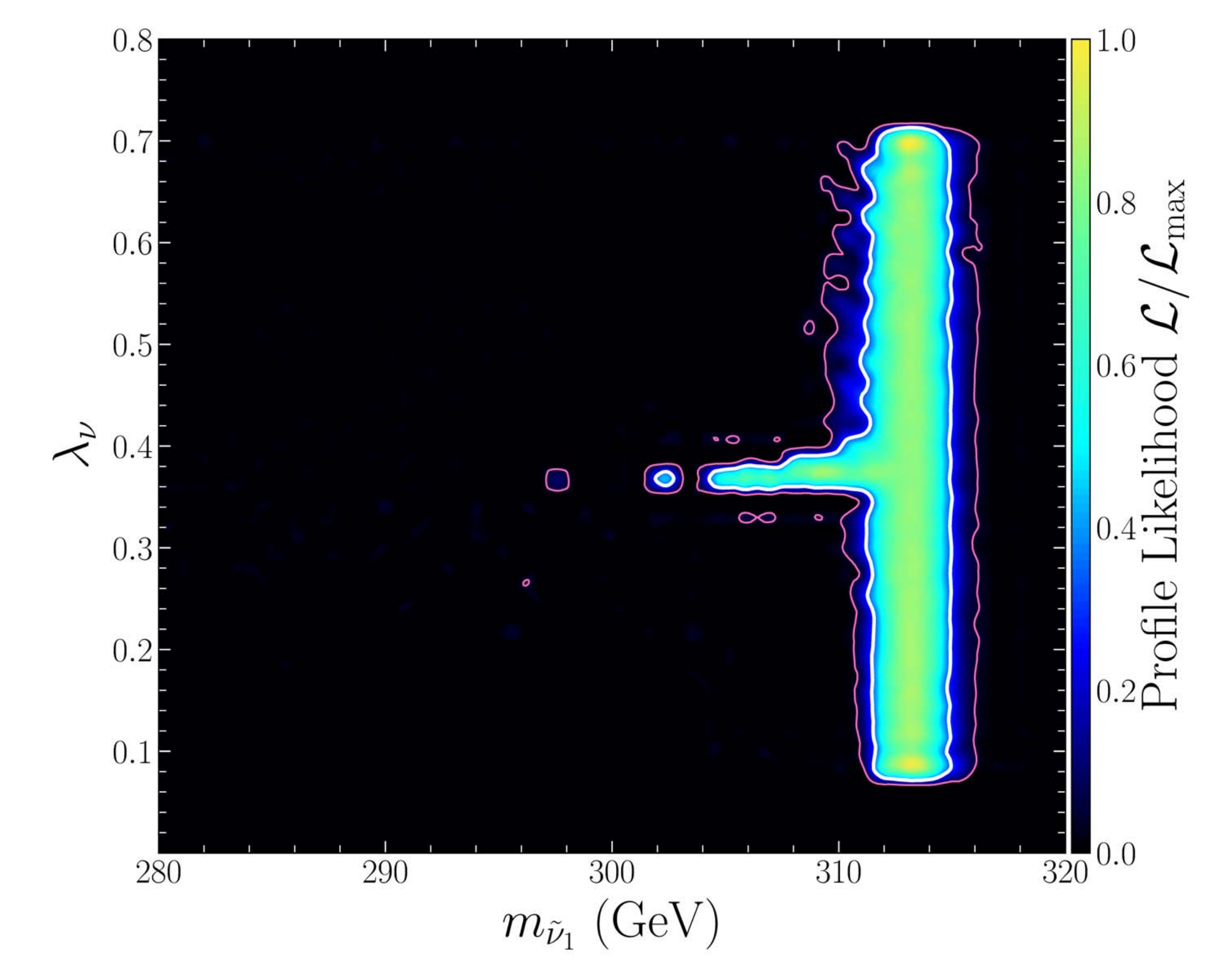}
        \includegraphics[width=0.90\textwidth]{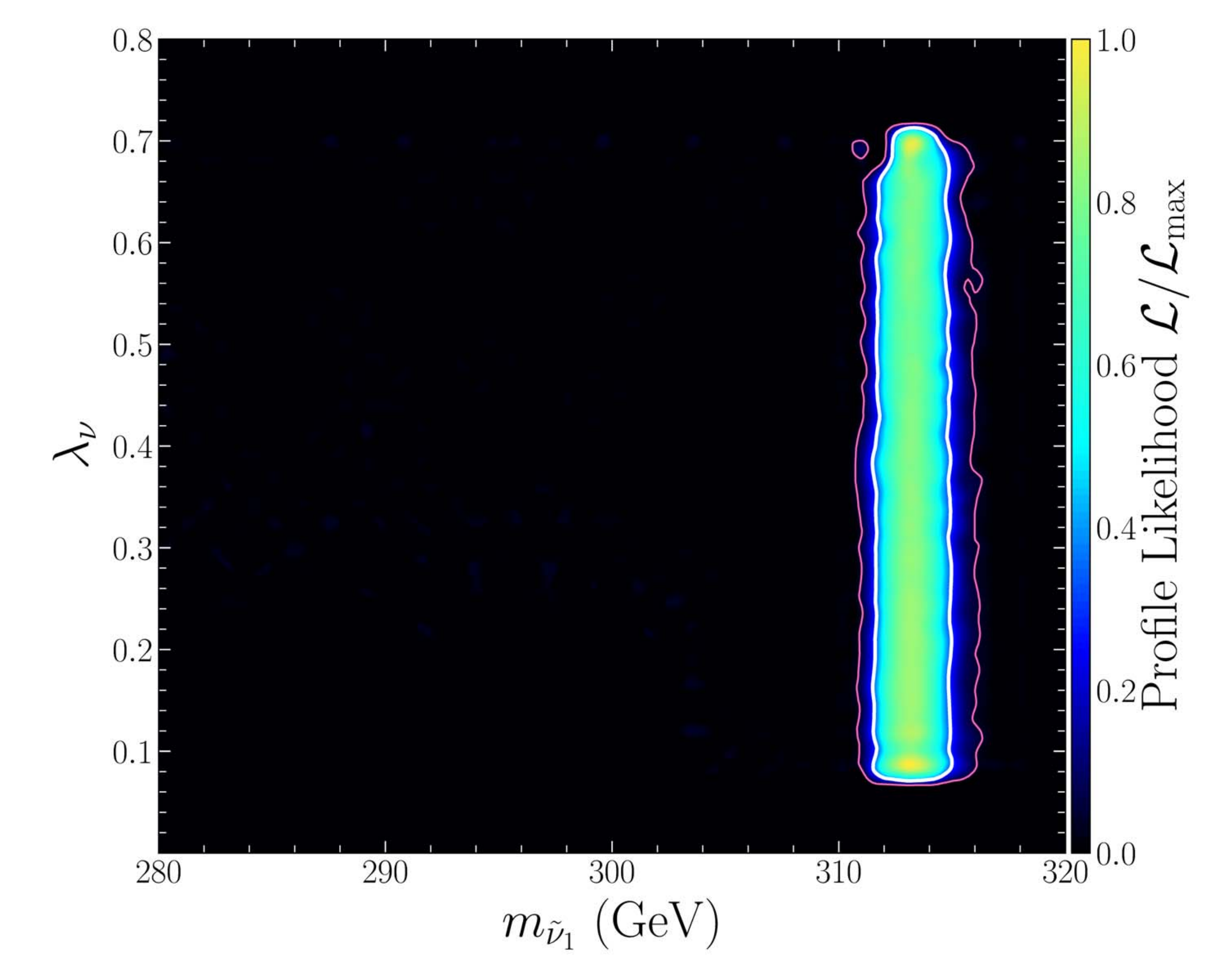}
        }
\caption{Same as Fig.\ref{fig1}, but for the massive $h_s$ scenario defined in Table \ref{table1}.  \label{fig6} }
\end{figure*}

\begin{figure*}[htbp]
		\centering
		\resizebox{0.92\textwidth}{!}{
        \includegraphics[width=0.90\textwidth]{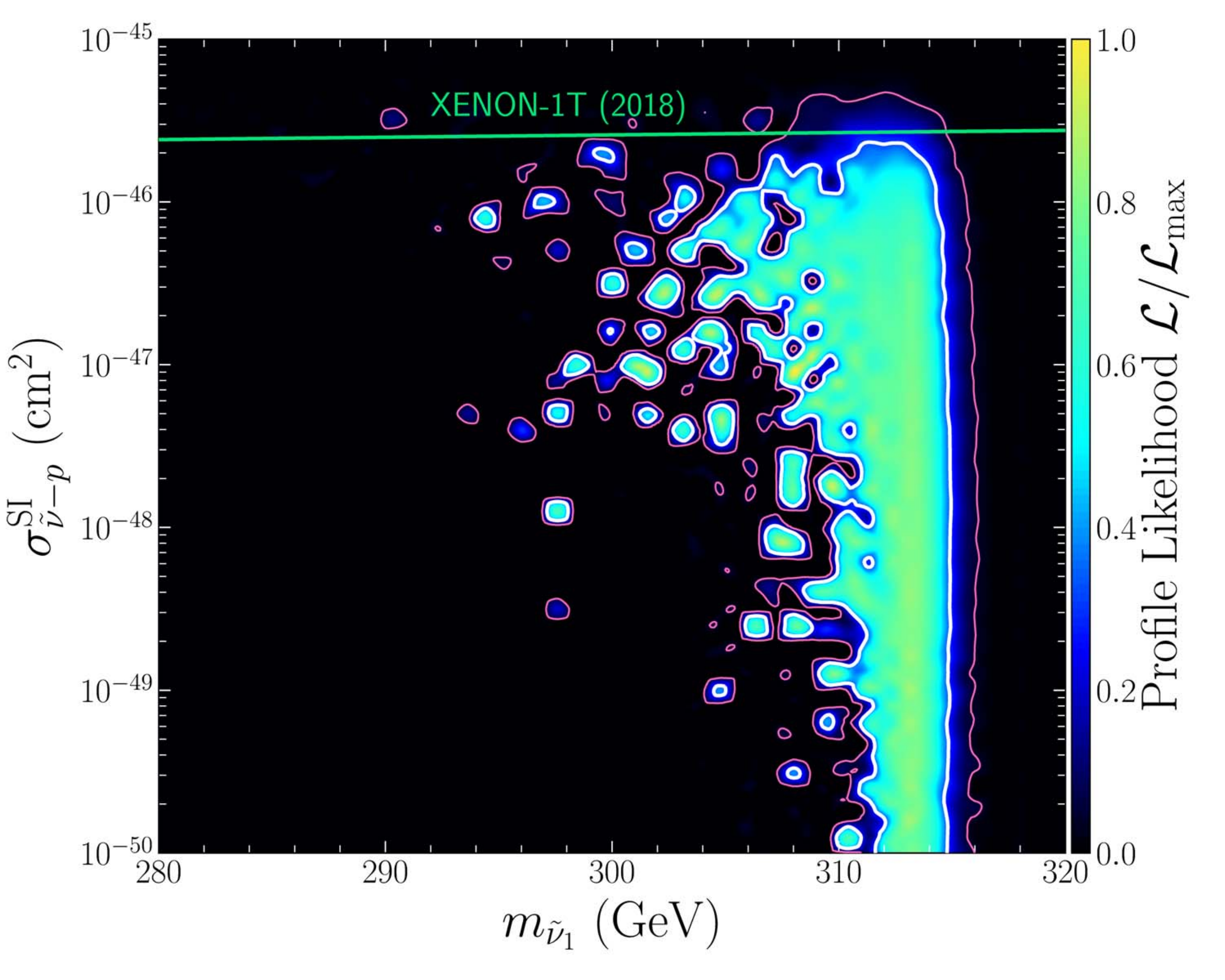}
        \includegraphics[width=0.90\textwidth]{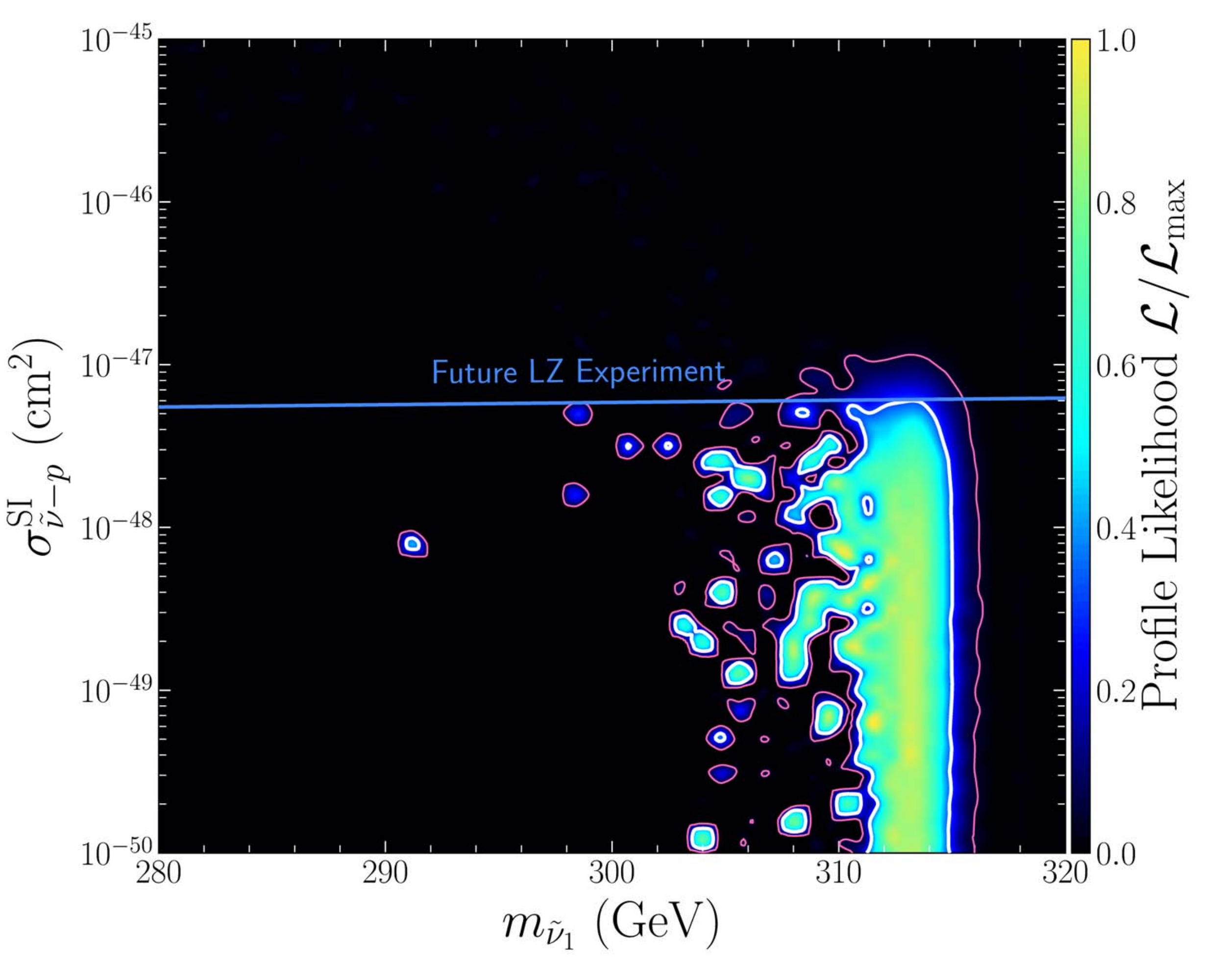}
        }

		\resizebox{0.92\textwidth}{!}{
        \includegraphics[width=0.90\textwidth]{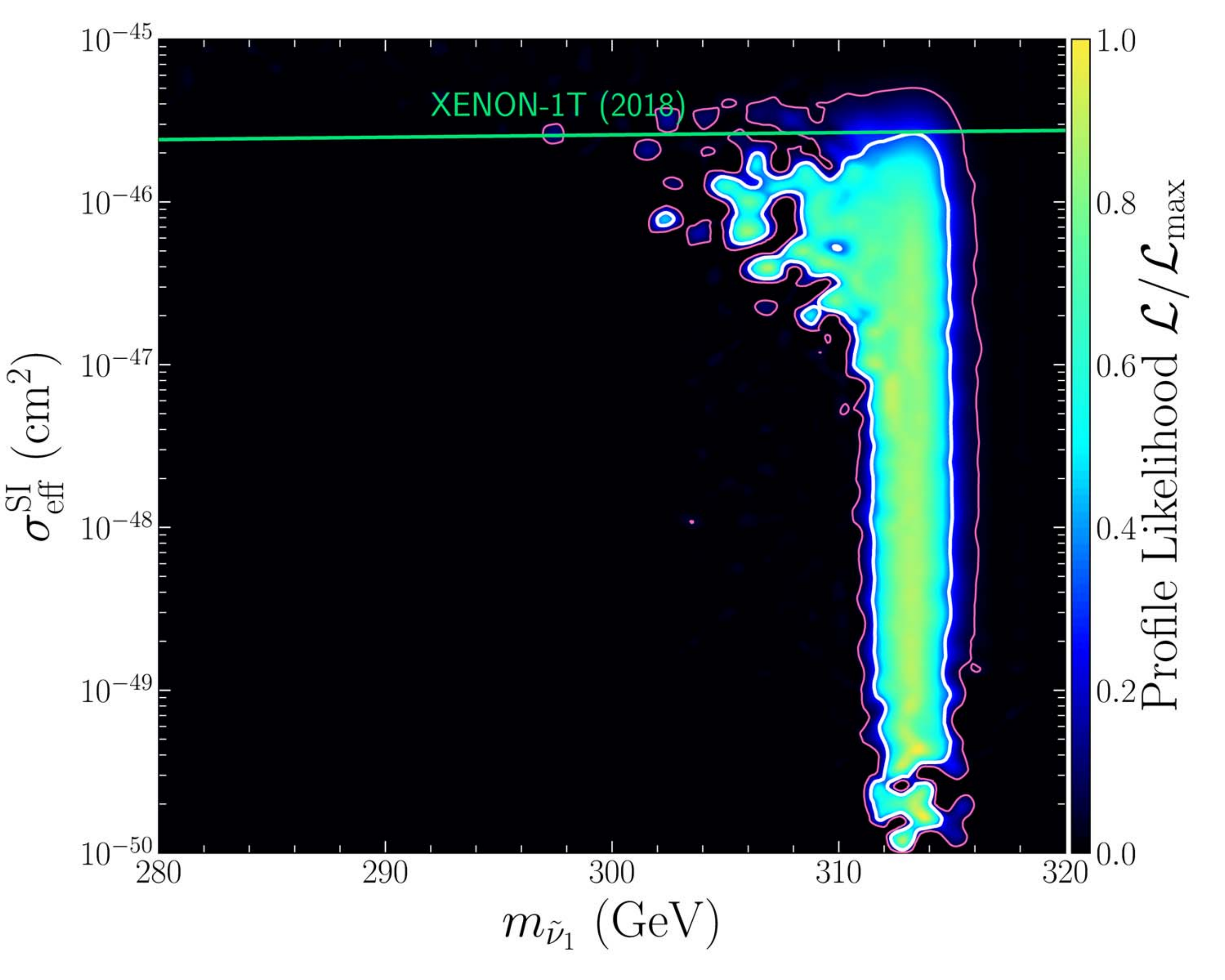}
        \includegraphics[width=0.90\textwidth]{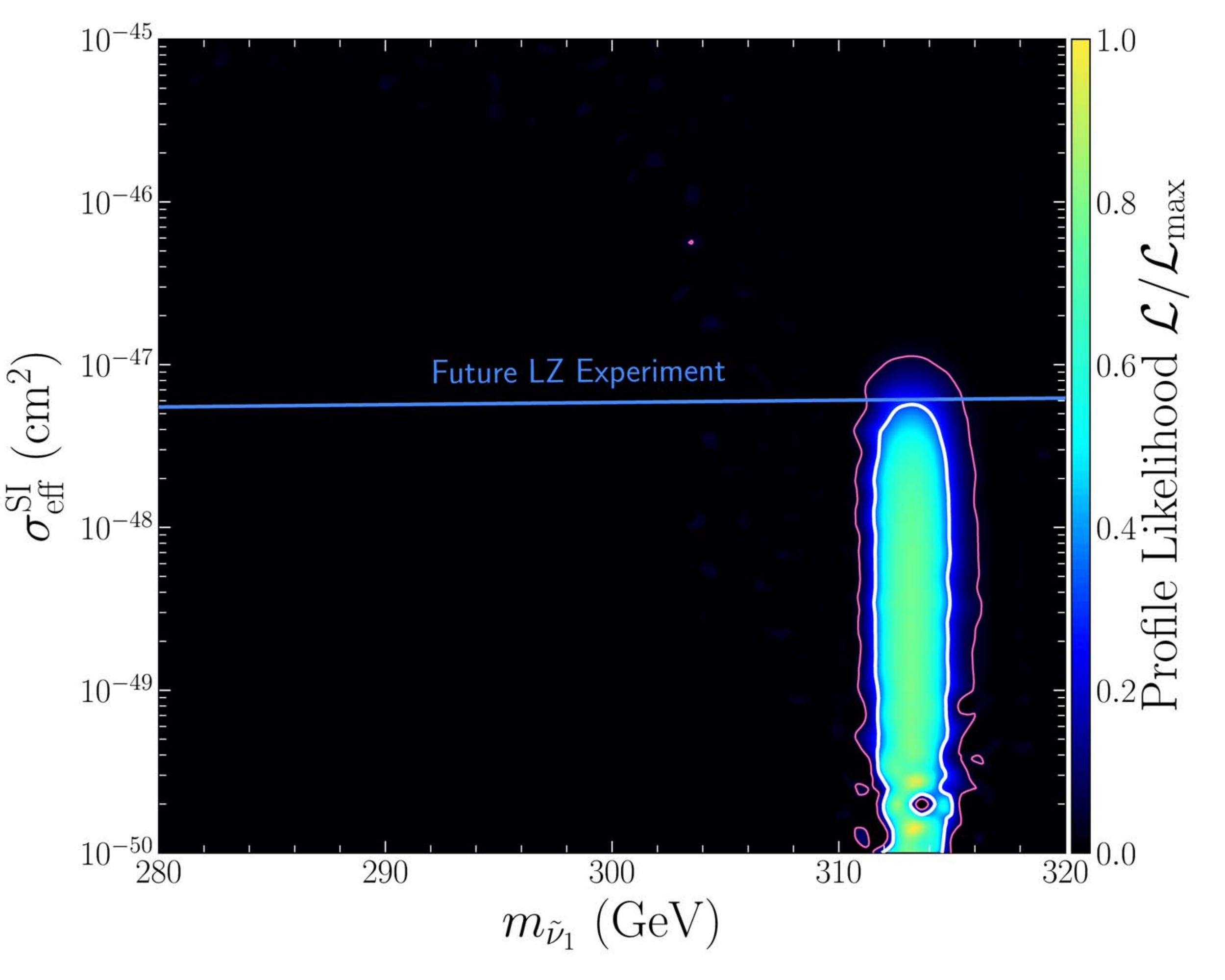}
        }
\caption{Same as Fig.\ref{fig2}, but for the massive $h_s$ scenario defined in Table \ref{table1}. \label{fig7} }
\end{figure*}

\begin{figure*}[htbp]
		\centering
		\resizebox{0.92\textwidth}{!}{
        \includegraphics[width=0.90\textwidth]{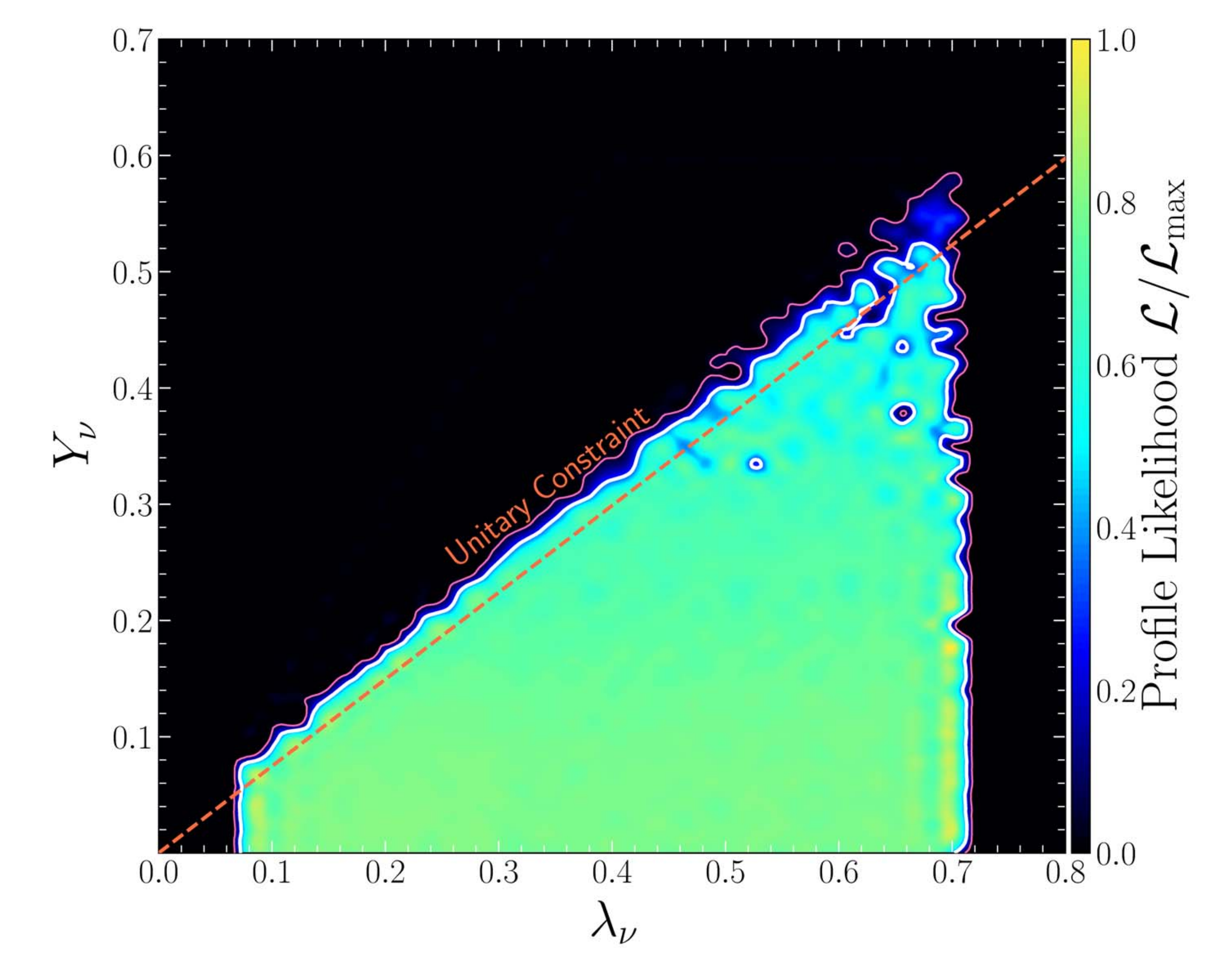}
        \includegraphics[width=0.90\textwidth]{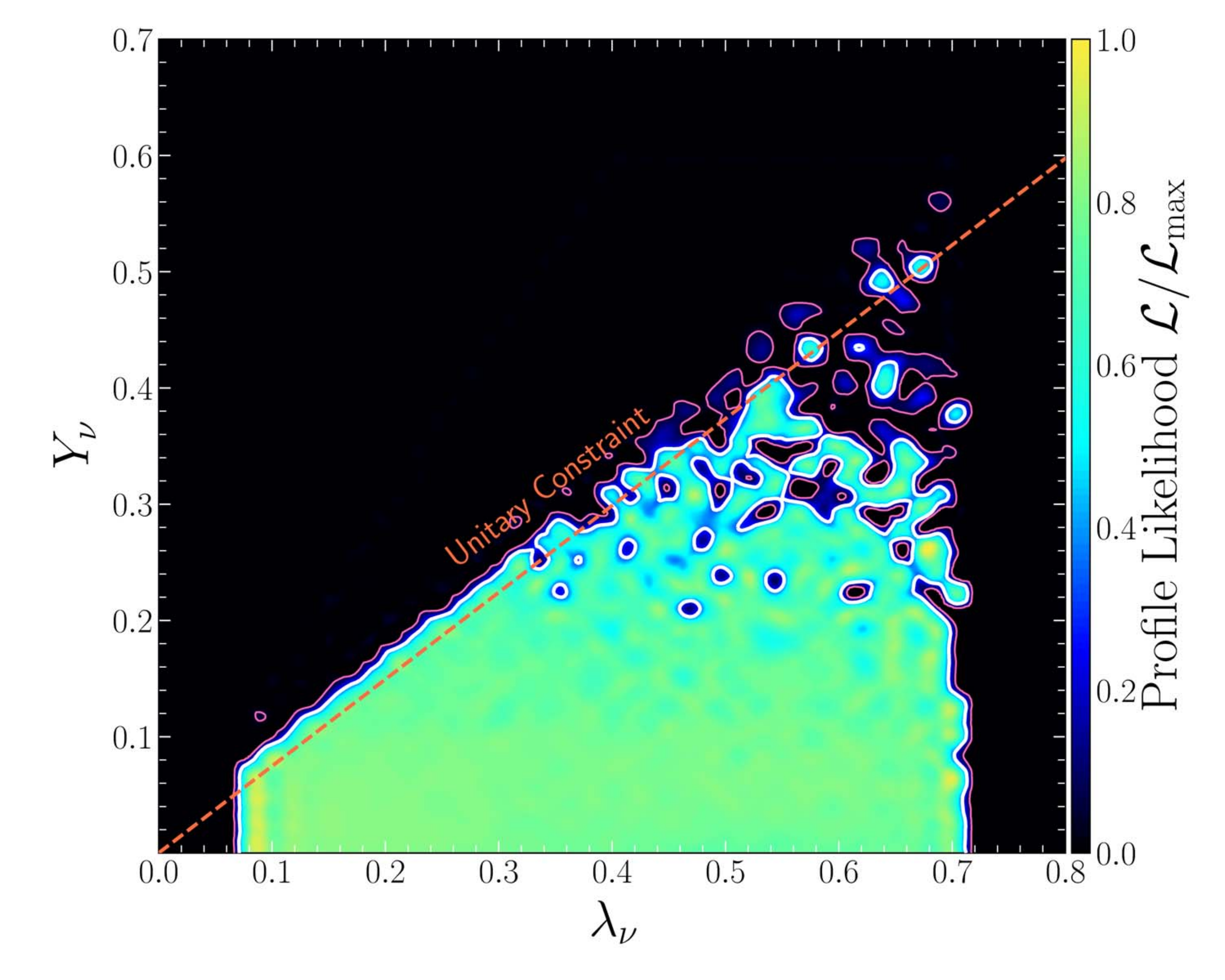}
        }

		\resizebox{0.92\textwidth}{!}{
        \includegraphics[width=0.90\textwidth]{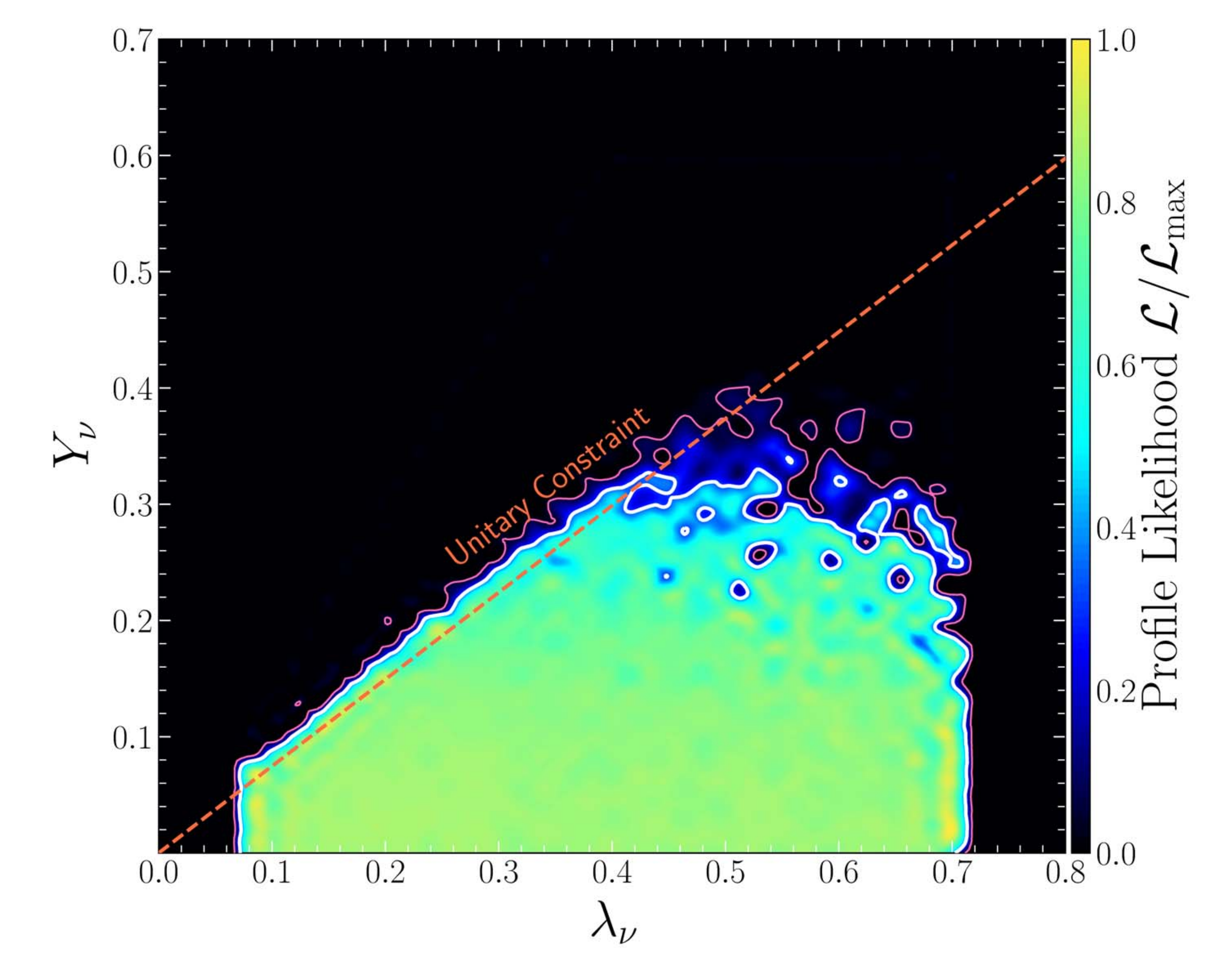}
        \includegraphics[width=0.90\textwidth]{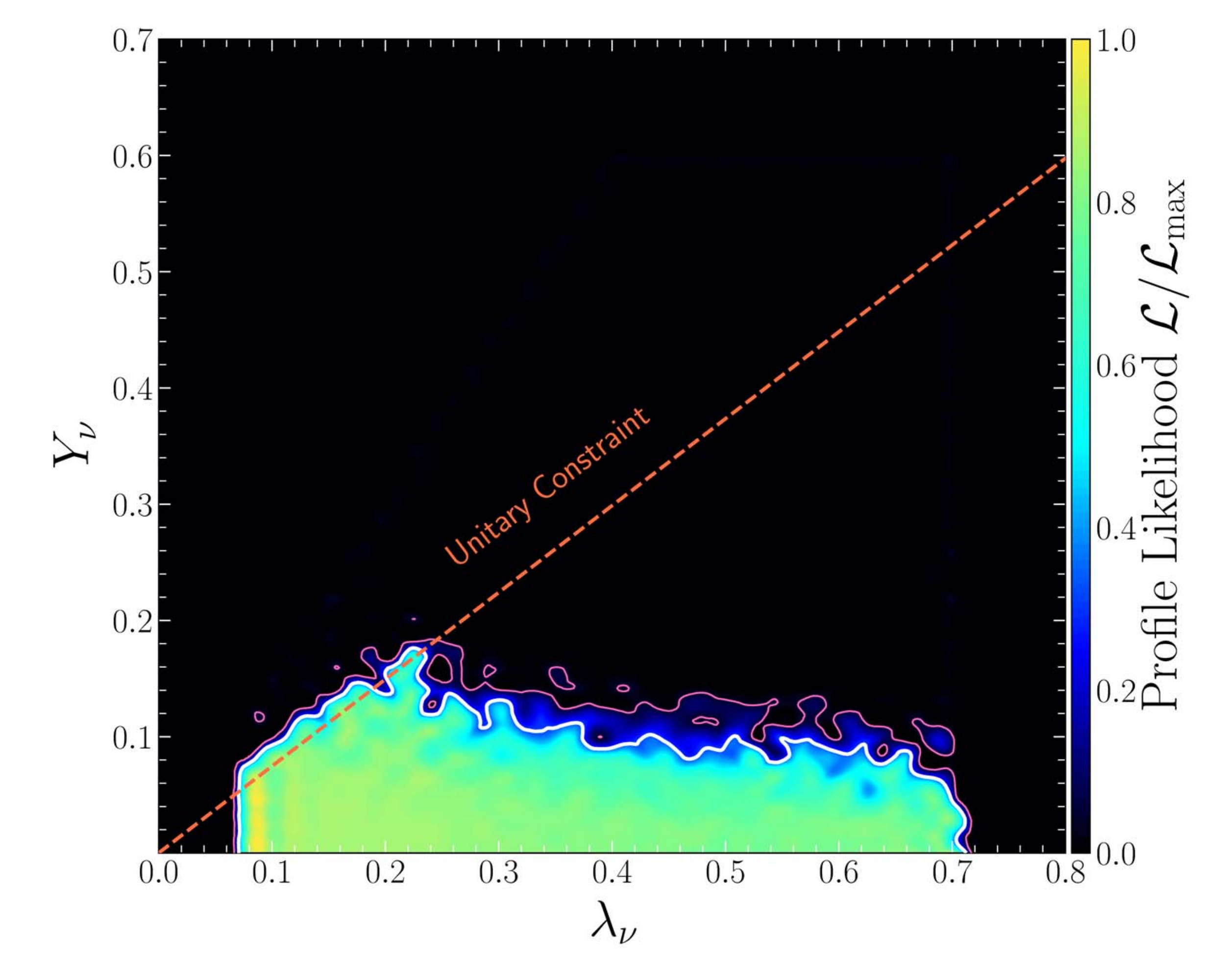}
        }
\caption{Same as Fig.\ref{fig3}, but for the massive $h_s$ scenario defined in Table \ref{table1}. \label{fig8}}
\end{figure*}

\begin{figure*}[htpb]
		\centering
		\resizebox{0.92 \textwidth}{!}{
        \includegraphics[width=0.90\textwidth]{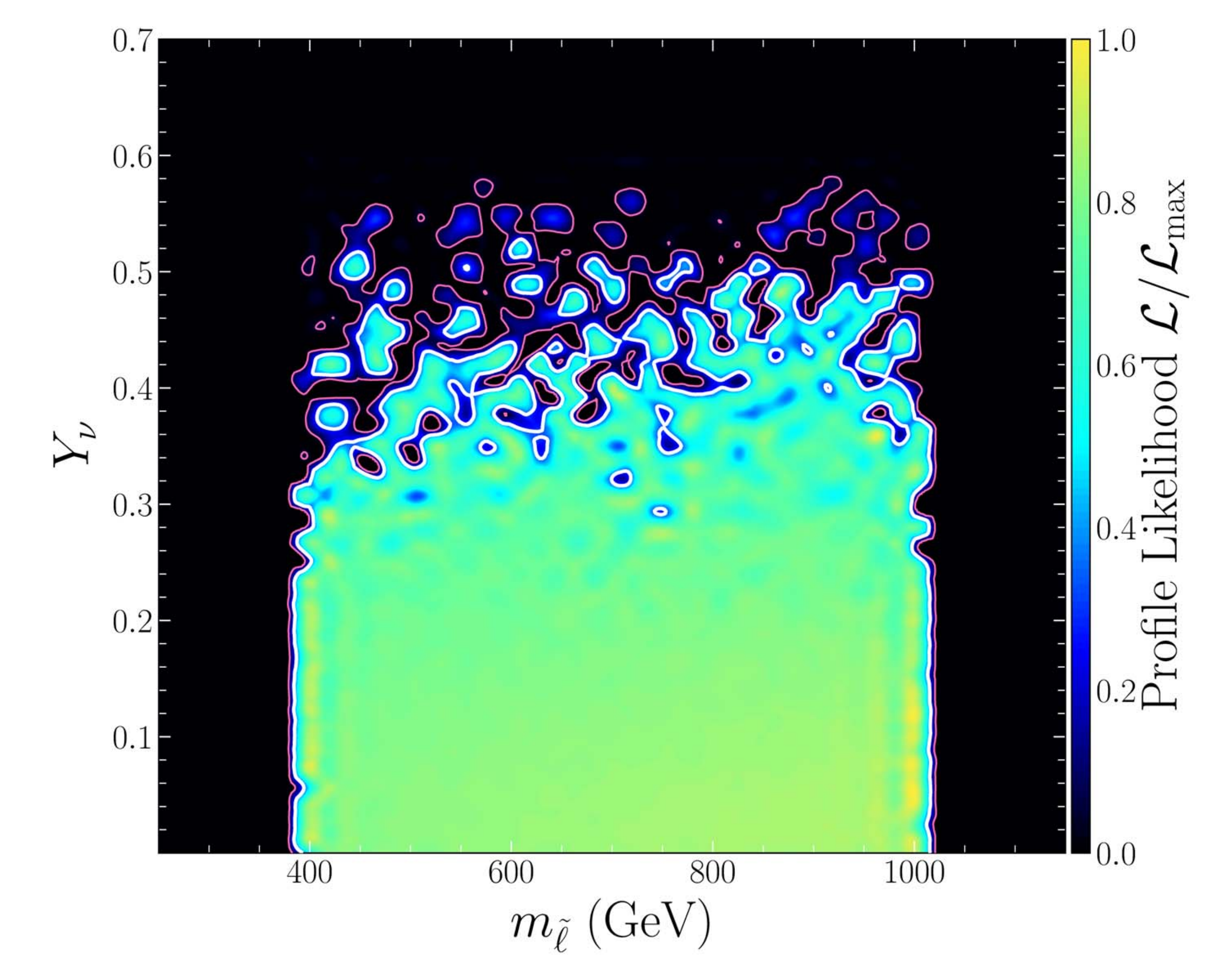}
        \includegraphics[width=0.90\textwidth]{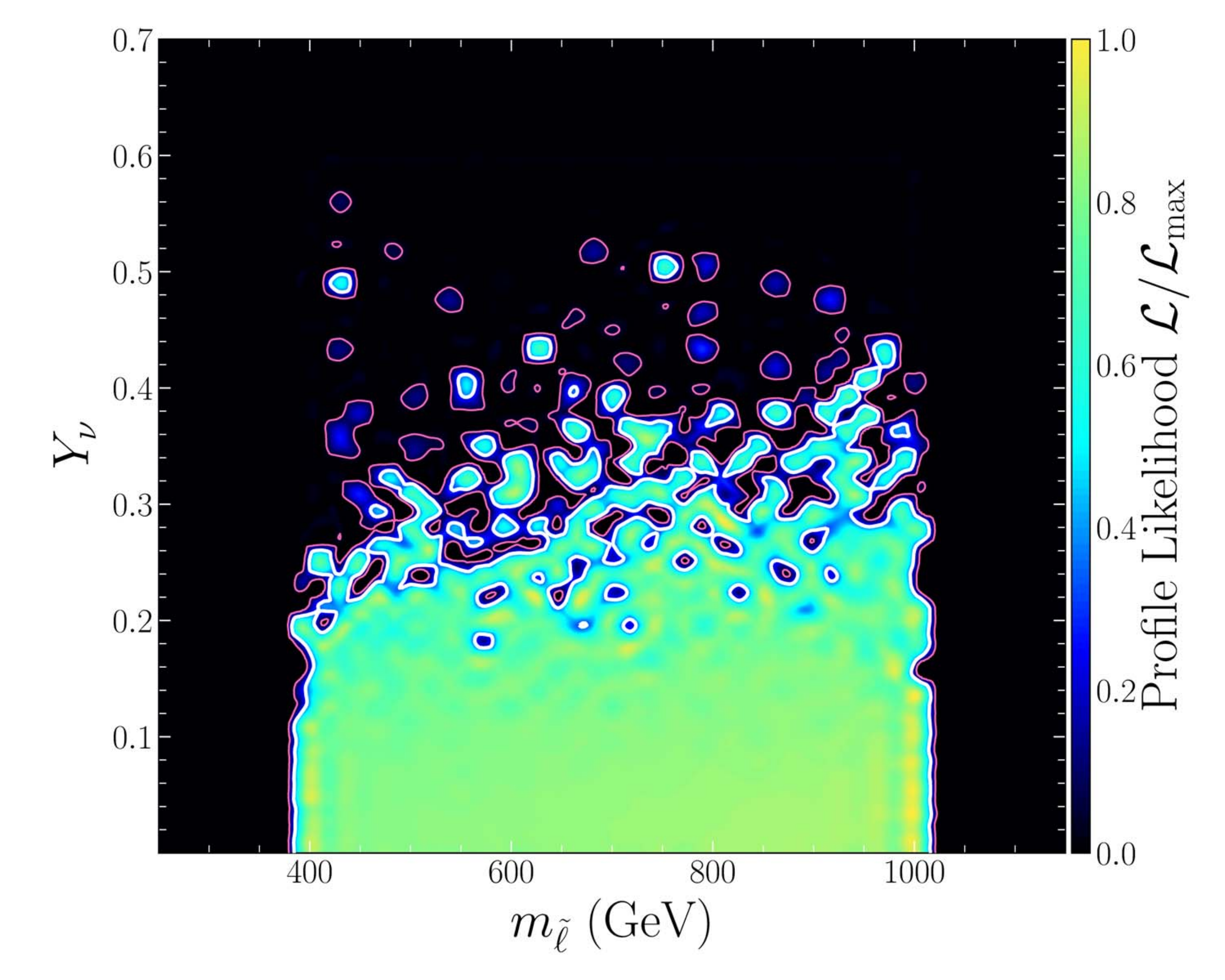}
        }

		\resizebox{0.92 \textwidth}{!}{
        \includegraphics[width=0.90\textwidth]{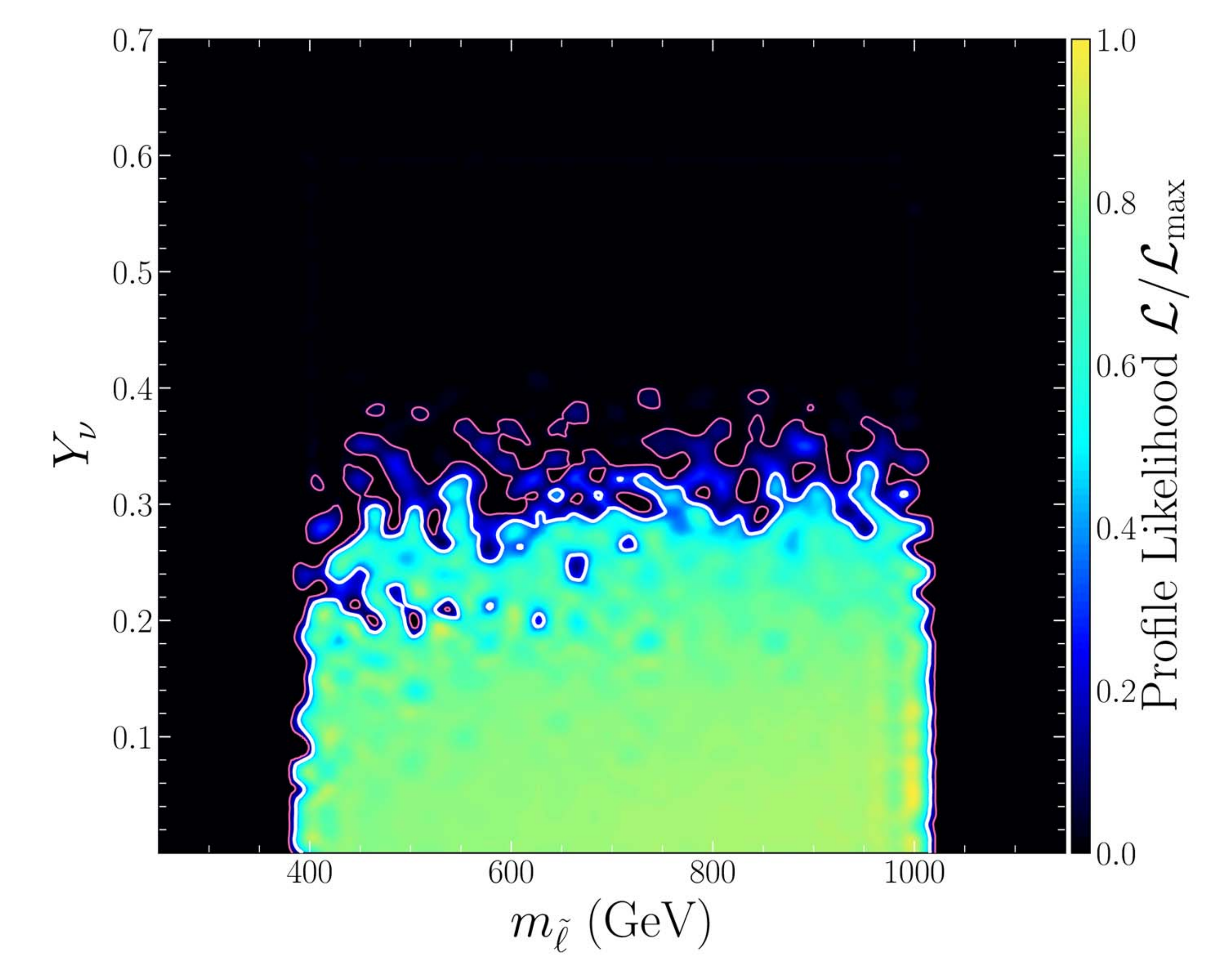}
        \includegraphics[width=0.90\textwidth]{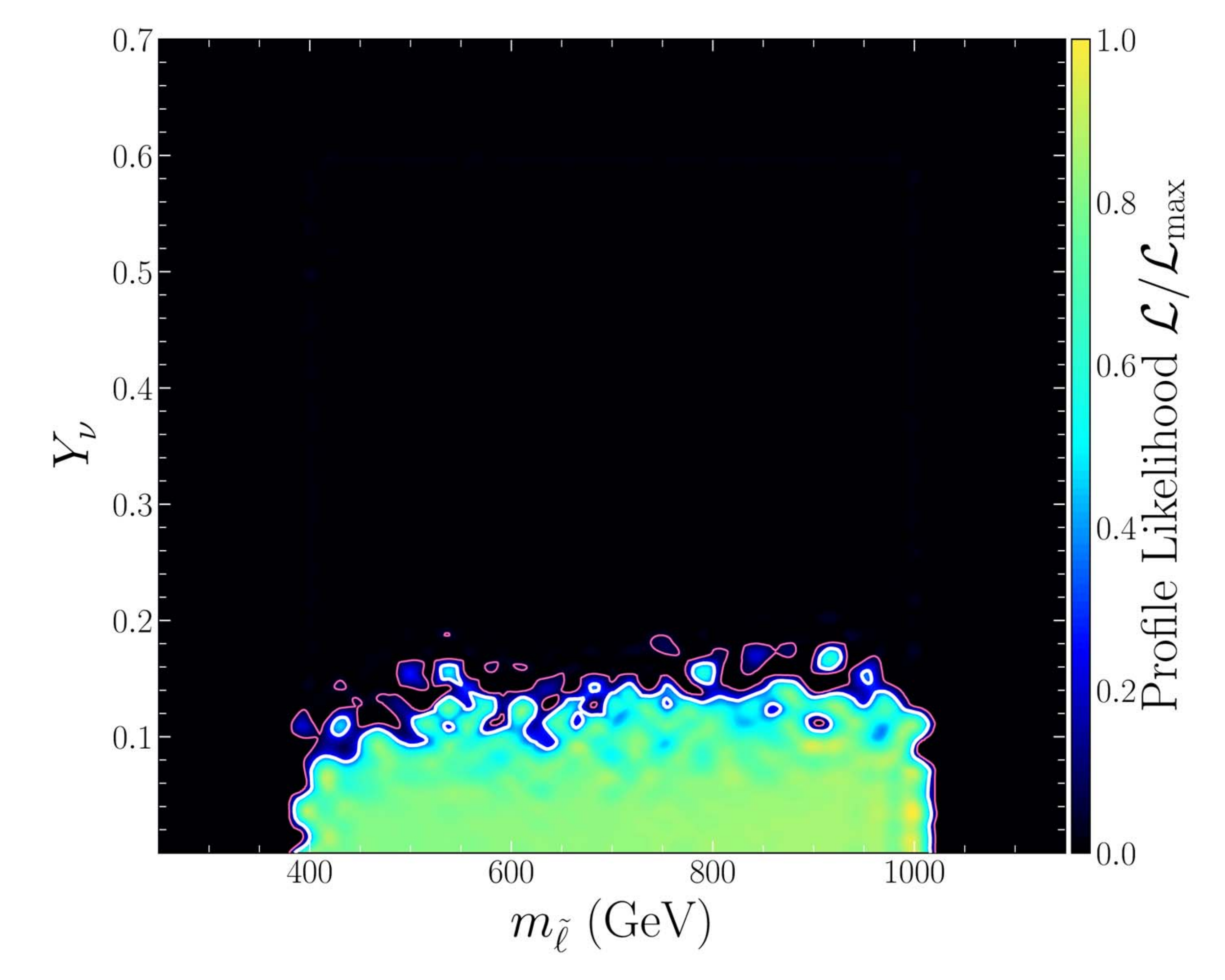}
        }
\caption{Same as Fig.\ref{fig4}, but for the massive $h_s$ scenario defined in Table \ref{table1}.  \label{fig9}}
\end{figure*}

\begin{figure*}[htpb]
		\centering
		\resizebox{0.92 \textwidth}{!}{
        \includegraphics[width=0.90\textwidth]{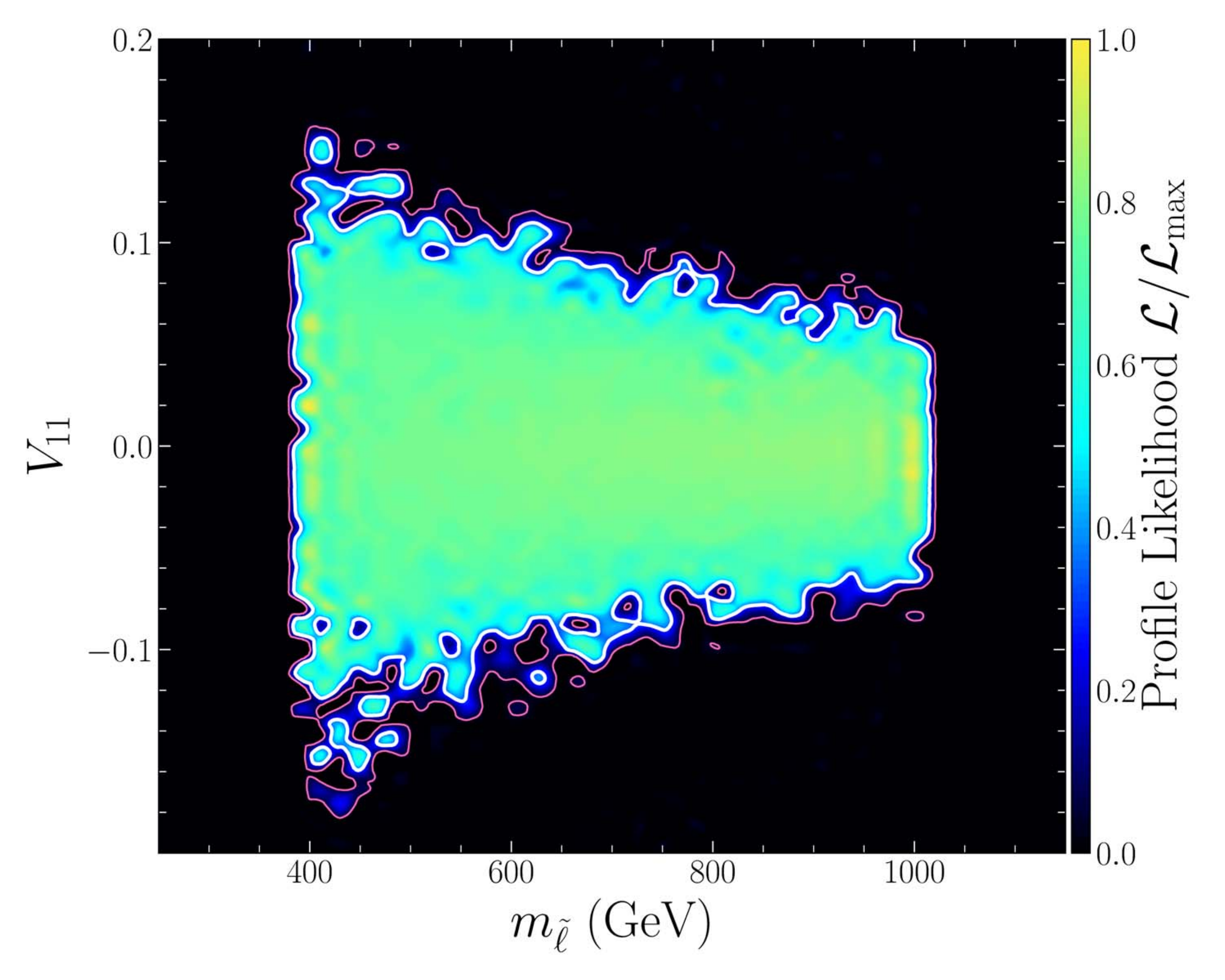}
        \includegraphics[width=0.90\textwidth]{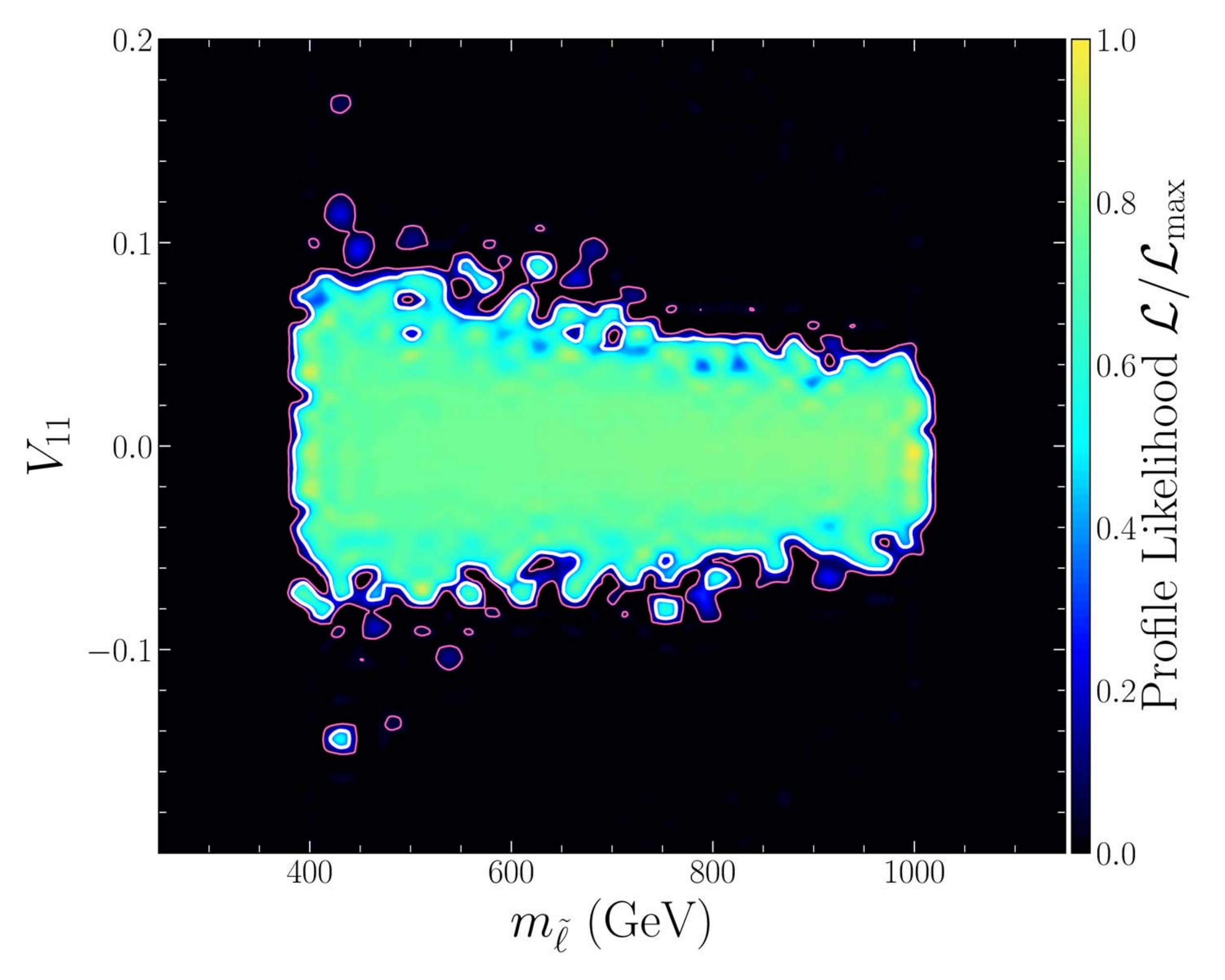}
        }

		\resizebox{0.92 \textwidth}{!}{
        \includegraphics[width=0.90\textwidth]{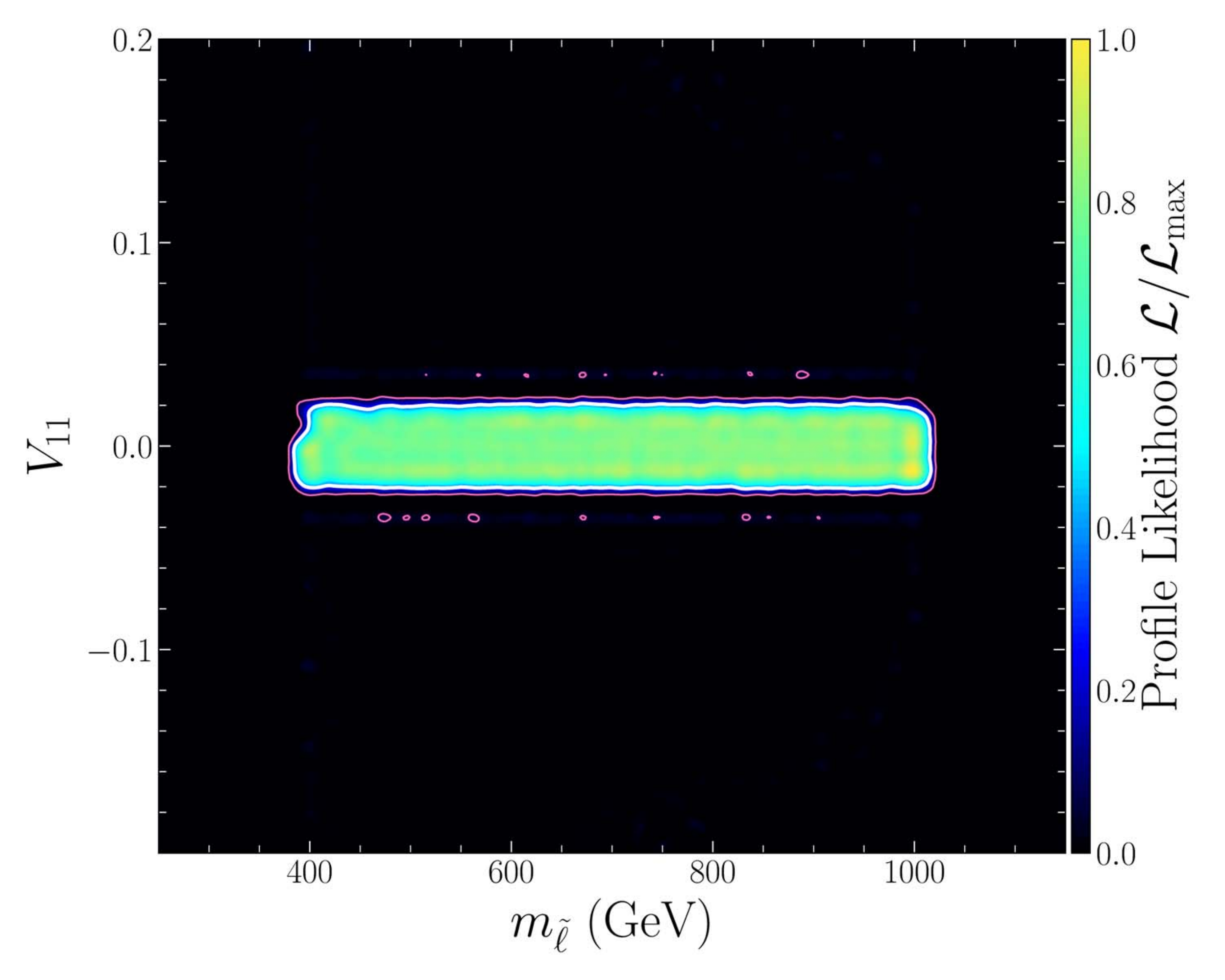}
        \includegraphics[width=0.90\textwidth]{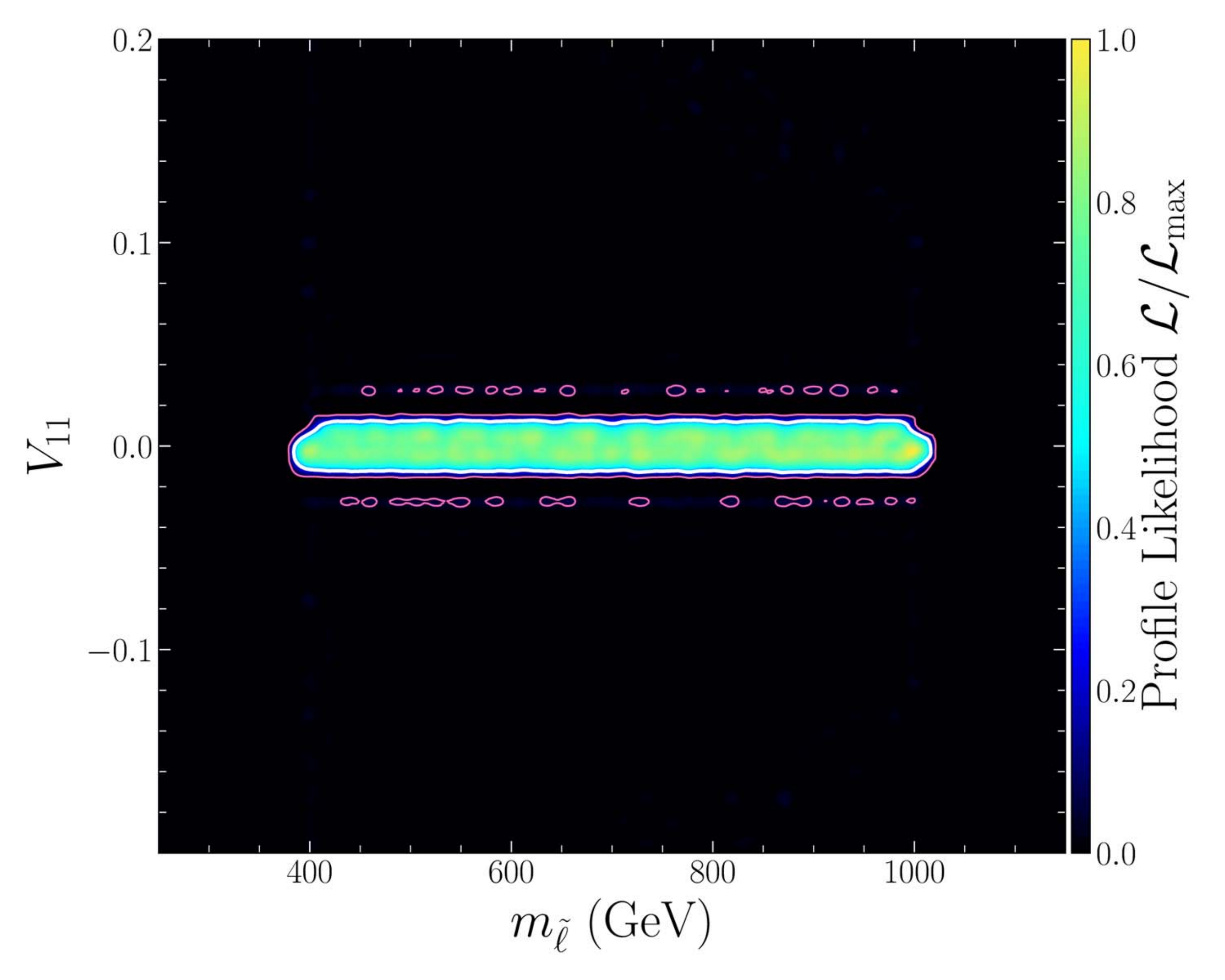}
        }
\caption{Same as Fig.\ref{fig5}, but for the massive $h_s$ scenario defined in Table \ref{table1}.  \label{fig10}}
\end{figure*}

\subsection{Results for the light $h_s$ scenario}

Given the information of the light $h_s$ scenario in Table \ref{table1}, the Higgs-mediated SI cross-section in Eq.~(\ref{Approximation}) is approximated by
\begin{eqnarray}
\sigma^{\rm SI}_{\tilde{\nu}_1-N} & \simeq & 4.2 \times 10^{-44}~{\rm cm^2} \times \nonumber \\
&& \left ( \frac{0.02 C_{\tilde{\nu}_1^\ast \tilde{\nu}_1 {\rm Re}[S]}}{m_{\tilde{\nu}_1}} + \frac{ C_{\tilde{\nu}_1^\ast \tilde{\nu}_1 {\rm Re}[H_u^0]} + 0.08 C_{\tilde{\nu}_1^\ast \tilde{\nu}_1 {\rm Re}[H_d^0]} }{m_{\tilde{\nu}_1}} \right )^2. \label{Simplified formula}
\end{eqnarray}
Since $C_{\tilde{\nu}_1^\ast \tilde{\nu}_1 {\rm Re}[S]}$ may be larger than $C_{\tilde{\nu}_1^\ast \tilde{\nu}_1 {\rm Re}[H_u^0]}$  and $C_{\tilde{\nu}_1^\ast \tilde{\nu}_1 {\rm Re}[H_d^0]}$ by two orders, the first term in the brackets can be comparable with the other contributions. To clarify the impact of the unitarity and DM DD experiments on the theory, we performed four independent scans over the following parameter space:
\begin{eqnarray}
& & 0 \leq Y_\nu, \lambda_\nu \leq 0.7, \quad \quad \quad 0 \leq m_{\tilde{\nu}}, m_{\tilde{x}} \leq 500~{\rm GeV}, \nonumber \\ &&
|A_{Y_\nu}|, |A_{\lambda_\nu} | \leq 1000 ~{\rm GeV}, \quad \quad  400~{\rm GeV} \leq m_{\tilde{l}} \leq 1000~{\rm GeV},  \label{Parameter-Space}
\end{eqnarray}
where $m_{\tilde{l}}$ denoted the common soft breaking mass of three-generation sleptons and its lower bound was motivated by the non-observation of slepton signals at the LHC Run-II. For the first scan, we fixed $B_{\mu_X} = - 100~{\rm GeV}^2$ and used the XENON-1T's bound on the SI cross-section to calculate the $\mathcal{L}_{\rm DD}$. The second scan was same as the first one except that we adopted the sensitivity of the LZ experiment. The last two scans differed from the previous ones only in that we set $B_{\mu_X} = 0$. As explained before, the setting induces an additional $Z$-mediated contribution to the DM-nucleon scattering so that the constraints of the DD experiments are strengthened.

With the samples obtained in the scans, we show different 2D PL maps in Figures \ref{fig1} to \ref{fig5}. Fig.~\ref{fig1} and \ref{fig2} plot the CIs on $\lambda_\nu - m_{\tilde{\nu}_1}$ and $\sigma^{\rm SI}_{\tilde{\nu}_1-p} - m_{\tilde{\nu}_1}$ planes. They show the following features:
\begin{itemize}
\item $m_{\tilde{\nu}_1}$ is concentrated on the range from $120$ to $181~{\rm GeV}$. Specifically, $m_{\tilde{\nu}_1}$ is close to $m_{\tilde{\chi}_1^0}$ for $172~{\rm GeV} \lesssim m_{\tilde{\nu}_1} \lesssim 181~{\rm GeV} $, and the DM achieves the correct density mainly through the $\tilde{\chi}_1^0 \tilde{\chi}_1^0$ annihilation (see discussions about the co-annihilation in Section 2.4 and details of the points in subsequent Table \ref{table2}). In this case, the density is insensitive to the parameter $\lambda_\nu$. Thus, $\lambda_\nu$ varies within a broad range from 0.15 to 0.6 in Fig.~\ref{fig1}, where the lower limit forbids the decay $h \to \nu_h \bar{\nu}_h$ kinematically, and the upper bound comes from the DM DD experiments (discussed below).  For the other mass range, the DM obtains the correct density mainly through the annihilations $\tilde{\nu}_1 \tilde{\nu}_1^\ast \to h_s h_s, h_s h, h h$. This requires $\lambda_\nu \gtrsim 0.26$, which can be understood from the discussion of Eq.~(\ref{relic density}).

\item Fig.~\ref{fig2} indicates that the SI cross-section of the DM-nucleon scattering may be as low as $10^{-49}~{\rm cm^2}$ over the entire mass range. It reflects that the theory has multiple mechanisms to suppress the scattering, which becomes evident by the approximation in Eq.~(\ref{Simplified formula}) and was recently emphasized in~\cite{Cao:2019qng}.

\item Although $\lambda_\nu > 0.6$ is allowed by the setting in Eq.~(\ref{Parameter-Space}), it is upper bounded by 0.56, 0.56, 0.50, and 0.45 for the $2 \sigma$ CIs of the four cases, respectively. Careful comparisons of the left and right panels revealed that it was due to the DD experiments' constraint on the co-annihilation region. Besides, we studied the Bayesian evidences $Z_i$ ($i=1,2,3,4$) of the four cases and found $\ln Z_1 = -55.6$, $\delta_{12} \equiv \ln Z_1 - \ln Z_2 = 1.0$, $\delta_{13} \equiv \ln Z_1 - \ln Z_3 = 0.54$, and  $\delta_{34} \equiv \ln Z_3 - \ln Z_4 = 1.43$. These results reveal at least two facts. On the one side, the Jeffreys' scale $\delta_{13}$~\cite{Bayes,Jeffreys} reflects that current XENON-1T experiment has no significant preference of the $B_{\mu_X} \neq 0$ case to the $B_{\mu_X} = 0$ case~\cite{Feroz:2008wr}. On the other side, $\delta_{12}$ and $\delta_{34}$ show that the Bayesian evidence (or equivalently the averaged $\mathcal{L}_{\rm DM}$) is reduced by a factor of more than $40\%$. It implies that a sizable portion of the parameter space will become disfavored once the future LZ experiment improves the XENON-1T's sensitivity by 50 times. This feature is also reflected in Fig.~\ref{fig1} and Fig.~\ref{fig2} by the sizable shrink of the $1\sigma$ CIs.
\end{itemize}

In order to better understand Fig. 1, we describe how we obtained it. From Eq.~(\ref{stat:Frequen}), the 2D PL $\mathcal{L} (\lambda_\nu, m_{\tilde{\nu}_1})$ is given by
\begin{eqnarray}
\mathcal{L} (\lambda_\nu, m_{\tilde{\nu}_1}) = \max_{Y_\nu, A_{\lambda_\nu}, \cdots} \mathcal{L}_{\rm DM} (\lambda_\nu, Y_\nu, A_{\lambda_\nu}, A_{Y_\nu}, m_{\tilde{\nu}}, m_{\tilde{x}}, m_{\tilde{l}}).  \label{DM-profile1}
\end{eqnarray}
In plotting the figure, we implemented the maximization over the parameters $Y_\nu$, $A_{\lambda_\nu}$, $A_{Y_\nu}$, $m_{\tilde{\nu}}$, $m_{\tilde{x}}$, and $m_{\tilde{l}}$
by three steps. First, we split the $\lambda_\nu-m_{\tilde{\nu}_1}$ plane into $80 \times 80$ equal boxes, i.e., we divided each dimension of the plane by 80 regular bins.
Second, we fit the samples obtained in the  scan into each box. Consequently, samples in each box correspond to roughly equal $\lambda_\nu$ and $m_{\tilde{\nu}_1}$, even though
the other parameters may differ significantly. Finally, we select the maximum likelihood value of the samples in each box as the PL value. These procedures imply that the CIs are not necessarily contiguous, instead
they usually distributed in isolated islands~\cite{Fowlie:2016hew,Cao:2018iyk}. Besides, we emphasize that $\chi^2_{\rm min} \simeq 0$ for the best point in the scans.  This is because the DM experiments are independent and consistent with each other, and the ISS-NMSSM can explain them well.

Next, we study 2D PL on $Y_\nu-\lambda_\nu$ plane. The results are shown in Fig.~\ref{fig3} where the red dashed line denotes the correlation $\lambda_{\nu} \mu/(Y_{\nu} \lambda v_u) = 9.4 $ or equivalently $\lambda_\nu = 2.9 Y_\nu$ from the unitarity constraint. This figure shows that $Y_\nu$ is maximized at 0.17 when $\lambda_\nu \simeq 0.52$ and it is upper bounded only by the unitarity. The reason is that the unitarity requires $\lambda_\nu \gtrsim 2.9 Y_\nu$, so the SI cross-section is much more sensitive to $\lambda_\nu$ than to $Y_\nu$. Consequently, the DD experiments set the upper bound of $\lambda_\nu$ and by contrast, the unitarity limits $Y_\nu$.

We also plot 2D PLs on $Y_\nu-m_{\tilde{l}}$ and $V_{11}-m_{\tilde{l}}$ planes in Fig.~\ref{fig4} and Fig.~\ref{fig5}, respectively. Fig.~\ref{fig4} indicates that the $2\sigma$ CI in each panel occupies a roughly rectangular area on the $Y_\nu - m_{\tilde{l}}$ plane. This result reflects that $\mathcal{L}_{\rm DM}$ is insensitive to parameter $m_{\tilde{l}}$. It can be understood from the following two aspects. One is that $\mathcal{L}_{\rm DM}$ relies on $m_{\tilde{l}}$ mainly through $V_{11}$ by the $\tilde{\nu}_1 \tilde{\nu}_1^\ast h_i$ coupling in Eq.~(\ref{Coupling-expression}). The other is that $m_{\tilde{l}}$ and $V_{11}$ are weakly correlated, which can be inferred by the expression of $m_{12}$ and $m_{13}$ in Eq.~(\ref{Matrix-elements}) and is shown numerically in Fig.~\ref{fig5} and Fig.~\ref{fig10}. Specifically, for the $B_{\mu_X} \neq 0$ case, both the annihilation and the scattering are insensitive to $V_{11}$ since its magnitude is small, and so is $\mathcal{L}_{\rm DM}$. This property determines that the allowed range of $Y_\nu$ is roughly independent of $m_{\tilde{l}}$, and thus explains the rectangular shape. For the $B_{\mu_X} = 0$ case, although the effective cross-section in Eq.~(\ref{Effective-Cross-Section}) is sensitive to $V_{11}$ by the formula in Eq.~(\ref{DM-neutron}), the XENON-1T experiment has required $|V_{11}| \lesssim 0.02$ and this upper bound is very insensitive to $m_{\tilde{l}}$. In this case, one may replace $m_{\tilde{l}}$ by $V_{11}$ as a theoretical input so that $\mathcal{L}_{\rm DM}$ does not depend on $m_{\tilde{l}}$ any more. This feature again leads to the conclusion that the allowed range of $Y_\nu$ is roughly independent of $m_{\tilde{l}}$. We add that the tight experimental constraint on the mixing $V_{11}$ for the $B_{\mu_X} = 0$ case was also discussed in \cite{Kakizaki:2016dza}. We also add that one may fix $m_{\tilde{l}}$ in performing global fit of the ISS-NMSSM to experimental data due to the insensitivity of  $\mathcal{L}_{\rm DM}$ to $m_{\tilde{l}}$. Such a treatment affects little the generality of the fit results.

In Table \ref{table2}, we present the details of two points to illustrate the scenario's features further. For the point P1, the DMs annihilated mainly by $\tilde{\nu}_1 \tilde{\nu}_1^\ast \to h_s h_s, h_s h$ to get the density. The process $\tilde{\nu}_1 \tilde{\nu}_1^\ast \to h h$ is unimportant because $|C_{h_s h h}|$ is significantly smaller than  $|C_{h_s h_s h_s}|$ and $|C_{h_s h_s h}|$, and also because the phrase space of the final state is relatively small. By contrast, the DMs got their right relic density mainly by the Higgsino pair annihilation for the point P2, and the mass splitting is $\Delta \equiv m_{\tilde{\chi}_1^0} - m_{\tilde{\nu}_1} \simeq 7~{\rm GeV}$. We confirmed that, due to the specific parameter setting of P2, there is cancellation between different contributions to the process  $\tilde{\nu}_1 \tilde{\nu}_1^\ast \to h_s h_s$, and consequently, its effect is negligibly small. Besides, both the points predict $Y_\nu \sim 0.01$. As a result, $V_{11}$'s magnitude is only a few thousandths, and the DM-neutron scattering rate is not much larger than the DM-proton scattering rate. We verified that, once we set $B_{\mu_X} = - 100~{\rm GeV}$, the two rates became roughly equal.

\subsection{Results for the massive $h_s$ scenario}

In the massive $h_s$ scenario, the Higgs-mediated SI cross-section is given by
\begin{eqnarray}
\sigma^{\rm SI}_{\tilde{\nu}_1-N} & \simeq & 4.2 \times 10^{-44}~{\rm cm^2} \times \left ( \frac{0.003 C_{\tilde{\nu}_1^\ast \tilde{\nu}_1 {\rm Re}[S]}}{m_{\tilde{\nu}_1}} + \frac{ C_{\tilde{\nu}_1^\ast \tilde{\nu}_1 {\rm Re}[H_u^0]} + 0.04 C_{\tilde{\nu}_1^\ast \tilde{\nu}_1 {\rm Re}[H_d^0]} }{m_{\tilde{\nu}_1}} \right )^2, \nonumber
\end{eqnarray}
when one takes the parameters in Table \ref{table1}. In large $\lambda_\nu$ and $Y_\nu$ case, e.g., $\lambda_\nu \gtrsim 0.4$ and $Y_\nu \gtrsim 0.4$, the typical sizes of
$C_{\tilde{\nu}_1^\ast \tilde{\nu}_1 {\rm Re}[S]}$ and $C_{\tilde{\nu}_1^\ast \tilde{\nu}_1 {\rm Re}[H_u^0]}$ are $100~{\rm GeV}$ and $10~{\rm GeV}$, respectively. Thus, the first term in the brackets is no longer more critical than the other terms, and the $\sigma^{\rm SI}_{\tilde{\nu}_1-N}$ for $m_{\tilde{\nu}_1} \simeq 300~{\rm GeV}$ may reach $10^{-46}~{\rm cm^2}$ only in optimal cases. Consequently, the XENON-1T experiment scarcely limit the $B_{\mu_X} \neq 0$ case. This situation is significantly different from the light $h_s$ scenario.

Similar to the analysis of the light $h_s$ scenario, we performed four independent scans over the parameter region in Eq.~(\ref{Parameter-Space}), and projected the PL onto different planes. The results are presented from Fig.~\ref{fig6} to Fig.~\ref{fig10} in a way similar to those for the light $h_s$ scenario. These figures indicate the following facts:
\begin{itemize}
\item Since the unitarity for the parameters in Table \ref{table1} requires only $\lambda_\nu \gtrsim 1.3~Y_\nu$, $Y_\nu$ may be comparable with $\lambda_\nu$ in size.
As a result, the SI cross-section is sensitive to both $\lambda_\nu$ and $Y_\nu$, which is different from the light $h_s$ scenario.

\item Since the $B_{\mu_X} \neq 0$ case is hardly limited by the XENON-1T experiment, both $\lambda_\nu$ and $Y_\nu$ may be larger than 0.4, which is shown in the upper left panel of Fig.~\ref{fig8}. However, with the experimental sensitivity improved or the $Z$-mediated contribution considered in the $B_{\mu_X} = 0 $ case, the DM DD experiments become powerful enough to limit $\lambda_\nu$ and $Y_\nu$. In this case, $Y_\nu \gtrsim 0.4$ may contradict the experiments, which is indicated by the other panels of Fig.~\ref{fig8}.

\item $\tilde{\nu}_1$ obtained the correct density through the co-annihilation with $\tilde{\chi}_1^0$, which is reflected by the range of $m_{\tilde{\nu}_1}$ in Fig.~\ref{fig6}. We will take the points P3 and P4 in Table \ref{table2} as examples to show more details of the annihilation later.

    We confirmed that $\delta_{12} \equiv \ln Z_1 - \ln Z_2 = 0.2 $, $\delta_{13} \equiv \ln Z_1 - \ln Z_3 = 0.62$ and  $\delta_{34} \equiv \ln Z_3 - \ln Z_4 = 0.56$ in the massive $h_s$ scenario. Similar to the analysis of the light $h_s$ scenario, the smallness of $\delta_{12}$ and $\delta_{34}$ reflects that the LZ experiment can not improve the constraint of the XENON-1T experiment on the scenario significantly, and the smallness of $\delta_{13}$ reflects that the XENON-1T experiment does not show significant preference of the $B_{\mu_X} \neq 0$ case to the $B_{\mu_X} = 0$ case.

\item Concerning the other features of the massive $h_s$ scenario, such as the suppression of the SI cross-section and the correlation of $m_{\tilde{l}}$ with $Y_\nu$ and $V_{11}$, they are similar to those of the light $h_s$ scenario. We do not discuss them anymore.
\end{itemize}

Next, let us study two representative points, P3 and P4, of the massive $h_s$ scenario in Table \ref{table2}. For the former point, it is the annihilation of the Higgsino pair that is responsible for the measured density, and the corresponding mass splitting is about $5~{\rm GeV}$. By contrast, the $\tilde{\nu}_1^0 \tilde{H}$ annihilation mainly accounts for the latter point density, and the mass splitting reaches about $19~{\rm GeV}$. The difference is caused by the fact that P4 takes a relatively large $Y_\nu$ and a lighter $m_{\tilde{l}}$, making the $\tilde{\nu}_1^0 \tilde{H}$ annihilation more critical.
Besides, it is notable that both the points predict $Y_\nu \gtrsim 0.18$ to induce a sizable $\tilde{\nu}_L$ component in $\tilde{\nu}_1$, e.g., $|V_{11}| > 0.01$.
Consequently, $Z$ boson can mediate a large DM-neutron scattering so that $\sigma_{\tilde{\chi}_1-n}^{\rm SI} \gg \sigma_{\tilde{\chi}_1-p}^{\rm SI}$. Such a significant difference disappears if one sets $B_{\mu_X} \neq 0$.

In summary, both $\lambda_\nu$ and $Y_\nu$ are more constrained in the light $h_s$ scenario than in the massive $h_s$ scenario. The unitarity always plays a vital role in limiting $Y_\nu$ except for the case shown in the last panel of Fig.~\ref{fig8}, where the LZ experiment may be more critical in limiting $Y_\nu$. We emphasize that the tight DD constraint on the $B_{\mu_X} \neq 0$ case of the light $h_s$ scenario arises from that $h_s$ is light and it contains sizable doublet components. In this case, the coupling  $C_{\tilde{\nu}_1^\ast \tilde{\nu}_1 {\rm Re}[S]}$ contributes significantly to the scattering rate.

Before we end this section, we emphasize that the parameter points discussed in this work are consistent with the LHC results in searching for sparticles. Specifically, for the parameters in Table \ref{table1}, it is evident that the LHC fails to detect gluinos and squarks because these particles are too massive. Concerning the Higgsino-dominated particles, they may be detectable at the 8~TeV and 13~TeV LHC since their production rates reach  $100~{\rm fb}$. We scrutinized the property of the points in $B_{\mu_X} = 0$ case and found that they all predict
\begin{eqnarray}
{\rm Br}(\tilde{\chi}_{1,2}^0 \to \tilde{\nu}_1 \bar{\nu}_\tau) =  {\rm Br}(\tilde{\chi}_{1,2}^0 \to \tilde{\nu}_1^\ast \nu_\tau) \simeq 50\%,  \quad \quad  {\rm Br}(\tilde{\chi}_{1}^\pm \to \tilde{\nu}_1^{(\ast)} \tau^\pm) \simeq 100\%,
\end{eqnarray}
due to the Yukawa interaction $Y_\nu \,\hat{l} \cdot \hat{H}_u \,\hat{\nu}_R$ in the superpotential. In this case, the most promising way to explore the two scenarios at the LHC is to search the Di-$\tau$ plus missing momentum signal through the process $ p p \to \tilde{\chi}_1^\pm \tilde{\chi}_1^\mp \to (\tau^\pm E_{\rm T}^{\rm Miss}) (\tau^\mp E_{\rm T}^{\rm Miss}) $~\cite{Cao:2017cjf,Cao:2018iyk}. So far, the ATLAS collaboration has finished three independent analyses of the signal based on $ 20.3~{\rm fb}$ data at the 8~TeV LHC~\cite{tausLHC1}, $ 36.1~{\rm fb}$ data at the 13~TeV LHC~\cite{Aaboud:2017nhr}, and $139~{\rm fb}$ data at the 13~TeV LHC~\cite{Aad:2019byo}, respectively. We repeated these analyses by elaborated Monte Carlo simulations, like what we did for the first two analyses in~\cite{Cao:2017cjf,Cao:2018iyk}. We found that the tightest constraint on the two scenarios comes from the last analysis, and its efficiency in detecting the signal decreases gradually as the gap between $m_{\tilde{\nu}_1}$ and $m_{\tilde{\chi}_1^\pm}$ becomes narrow. As far as the light and massive $h_s$ scenarios are concerned, the analysis can not exclude at $95\%$ confidence level the points satisfying $m_{\tilde{\nu}_1} \gtrsim 100~{\rm GeV}$ and $m_{\tilde{\nu}_1} \gtrsim 200~{\rm GeV}$, respectively. So we conclude that the LHC analyses do not affect the results presented in this work.

\section{\label{Section-Conclusion}Conclusion}

Motivated by the increasingly tight limitation of the DM DD experiments on the traditional neutralino DM in the natural MSSM and NMSSM, we extended the NMSSM by the inverse seesaw mechanism to generate the neutrino mass in our previous studies~\cite{Cao:2017cjf,Cao:2019ofo,Cao:2019qng}, and studied the feasibility that the lightest sneutrino acts as a DM candidate. A remarkable conclusion for the theory is that experimental constraints from both the collider and DM search experiments are relaxed significantly. Consequently, large parameter space of the NMSSM that has been experimentally excluded resurrects as physical points in the extended theory. In particular,  the higgsino mass may be around 100 GeV to predict Z-boson mass naturally. This feature makes the extension attractive and worthy of a careful study.

We realized that sizable neutrino Yukawa couplings $\lambda_\nu$ and $Y_\nu$ contributed significantly to the DM-nucleon scattering rate. Thus, the recent XENON-1T experiment could limit them. We also realized that the unitarity in the neutrino sector set a specific correlation between the couplings $\lambda_\nu$ and $Y_\nu$, which in return limited the parameter space of the ISS-NMSSM. Since these issues were not studied before, we investigated the impact of the leptonic unitarity and current and future DM DD experiments on the sneutrino DM sector in this work. Specially, we considered the light and massive $h_s$ scenarios after noticing that the singlet dominated Higgs plays a vital role in both the DM annihilation and the DM-nucleon scattering. For each scenario, we studied the $B_{\mu_X} \neq 0$ and $B_{\mu_X} = 0$ case separately. Their difference comes from that Z boson can mediate the DM-nucleon scattering for the $B_{\mu_X} = 0$ case, and thus, the experimental constraints on it are much tighter.

In this work, we encoded the experimental constraints in a likelihood function and performed sophisticated scans over the vast parameter space of the model by the Nested Sampling method. The results of our study are summarized as follows:
\begin{itemize}
\item The XENON-1T experiment set an upper bound on the couplings $\lambda_\nu$ and $Y_\nu$, and the future LZ experiment will improve the bound significantly. The limitation is powerful when $h_s$ is light and contains sizable doublet components.
\item As an useful complement to the DM DD experiments, the unitarity always plays a vital role in limiting $Y_\nu$. It becomes more and more powerful when $v_s$ approaches $v$ from top to bottom.
\item The parameter space favored by the DM experiments shows a weak dependence on the left-handed slepton soft mass $m_{\tilde{l}}$. This property implies that one may fix $m_{\tilde{l}}$ in surveying the phenomenology of the ISS-NMSSM by scanning intensively its parameters and considering various experimental constraints. This treatment does not affect the comprehensiveness of the results.
\item The DM experiments tightly limit the left-handed sneutrino component in the sneutrino DM, e.g.,  if one considers the XENON-1T experiment's results, $|V_{11}| \lesssim 0.15$ for the $B_{\mu_X} \neq 0$  case and  $|V_{11}| \lesssim 0.02$   for the $B_{\mu_X} = 0$ case; these upper bounds become 0.10 and 0.01, respectively, if one adopts the LZ experiment's sensitivity.
\end{itemize}

Finally, we briefly discuss the phenomenology of the ISS-NMSSM. The sparticles's signal in this theory may be distinct from those in traditional supersymmetric theories, and so is the strategy to look for them at the LHC. This feature can be understood as follows: since the sneutrino DM carries a lepton number, and in most cases has feeble interactions with particles other than the singlet-dominated Higgs boson and the massive neutrinos, the sparticle's decay chain is usually long, and its final state contains at least one $\tau$ or $\nu_\tau$. In addition, the decay branching ratio depends not only on particle mass spectrum but also on new Higgs couplings, such as $Y_\nu$ and $\lambda_\nu$. As a result, sparticle's phenomenology is quite complicated~\cite{Cao:2019qng,Cao:2019ofo}. Depending on the mechanism by which the DM obtained the correct density, one usually encounters the following two situations:
\begin{itemize}
\item The DM co-annihilated with the Higgsino-dominated particles.  This situation requires the mass splitting $\Delta\equiv m_{\tilde{\chi}_1^0} - m_{\tilde{\nu}_1}$ to be less than about $10~{\rm GeV}$. Consequently, the Higgsino-dominated particles usually appear as missing momentum at the LHC due to the roughly degenerate mass spectrum. As  pointed out in~\cite{Cao:2019qng}, this situation's phenomenology may mimic that of the NMSSM with the Higgsino-dominated $\tilde{\chi}_1^0$ as a DM candidate.
\item The singlet-dominated particles $\tilde{\nu}_1$, $h_s$, $A_s$, and $\nu_h$ compose a secluded DM sector where the DM was mainly annihilated by any of the channels $\tilde{\nu}_1 \tilde{\nu}_1^\ast \to A_s A_s, h_s h_s, \nu_h \bar{\nu}_h$. It communicates with the SM sector by the Higgs-portal or the neutrino-portal. As we introduced before, this situation constrains the Yukawa coupling $\lambda_\nu$ tightly in getting the measured density, but it has no limitation on the splitting between $m_{\tilde{\nu}_1}$ and the Higgsino mass. As mentioned before, the signals of the sparticles in this situation are complicated. However, systematic researches on this subject are still absent.
\end{itemize}

We suggest experimentalists to look for the $2 \tau$ plus missing momentum signal of the process $ p p \to \tilde{\chi}_1^\pm \tilde{\chi}_1^\mp \to (\tau^\pm E_{\rm T}^{\rm Miss}) (\tau^\mp E_{\rm T}^{\rm Miss}) $ in testing the theory. Unlike the colored sparticles that may be very massive, light Higgsinos are favored by natural electroweak symmetry breaking. As a result, they are expected to be richly produced at the LHC. For the secluded DM case, ATLAS analyses have excluded some parameter space discussed at the end of the last section. With the advent of the LHC's high luminosity phase, more parameter space will be explored. For example, we once compared the ATLAS analyses of the signal at the 13 TeV LHC with $36.1~{\rm fb^{-1}}$ and $139~{\rm fb^{-1}}$ data~\cite{Aaboud:2017nhr,Aad:2019byo}. We found the excluded region on $m_{\tilde{\nu}_1} - m_{\tilde{\chi}_1^\pm}$ plane expanded from $m_{\tilde{\chi}_1^0} \lesssim 45~{\rm GeV}$ to $m_{\tilde{\chi}_1^0} \lesssim 110~{\rm GeV}$ for $m_{\tilde{\chi}_1^\pm} = 200~{\rm GeV}$, and from $m_{\tilde{\chi}_1^0} \lesssim 120~{\rm GeV}$ to $m_{\tilde{\chi}_1^0} \lesssim 200~{\rm GeV}$ for $m_{\tilde{\chi}_1^\pm} = 300~{\rm GeV}$. Concerning the co-annihilation case, it is hard for the LHC to detect the signal due to the compressed spectrum, but the future International Linear Collider may be capable of doing such a job (see, for example, the study in~\cite{Baer:2019gvu} for the compressed spectrum case). We emphasize that, different from the prediction of the MSSM, $m_{\tilde{\chi}_1^\pm}$ may be significantly larger than $m_{\tilde{\chi}_1^0}$ in the ISS-NMSSM due to the mixing of $\tilde{H}_{u,d}$ with $\tilde{S}$ in Eq.~(\ref{eq:MN}). As a result, the splitting between $m_{\tilde{\chi}_1^\pm}$ and $m_{\tilde{\nu}_1}$ can reach $20~{\rm GeV}$ (see the points in Table \ref{table2}), and it becomes even larger as the parameter $\lambda$ increases. This feature is beneficial for the signal's detection.

\section*{Acknowledgement}

This work is supported by the National Natural Science Foundation of China (NNSFC) under grant No. 11575053 and 12075076.


\end{document}